\documentclass[twocolumn]{aastex631}

\shorttitle{Density Structure of the Milky Way Disk}
\shortauthors{Imig et al. 2025}
\graphicspath{{./}{}}
\usepackage{animate}
\usepackage{amsmath}
\usepackage{longtable}
\usepackage{xcolor}
\usepackage{soul}
\usepackage[normalem]{ulem}
\usepackage [autostyle, english = american]{csquotes}
\MakeOuterQuote{"}

\begin{document}

\title{A Galactic Self-Portrait: Density Structure and Integrated Properties of the Milky Way Disk}

\correspondingauthor{Julie Imig}
\email{jimig@stsci.edu}

\author[0000-0003-2025-3585]{Julie Imig}
\affiliation{Space Telescope Science Institute, 3700 San Martin Drive, Baltimore, MD 21218, USA}
\affiliation{Department of Astronomy, New Mexico State University, P.O.Box 30001, MSC 4500, Las Cruces, NM, 88003, USA}

\author[0000-0002-9771-9622]{Jon A. Holtzman}
\affiliation{Department of Astronomy, New Mexico State University, P.O.Box 30001, MSC 4500, Las Cruces, NM, 88003, USA}

\author[0000-0001-6761-9359]{Gail Zasowski}
\affiliation{Department of Physics and Astronomy, University of Utah, 115 S. 1400 E., Salt Lake City, UT 84112, USA}

\author[0000-0001-5258-1466]{Jianhui Lian}
\affiliation{South-Western Institute for Astronomy Research, Yunnan University, Kunming, Yunnan 650091, People's Republic of China}

\author[0000-0002-9119-292X]{Nicholas F. Boardman}
\affiliation{School of Physics and Astronomy, University of St Andrews, North Haugh, St Andrews KY16 9SS, UK}

\author[0000-0003-4761-9305]{Alexander Stone-Martinez}
\affiliation{Department of Astronomy, New Mexico State University, P.O.Box 30001, MSC 4500, Las Cruces, NM, 88003, USA}

\author[0000-0001-8108-0935]{J. Ted Mackereth}
\affiliation{Department of Astronomy and Astrophysics, University of Toronto, 50 St. George Street, Toronto, ON, M5S 3H4, Canada}

\author[0000-0001-8302-0565]{Moire K. M. Prescott}
\affiliation{Department of Astronomy, New Mexico State University, P.O.Box 30001, MSC 4500, Las Cruces, NM, 88003, USA}

\author[0000-0002-1691-8217]{Rachael L. Beaton}
\affiliation{Space Telescope Science Institute, 3700 San Martin Drive, Baltimore, MD 21218, USA}
\affiliation{Department of Astrophysical Sciences, Princeton University, Princeton, NJ 08544, USA}
\affiliation{The Observatories of the Carnegie Institution for Science, 813 Santa Barbara St., Pasadena, CA~91101}

\author[0000-0003-4573-6233]{Timothy C. Beers}
\affiliation{Department of Physics and Astronomy and JINA Center for the Evolution of the Elements (JINA-CEE), University of Notre Dame, Notre Dame, IN 46556 USA}

\author[0000-0002-3601-133X]{Dmitry Bizyaev}
\affiliation{Apache Point Observatory and New Mexico State University, P.O. Box 59, Sunspot, NM, 88349-0059, USA}
\affiliation{Sternberg Astronomical Institute, Moscow State University, Moscow, 119234, Russia}

\author[0000-0003-1641-6222]{Michael R. Blanton}
\affiliation{Center for Cosmology and Particle Physics, Department of Physics, 726 Broadway, Room 1005, New York University, New York, NY 10003, USA}

\author[0000-0001-6476-0576]{Katia Cunha}
\affiliation{Steward Observatory, University of Arizona, Tucson, AZ 85721, USA}

\author[0000-0003-3526-5052]{Jos\'e G. Fern\'andez-Trincado}
\affiliation{Instituto de Astronom\'ia, Universidad Cat\'olica del Norte, Av. Angamos 0610, Antofagasta, Chile
}

\author[0000-0001-8245-779X]{Catherine E. Fielder}
\affiliation{Steward Observatory, University of Arizona, Tucson, AZ, 85721, USA}
\affiliation{Department of Physics and Astronomy, University of Pittsburgh, Pittsburgh, PA 15260, USA}

\author[0000-0001-5388-0994]{Sten Hasselquist}
\affiliation{Space Telescope Science Institute, 3700 San Martin Drive, Baltimore, MD 21218, USA}

\author[0000-0003-2969-2445]{Christian R. Hayes}
\affiliation{NRC Herzberg Astronomy and Astrophysics Research Centre, 5071 West Saanich Road, Victoria, B.C., Canada, V9E 2E7}

\author[0000-0003-0434-0400]{Misha Haywood}
\affiliation{GEPI, Observatoire de Paris, PSL Research University, CNRS, Sorbonne Paris Cité, 5 place Jules Janssen, 92190 Meudon, France}

\author[0000-0002-4912-8609]{Henrik J\"onsson}
\affil{Materials Science and Applied Mathematics, Malm\"o University, SE-205 06 Malm\"o, Sweden}

\author[0000-0003-1805-0316]{Richard R. Lane}
\affiliation{Centro de Investigación en Astronomía, Universidad Bernardo O'Higgins, Avenida Viel 1497, Santiago, Chile}

\author[0000-0003-2025-3147]{Steven R. Majewski}
\affiliation{Department of Astronomy, University of Virginia, Charlottesville, VA 22904, USA}

\author[0000-0001-8237-5209]{Szabolcs M{\'e}sz{\'a}ros}
\affiliation{ELTE E\"otv\"os Lor\'and University, Gothard Astrophysical Observatory, 9700 Szombathely, Szent Imre H. st. 112, Hungary}
\affiliation{MTA-ELTE Lend{\"u}let "Momentum" Milky Way Research Group, Hungary}

\author[0000-0002-5627-0355]{Ivan Minchev}
\affiliation{Leibniz-Institut fur Astrophysik Potsdam (AIP), An der Sternwarte 16, D-14482 Potsdam, Germany}

\author[0000-0002-1793-3689]{David L. Nidever}
\affiliation{Department of Physics, Montana State University, P.O. Box 173840, Bozeman, MT 59717, USA}

\author[0000-0003-4752-4365]{Christian Nitschelm}
\affiliation{Centro de Astronom{\'i}a (CITEVA), Universidad de Antofagasta, Avenida Angamos 601, Antofagasta 1270300, Chile}

\author[0000-0002-4989-0353]{Jennifer Sobeck}
\affiliation{Department of Astronomy, University of Washington, Box 351580, Seattle, WA 98195, USA}

\begin{abstract}

The evolution history of the Milky Way disk is imprinted in the ages, positions, and chemical compositions of individual stars. In this study, we derive the intrinsic density distribution of different stellar populations using the final data release of the Apache Point Observatory Galactic Evolution Experiment (APOGEE) survey. A total of 203,197 red giant branch stars are used to sort the stellar disk ($R \leq 20$ kpc) into sub-populations of metallicity ($\Delta$[M/H]$= 0.1$ dex), age ($\Delta \log(\frac{\textrm{age}}{\textrm{yr}})= 0.1$), and $\alpha$-element abundances ([$\alpha$/M]). We fit the present-day structural parameters and density distribution of each stellar sub-population after correcting for the survey selection function. The low-$\alpha$ disk is characterized by longer scale lengths and shorter scale heights, and is best fit by a broken exponential radial profile for each population.
The high-$\alpha$ disk is characterized by shorter scale lengths and larger scale heights, and is generally well-approximated by a single exponential radial profile.
These results are applied to produce new estimates of the integrated properties of the Milky Way from early times to the present day. 
We measure the total stellar mass of the disk to be $5.27^{+0.2}_{-1.5} \times 10^{10}$ M$_\odot$ and the average mass-weighted scale length is $R_{d} = 2.37 \pm 0.2$ kpc.
The Milky Way's present-day color of $(g-r) = 0.72 \pm 0.02$ is consistent with the classification of a red spiral galaxy, although it has only been in the "green valley" region of the galaxy color-mass diagram for the last $\sim 3$ Gyr.

\end{abstract}


\section{Introduction} \label{sec:intro}

The present-day kinematic and chemical structure of the Milky Way's stellar disk reflects the star formation history and evolution of our Galaxy. Fo r this reason, observational Galactic archaeology remains a cornerstone of galactic astronomy, placing key constraints on models of the formation and evolution of disk galaxies. The advent of large stellar spectroscopic surveys allows for the positions, stellar parameters, chemical compositions, and ages of hundreds of thousands of individual stars to be precisely measured in projects like APOGEE \citep{Majewski2017},  Gaia-ESO \citep{Gilmore_GaiaESO}, LAMOST \citep{LAMOST_Luo2015}, GALAH \citep{GALAH_Buder2018}, and RAVE \citep{RAVE_Steinmetz2020}. {Large photometric surveys like Gaia \mbox{\citep{GaiaDR2_Brown2018,GaiaDR3_Brown2021}} have observed billions of stars.} At the same time, large samples of other galaxies have revealed various fundamental relations across the galaxy population as a whole, accessible through spatially-resolved integral field unit (IFU) surveys like MaNGA \citep{MaNGAOverview}, CALIFA  \citep{CALIFA}, and SAMI \citep{Croom2021}.

Translating between the realms of Galactic and extragalactic astronomy, however, remains a challenge. 
While impressive sample sizes of other galaxies can be used to explore broad trends and relations within the galaxy population, we only have the one Milky Way, making it difficult to place within an extragalactic context. Our inside perspective of the Milky Way is both a benefit and a hindrance; the entire evolutionary history of the Milky Way can be accessed but it can not be directly compared to other galaxies. 

One of the largest obstacles for comparing Galactic and extragalactic observations is the selection function of astronomical surveys. Observational limitations severely influence what stars can be observed and where in the Galaxy. Despite the large numbers of stars observed in a survey, the sample is not representative of the entire Milky Way unless these selection biases are first corrected. The selection function, while important, is nontrivial to calculate 
leading many studies to ignore it entirely. The selection function depends on the targeting strategies and instrumentation specifications unique to each survey, requires some assumptions based on stellar evolution theory, and quantitative understanding of the complex distribution of dust in the Milky Way.

Once the survey selection function is accounted for, the effective sampling of the Milky Way expands from the limited solar vicinity into a representation of the entire Galaxy. These types of analyses are useful for measuring the Milky Way's structural parameters, including the disk scale length and scale height \citep[e.g.,][]{Bovy2012,Mackereth2017,Wang_2018,Yu2021,Lian2022}, and for estimating the total stellar mass of the different galactic components \citep[e.g.,][]{Bovy_2013, Mackereth_2020, Horta_2020}.

Density profile decompositions of the Milky Way disk near the Solar neighborhood led to the discovery of two geometric components; the "thin disk" and the "thick disk", which are characterized by a difference in exponential scale height as the name implies \citep[e.g.,][]{Yoshii1982, Gilmore1983,Robin2003,Juric_2008}. The two disks also differ in their dynamical signatures, with stars belonging to the thin disk generally having faster rotational velocities and less heated vertical orbits than thick disk members \citep[e.g.,][]{Soubiran2003,Juric_2008, Robin_2017, Mackereth_2019, Robin_2022}. Furthermore, the two disks have different chemical compositions; thin disk stars are generally younger, more metal-rich, and depleted in $\alpha$-element\footnote{$\alpha$-elements are elements with an atomic number multiple of 4 (the mass of a Helium nucleus, an $\alpha$-particle), e.g., O, Mg, S, Ca} abundances compared to thick disk members \citep[e.g.,][]{Furhmann1998, Bensby2005, Reddy2006, Bensby_2011, Cheng_2012, Bovy2012, Bovy2016b, Nidever2014, Hayden2015, Mackereth2017, Vincenzo2021, Katz_2021}. The origin of this structural, kinematical, and chemical bimodality in the disk remains unclear, and there is some debate on whether the two disks should be classified as different galactic components at all \citep[e.g.,][]{Bensby_2007,Bovy_2012,Kawata_2016,Hayden_2017,Anders_2018}. Nevertheless, due to these differences the two disk components are often considered independently of one another as two distinct components when fitting for the structural parameters of the Milky Way.

The vertical density profile of the Milky Way disk is also exponential. Near the solar neighborhood, the scale height of the thin disk is usually measured around 0.3 - 0.4 kpc, and the thick disk has a larger scale height around 0.75 - 1.3 kpc \citep[e.g.,][]{Gilmore1983,Bensby_2011,Cheng_2012,Lian2022}. This quantity does seem to have a radial dependence, with the scale height of some populations increasing with radius in a flared disk. There is some debate on which populations flare most strongly, with some studies finding the strongest flaring among the low-$\alpha$ thin disk populations \citep[e.g.,][]{Bovy2016b,Mackereth2017,Robin_2022} and others finding the strongest flaring in the high-$\alpha$ thick disk populations \citep[e.g.,][]{Yu2021,Lian2022}. This discrepancy has different implications for the formation history of the Milky Way, as some models predict that a flared thick disk arises under any realistic disk formation scenario \citep{Minchev_2015}, while others suggest that a strongly flared thick disk is often associated with more violent merger histories \citep{Garcia2020}.

Despite these advances in understanding the structural form of the Milky Way's stellar disk(s), these measurements are inherently different quantities than those observed in other disk galaxies. The Milky Way's structural parameters are measured through star counts, or its mass, while the structural parameters of other galaxies can only be measured through their light profiles \citep{Fathi_2010,Lange_2014}. This difference could explain the apparent discrepancies between the Milky Way and some of its peers, such as its scale length being too small for its mass \citep[e.g.,][]{Boardman_2020}.

In this paper, we explore the physical structure of the Milky Way disk and how its properties compare in integrated light. Section \ref{sec:data} introduces the data used in this analysis from the APOGEE survey. Section \ref{sec:method} summarizes the density structure fitting methodology we use, originally developed in previous literature \citep[e.g.,][]{Bovy_2012,Bovy_2012b,Rix2013,Bovy2016b,Mackereth2017}. Section \ref{sec:results} presents the best-fit structural parameters of each stellar population from the density fitting. These results are applied in section \ref{sec:integrated} to estimate some integrated properties of the Milky Way, including (g-r) colors and total star formation history. Section \ref{sec:discussion} discusses the implications of our results, and we conclude in Section \ref{sec:conclusions}.

\section{Data} \label{sec:data}

\subsection{APOGEE}\label{sec:data:APOGEE}

The Apache Point Observatory Galactic Evolution Experiment \citep[APOGEE;][]{Majewski2017} is a high-resolution ($R\sim22,500$) near-infrared ($1.51-1.70$ $\mu$m) spectroscopic survey containing observations of 657,135 unique stars released as part of the SDSS-IV survey \citep{Blanton_2017}. The spectra were obtained using the APOGEE spectrograph \citep{apogeespectrographs_Wilson2019} mounted on the $2.5$m Sloan Foundation telescope \citep{apo25m_Gunn2006} at Apache Point Observatory to observe the Northern Hemisphere (APOGEE-N). The survey was later expanded to include the southern hemisphere (APOGEE-S) using a second APOGEE spectrograph on the $2.5$ m Ir{\'e}n{\'e}e du Pont telescope \citep{lco25m_Bowen1973} at Las Campanas Observatory to observe the Southern Hemisphere (APOGEE-S). The targeting strategies for previous data releases of the APOGEE survey are described in \citet{Zasowski2013,Zasowski2017}, and for the final data release in \citet{Beaton_2021} (for APOGEE-N) and \citet{Santana_2021} (for APOGEE-S). The final version of the APOGEE catalog was published in December 2021 as part of the 17th data release of the Sloan Digital Sky Survey \citep[DR17;][]{SDSSdr17} and is available publicly online through the SDSS Science Archive Server and Catalog Archive Server\footnote{Data Access Instructions: \url{https://www.sdss.org/dr17/irspec/spectro_data/}}.

The APOGEE data reduction pipeline is described in \cite{Nidever_2015_ApogeeDataReduction}. Stellar parameters and chemical abundances in APOGEE were derived within the APOGEE Stellar Parameters and Chemical Abundances Pipeline \citep[ASPCAP;][J.A. Holtzman et al. in prep.]{Holtzman_2015,aspcap,Holtzman_2018,Jonsson_2020}. ASPCAP derives stellar atmospheric parameters, radial velocities, and as many as 20 individual elemental abundances for each APOGEE spectrum by comparing each to a multi-dimensional grid of theoretical model spectra \citep{Meszaros2012,Zamora2015} and corresponding line lists \citep{Shetrone_2015,Smith_2021}, employing a $\chi^2$ minimization routine with the code {\texttt{FERRE}} \citep{AllendePrieto_2006} to derive the best-fit parameters for each spectrum. We highlight that several elements were updated in DR17 to include non-LTE effects in the stellar atmosphere \citep{Osorio_2020}. ASPCAP reports typical precision in overall metallicity measurements within 0.01 dex \citep{aspcap2018}. In this study, we adopt the calibrated values for surface gravity ($\log g$), metallicity ([M/H]), and $\alpha$-element abundances ([$\alpha$/M]) from ASPCAP. 

{The ASPCAP abundances for metallicity [M/H] and alpha-element abundances [$\alpha$/M] are subtly different from the ASPCAP abundances [Fe/H] and [Mg/Fe]. To understand the difference, the ASPCAP pipeline can be summarized in two steps. First, the observed spectrum is compared to the model spectra grid to derive the fundamental atmospheric parameters, which include effective temperature, surface gravity, total scaled-solar general metallicity ([M/H]) and total alpha-element abundances ([$\alpha$/M]). These total abundances contain contributions from multiple elements; for example, [$\alpha$/M] encompasses six different elements (O, Mg, Si, S, Ca, and Ti). During the second step of ASPCAP, the pipeline extracts individual element abundances, one at a time, using small wavelength windows that have been optimized for each element. The measurements for [Fe/H] and [Mg/Fe] come from the second step. In this study, we adopt the \textit{total} values for metallicity ([M/H]) and $\alpha$-element abundances ([$\alpha$/M]). While the individual elemental abundances are more precisely measured, the total measurements are more directly applicable to how isochrone tracks and simple stellar population models are scaled, enabling us to pursue a primary goal of this study of comparing the MW to extragalactic populations. In practice, [M/H] is very similar to [Fe/H] and [$\alpha$/M] is very similar to [Mg/Fe], typically differing by less than 0.01 and 0.02, respectively; these differences are insignificant compared to our bin size of $\Delta$[M/H]=0.1.}

\subsection{Age and distance estimates}\label{sec:data:distmass}

In addition to the chemical abundances from ASPCAP, we adopt stellar age and distance estimates from the {\texttt{distmass}} value added catalog \citep{StoneMartinez_2023}.\footnote{distmass VAC: \url{https://www.sdss4.org/dr17/data_access/value-added-catalogs/?vac_id=distmass:-distances,-masses,-and-ages-for-apogee-dr17}}.

The {\texttt{distmass}} distances were derived through a neural network that was trained to estimate a star's luminosity based on its ASPCAP parameters, using Gaia distances and star cluster distances to provide the training labels. Distance estimates from the {\texttt{distmass}} catalog are typically precise within 10\%. These distances, along with the right ascension (RA) and declination (DEC) were used to derive Galactocentric positions for each star, with the reference location of the Sun defined to be $R_{\odot} = 8.3$ kpc with a height of $z_{\odot} = 0.027$ kpc above the plane \citep{BlandHawthorn2016}.

The stellar ages were similarly derived using a second neural network, working under the assumption that for evolved red giant stars, the chemical abundances of carbon and nitrogen provide information on the stellar mass due to the mass-dependence of stellar mixing \citep[e.g.,][]{Iben_1965,Salaris_2005,Shetrone_2019}. The neural network was trained on the ASPCAP parameters ($\log g$, [M/H], [C/Fe], [N/Fe]) of stars with known asteroseismic masses from the APOGEE-{\it Kepler} overlap survey \citep[APOKASC; ][Pinsonneault et al. 2023 in prep.]{Pinsonneault_2018}; then it predicts the masses for all giant stars from DR17. Knowing the masses for evolved stars, ages can be derived through stellar evolution theory that predicts a star's location on an isochrone. In the {\texttt{distmass}} catalog, isochrones from \cite{Choi_2016} were adopted to make this conversion from derived mass to stellar age. These isochrones cover a range of ages ($5.0 \leq \log_{10}($age [yr]$) \leq 10.3$), metallicities ($-2.0 \leq [Z/H] \leq 0.5$), and masses ($0.1 \leq M/M_{\odot} \leq 300$). The stellar age estimates in {\texttt{distmass}} report a typical uncertainty of $\pm 0.15 \log_{10}$(age [yr]), {or around $\sim3$ Gyr for a 10 Gyr old star}. {The \mbox{\cite{Choi_2016}} isochrones are solar-scaled in $\alpha$-element abundances, which could affect the age estimates. \mbox{\citet[][their Section 6.1.2]{Clontz_2024}} showed that a difference of 0.1 dex in $\alpha$-element abundances could change the derived age by up to 0.27 Gyr. While not completely negligible, this bias is smaller than the reported uncertainties in {\texttt{distmass}} and smaller than our bin size, and is therefore not expected to significantly impact our results.}

\begin{figure*}[ht]
    \centering
    \includegraphics[width=\textwidth]{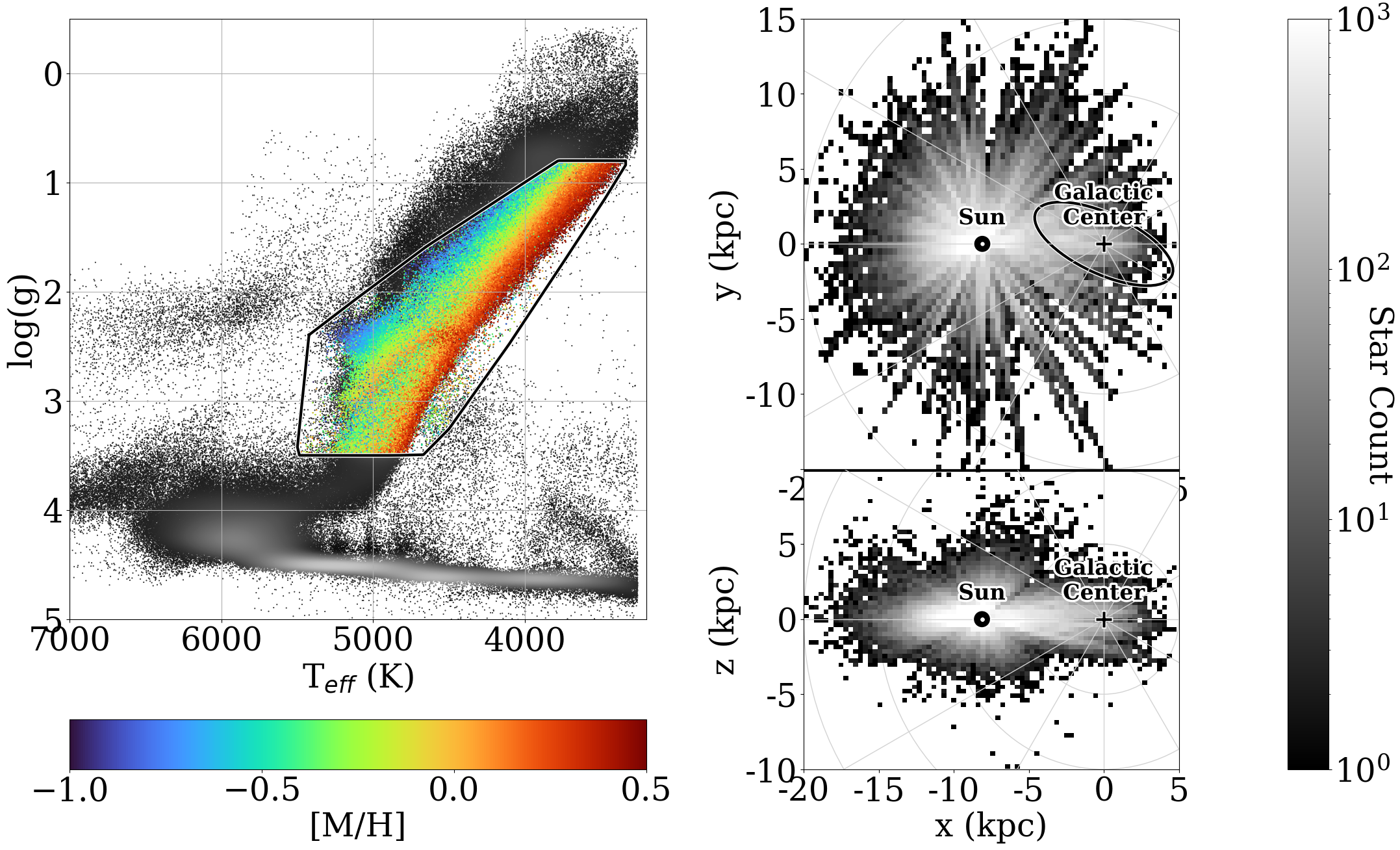}
    \caption{Summary of our red giant branch stellar sample from APOGEE. \textbf{Left:} The $T_{\rm eff}$-log$g$ distribution of our sample (with point color corresponding to metallicity) compared to the entire APOGEE catalog (in grayscale, density on the diagram). \textbf{Right:} The spatial distribution of our RGB sample, displayed as a face-on view of the Galactic disk (top), and an edge-on view (bottom). The plus (+) marks the location of the Galactic Center, the large circle ($\odot$) is the Sun. The ellipsoid is the approximate location of the Galactic bar.}
    \label{fig:data_sample}
\end{figure*}

\citet{StoneMartinez_2023} evaluate the {\texttt{distmass}} ages and their uncertainties by comparing to previous literature and find good agreement. Compared with a sample of stellar cluster members and their independent age estimates from main sequence turnoff fitting, \citet{StoneMartinez_2023} reports that the {\texttt{distmass}} ages are accurate within $\pm1\sigma = 0.16 $ in $\log_{10}$(age) across 12 different star clusters with ages $ 9.2 \leq \log_{10}$(age) $\leq 9.7$. Compared to a larger sample of field stars from the {\texttt{astroNN}} catalog \mbox{\citep{Leung_2018,Mackereth_2019}}, the {\texttt{distmass}} ages show a typical spread of $\pm1\sigma = 0.11 $ in $\log_{10}$(age). Both of these comparisons are consistent within the reported {\texttt{distmass}} age uncertainties.

For the remainder of this paper, we adopt the {\texttt{distmass}} results trained on the uncorrected "SS" ages from APOKASC 3 {\mbox{\citep{Pinsonneault_2025}}}, corresponding to the column named "{\texttt{AGE$\_$UNCOR$\_$SS}}" in the {\texttt{distmass}} catalog. {Stellar ages remain difficult to precisely constrain for large samples of stars, and there are some known systematic uncertainties associated with ages derived from asteroseismic training sets: For example, the age estimates are both less accurately and less precisely constrained for giants with $\log g \leq 2.0$, driven by the lack of asteroseismic data for the training set in this range. However, these systematics are not unique to the \mbox{\texttt{distmass}} catalog, and culling the sample by this value would severely limit the galactic distances we are able to probe in this study.} We refer the reader to \citet{StoneMartinez_2023} and \citet{Imig_2023} (their Appendix A) for a more detailed discussion on the various caveats associated with this stellar age catalog.

\begin{table*}
    \centering
    \begin{tabular}{llll}
         Catalog & Parameter & Range & Notes \\
         \hline \hline
         allStar & $\log g$ & $0.8 < \log g < 3.5 $ & restrict sample to red giant stars \\
         allStar & Targeting Bits & EXTRATARG$ = 0$ & main survey targets; no duplicates\\
         allStar & ASPCAP Bits & ASPCAPFLAG $\notin 23$& removes STAR$\_$BAD stars with bad ASPCAP parameters\\
         distmass & DISTMASS Bits & BITMASK $\notin 2$ & age estimates reliable; within training set range\\ &        
    \end{tabular}
    \caption{Summary of the various quality refinements made to our sample described in Section \ref{sec:data:selection}.}
    \label{tab:sample_refinement}
\end{table*}

\subsection{Sample Refinement}
\label{sec:data:selection}

A number of data and quality cuts were applied to the APOGEE data to refine our sample prior to the fitting. First, we restrict the sample to the "main survey targets", which are stars that were targeted randomly within the selection criteria and removes duplicates, using the {\texttt{EXTRATARG}} $ == 0$ flag in the APOGEE {\texttt{allStar}} catalog. The motivation for this cut, and more details on the APOGEE targeting strategies, selection function, and how it influences our methodology are discussed further in Section \ref{sec:method:selfunc}. In short, this cut is necessary to scale from the stars that APOGEE observed to the true distribution of stars in the Galaxy; the selection function allows for this correction, but only for stars that were targeted randomly.

Next, we restrict the sample to red giant stars between the limits $0.8 \leq \log g \leq 3.5$. Red giant stars are luminous, allowing the sample to probe far distances in the Galaxy. The stellar parameters in APOGEE are known to have systematic uncertainties that vary with $\log g$ and $T_{\rm eff}$ \citep[e.g.,][]{aspcap2018,Eilers2021}, the effects of which can be mitigated by limiting the sample to a small range in $\log g$. The average signal-to-noise ratio of our sample is {$\rm S/N = 222$ \mbox{\citep[per half resolution element;][]{Nidever_2015_ApogeeDataReduction}}.}

We also remove any stars with unreliable ASPCAP stellar parameters using the {\texttt{STAR$\_$BAD}} ASPCAP bit. This flag is triggered when the derived parameters for a star are designated a bad fit by its high $\chi^2$ value, when the derived temperature does not match the star's observed color, when any individual stellar parameter measurement is flagged as bad, or when the derived parameters lie on an edge of the synthetic spectral grid and are unphysical.

We make an additional cut for the stellar age estimates using the {\texttt{distmass}} quality flag bit 2, indicating that a star's atmospheric parameters are within the region covered by the training set, meaning the neural network is not allowed to extrapolate and the stellar age estimates are reliable \citep{StoneMartinez_2023}. This cut is partially redundant with the previous $\log g$ restriction, as the availability of asteroseismic data used to train the neural network is limited to $\log g \geq 0.8$ and age estimates are only available for giant stars due to the assumptions used in converting from stellar mass to age. Notably, this also removes all stars with metallicity [M/H] $\leq -0.7$, imposed by the parameter range covered by the training set. Metal-poor stars are expected to have extra mixing effects \citep[][e.g.,]{Shetrone_2019, Roberts_2024} that were not learned by the neural network because there are no metal-poor stars in the training set.

However, metal-poor ([M/H] $\leq -0.7$) stars make up a decent fraction of the high-$\alpha$ population (roughly $~10\%$), and therefore we do not want to exclude them from our sample. For the high-$\alpha$, metal-poor stars, we therefore assign an age estimate under the assumption that they follow the same age distribution as the rest of the high-$\alpha$ stars in the sample. This assumption is motivated by the predictions of chemical evolution models \citep[e.g.,][]{Chiappini1997, Kobayashi2006} as well as previous age-resolved observational studies of the Milky Way \citep[e.g.,][]{Haywood_2013,Feuillet2018,Lian2022} which generally agree that high-$\alpha$ stars are uniformly old independent of their metallicity. We adopt an assumed age estimate for these stars by sampling all high-$\alpha$ stars in the $-0.4 < \textrm{[M/H]} <-0.7$ range and use their ages from \verb|distmass| to determine the target age distribution. The metal-poor ([M/H] $\leq -0.7$) stars are then assigned an assumed age by re-sampling them to follow the target age distribution. 

A summary of the sample restrictions implemented in this paper is presented in Table \ref{tab:sample_refinement}. A Kiel diagram and the spatial distribution of our sample after these refinements is shown in Figure \ref{fig:data_sample}. {Our sample does not cover the entire width or height of the Red Giant Branch; these limits are imposed by the range of parameter space covered by the training set of the age catalog. However, the age and metallicity limits of our final sample represent $\geq 95\%$ of the stars in the Milky Way \mbox{\citep[e.g.,][]{Sharma_2011}}.} The {total} number of stars in our sample is \textcolor{black}{203,197}.

\subsection{Population Binning} \label{sec:data:populations}

To explore the history and evolution of the Milky Way, we further split our sample into \textbf{m}ono-\textbf{a}ge, mono-\textbf{a}bundance stellar \textbf{p}opulations (referred to here as "MAAPs") by groups of similar age, metallicity ([M/H]), and $\alpha$-element abundances ([$\alpha$/M]) to approximate simple stellar populations.

We define two bins in $\alpha$-element abundances, separated in the [$\alpha$/M]-[M/H] plane with the equation from \cite{Patil_2023} (their eq. 24): 

\begin{equation}
    \label{eq:alpha_split}
    \begin{split}
    {\rm [\alpha/M]} = 0.1754*{\rm [M/H]}^{3} + 0.1119*{\rm [M/H]}^{2}
    \\
    - 0.1253*{\rm [M/H]} + 0.1353 - \textit{0.05}
    \end{split}
\end{equation}

\begin{figure}
    \centering
    \includegraphics[width=0.5\textwidth]{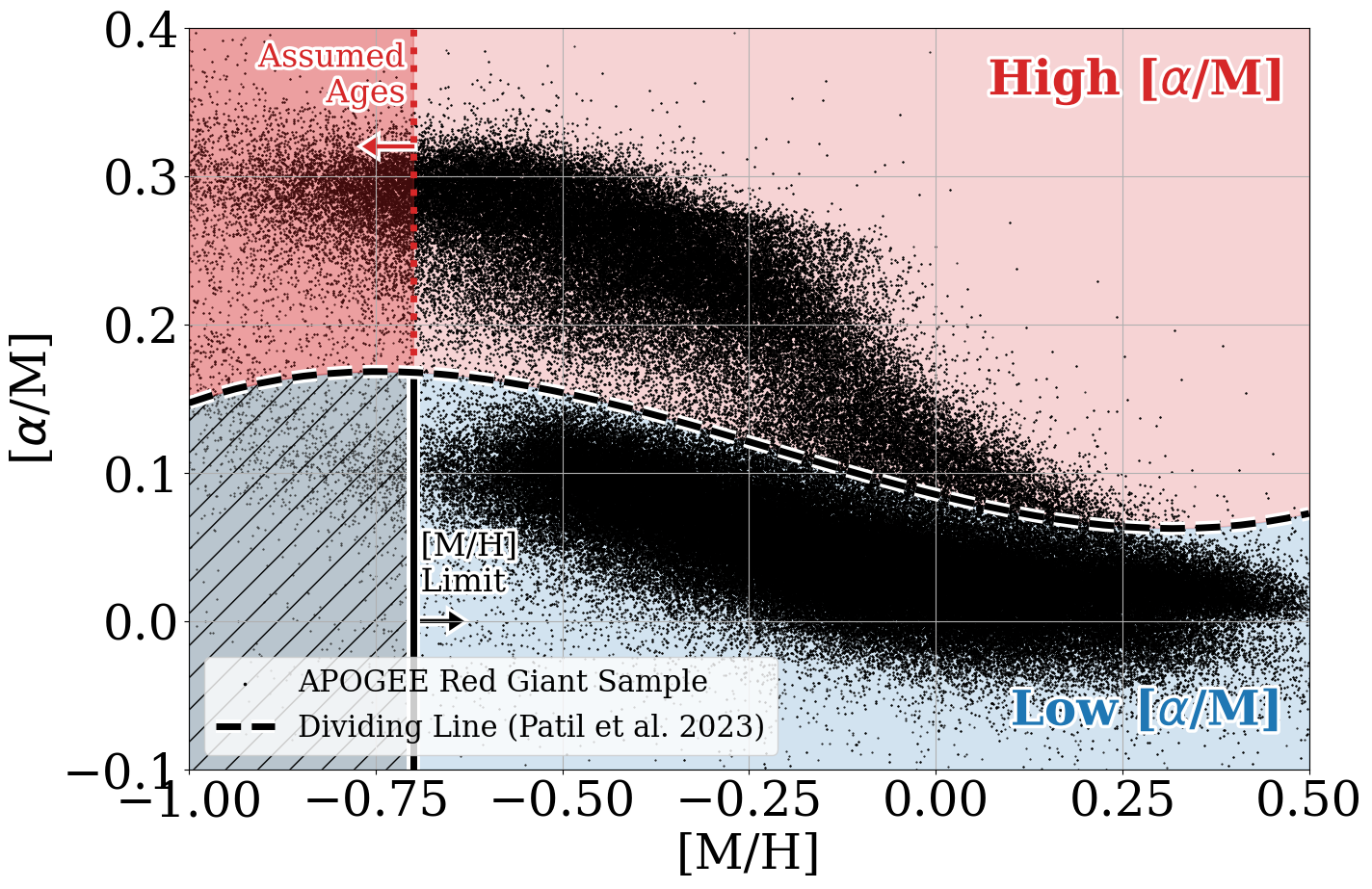}
    \caption{The metallicity ([M/H]) and alpha-element abundance ([$\alpha$/M]) distribution covered by our sample, showing the designated separation between low-$\alpha$ and high-$\alpha$ samples defined in Equation \ref{eq:alpha_split}. The vertical line at [M/H]$ = -0.7$ dex marks the lower metallicity limit of the \texttt{distmass} stellar age estimates.}
    \label{fig:alphacuts}
\end{figure}

This equation was derived using statistical techniques to separate the two populations in copula space, using a coordinate transformation to extract statistical structure from data. We added a small $-0.05$ dex offset (noted in italics) to account for abundance differences because \cite{Patil_2023} used [Mg/Fe] instead of [$\alpha$/M].

This separation is shown in Figure \ref{fig:alphacuts}. The two bins are hereafter referred to the low-$\alpha$ and high-$\alpha$ populations. Note that the low-$\alpha$ populations are not truly depleted in [$\alpha$/M] and may be more accurately described as a "solar-$\alpha$" group, but we adopt the monikers  low-$\alpha$ and high-$\alpha$ populations for simplicity.

In metallicity, we split the sample into 15 equal-sized bins ranging between $-1.0 \leq$ [M/H] $\leq 0.5$ with spacing of $\Delta$ [M/H] $=  0.1$ dex. This is significantly larger than the typical ASPCAP uncertainty in [M/H], which is generally on the order of $ 0.01$ dex. 

In stellar age, we define 12 bins between $9.0 \leq$ log$_{10}$(age [yr]) $\leq 10.2$ (corresponding to $ 1.0 \leq$ age $\leq 15.8$ Gyr) with spacing of $\Delta$ log$_{10}$(age) $=  0.1$. The age bins are handled in log space because it is more representative of the expected uncertainties associated with the stellar age estimates, which are typically around log$_{10}$(age) $= 0.15$. The age bin size is {smaller} than the age uncertainties, so there will be some expected overlap between bins and the results may be artificially broadened in this direction. The youngest age bin goes only down to 1.0 Gyr, as there are not enough young stars in our red giant sample to extend the range lower. In {\mbox{\citet{zasowski2025}}}, younger MAAPs are generated and shown not to significantly change the main conclusions of this work. 

\begin{figure*}
    \centering
    \includegraphics[width=\textwidth]{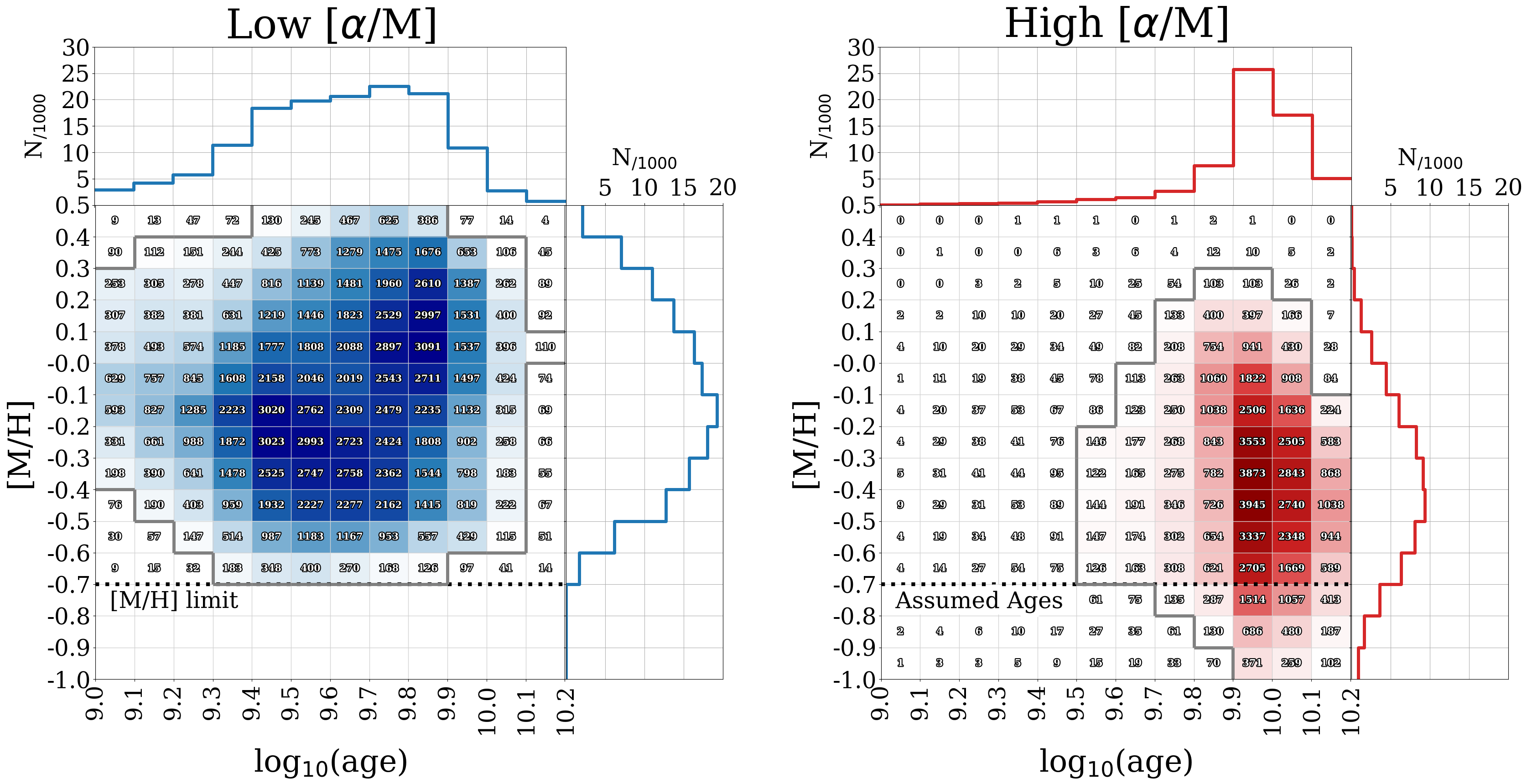}
    \caption{The number of stars in our sample for each stellar population bin, separated by the low-$\alpha$ (blue; left) and high-$\alpha$ (red; right) samples and different bins in stellar age (x-axis) and metallicity (y-axis). The number of stars in each population is printed in the corresponding bin and represented by color intensity. The dark gray outline highlights bins with more than 100 stars, which is the limit for our parameter fitting. Flattened histograms of each axis are shown above (age distribution) and to the right (metallicity distribution) for each panel.}
    \label{fig:bincounts}
\end{figure*}

The number of stars from our sample in each resulting mono-age mono-abundance stellar population is shown in Figure \ref{fig:bincounts}. For the remainder of this paper, each bin is considered a single stellar population and worked with independently of the rest unless otherwise specified.

\section{Method} \label{sec:method}

In the methodology developed by \citet{Bovy2012,Bovy_2013,Rix2013, Bovy2016a,Bovy2016b}, the probability that any given star was observed in an astronomical survey is directly proportional to the overall density of the Galaxy at the star's location, after accounting for the survey selection function which includes the distribution of dust. Using this technique, the density structure and stellar mass contribution can be recovered for various components of the Galaxy \citep[e.g.,][]{Bovy2012,Rix2013,Bovy2016a, Bovy2016b,Mackereth2017,Mackereth_2020,Yu2021,Lian2022}, revealing the history and evolution of the Milky Way through the present-day distribution of its stellar populations.

In this study, we follow this approach and fit for the density distribution of each stellar population bin defined in Section \ref{sec:data:populations} independently. This allows us to explore how the structural parameters of the Milky Way vary with stellar age, metallicity, and $\alpha$-element abundance. In this section, we provide an overview of the methodology in Section \ref{sec:method:densityfitting}, describe the APOGEE selection function in \ref{sec:method:selfunc}, present our density models and corresponding parameters in \ref{sec:method:densitymodels}, and summarize the practical computational application of this procedure in \ref{sec:method:application}.

\subsection{Density Fitting Methodology} \label{sec:method:densityfitting}

In a randomly-selected sample of stars, the probability that any given star was observed is proportional to the overall stellar number density of the Galaxy at the star's location. Therefore, the number counts of stars observed in a stellar survey like APOGEE can be used to fit a density model with generic parameters $\theta$ to recover the underlying density distribution for each sub-population of stars.

\cite{Bovy_2012b,Rix2013,Bovy2016a, Bovy2016b, Mackereth2017} have developed a methodology to determine the underlying density profile of a population of stars based on this probability. The rate at which stars have been observed in a stellar survey like APOGEE, as a function of position, color, and metallicity, is well-modeled by an inhomogeneous Poisson point process. The remainder of this section restates the methodology from these studies, which we adopt to fit the density parameters of the stellar populations in this paper. The expected rate of observation in number counts $\lambda(O|\theta)$ for a given stellar number density model $\nu_*$ with parameters $\theta$ is expressed as:

\begin{equation}
\label{eq:rate}
\begin{split}
    \lambda(O|\theta) = & \ \nu_{*}(X,Y,Z\ |\  \theta) \times |J(X,Y,Z;\ l,b,D)| \\
    & \times \rho(H, [J-K_{S}]_{0}, [\text{Fe/H}]\ |\ X,Y,Z)\\
    & \times S(l,b,H)
\end{split}
\end{equation}

\noindent where $\nu_{*}(X,Y,Z\ |\  \theta)$ is the number density of stars (with units of stars kpc$^{-3}$) in Galactocentric rectangular coordinate. This is the quantity we wish to match to the observed number counts in APOGEE. The exact form of the density model we adopt is presented later in Section \ref{sec:method:densitymodels}, but we continue this discussion with a generalized density model with arbitrary parameters $\theta$. The second term, $|J(X,Y,Z;\ l,b,D)|$ is the Jacobian transformation from rectangular Galactocentric coordinates $(X,Y,Z)$ to the observable Galactic latitude, longitude, and distance coordinates $(l,b,D)$. The third term, $\rho(H, [J-K_{S}]_{0}, [\text{Fe/H}]\ |\ X,Y,Z)$ is the assumed density of stars in magnitude, color, and metallicity space from stellar evolution theory. Finally, $ S(l,b,H)$ is the survey selection function, the fraction of the total underlying population of stars that was observed by APOGEE for each magnitude, color, and metallicity bin. The selection function includes the effects of dust extinction, and is discussed in more detail in Section \ref{sec:method:selfunc}.

Finding the density model that best matches the observed data becomes a maximum-likelihood problem matching the rate function to the observed number counts. For each population of stars, the likelihood $\mathcal{L}(\theta)$ that an underlying density model with parameters $\theta$ produced the observed number counts is written as: 

\begin{equation}
    \label{eq:likelihood}
    \ln{\mathcal{L}(\theta)} = \sum_{i}\bigg[ \ln{\nu_{*}(X_{i},Y_{i},Z_{i}\ |\ \theta)} - \ln{\int{dO d\lambda(O|\theta)}} \bigg]
\end{equation}

\noindent where $\nu_{*}(X_{i},Y_{i},Z_{i}\ |\ \theta)$ is evaluated at the position of each observed star $i$ and summed over all observations. The second term, $\int{dO d\lambda(O|\theta)}$, is the "effective volume" of the survey, or the predicted total number of stars that APOGEE would have observed in a galaxy with density model parameters $\theta$, given the survey selection function and the effects of dust. This term is necessary to regularize the likelihood $\mathcal{L}(\theta)$; without it, the density model with the highest likelihood would always be the model with the highest overall density. The effective volume is an intrinsic property of the survey for each given density model, and is independent of the observed number counts. For a pencil-beam style survey like APOGEE, which observed stars along targeted lines-of-sight (or "fields"), the effective volume can be expressed as:

\begin{equation}
    \label{eq:effvolume}
    \begin{split}
    \int{dO d\lambda(O|\theta)} = & \\ \sum_{\text{fields}} \Omega_{f} &  \int{dD D^2}\nu_{*}([X,Y,Z](D,\text{field})|\theta)\\
    & \times \mathfrak{S}(\text{field},D)
    \end{split}
\end{equation}

\noindent where $\Omega_{f}$ is the solid angle covered by each APOGEE field. The number density $\nu_{*}([X,Y,Z](D,\text{field})|\theta)$ is the density model evaluated along the line-of-sight (distance $D$) for a given field, the integral of which gives the total number of stars expected in the field for that set of $\theta$. Finally, that number is multiplied by the effective survey selection function $\mathfrak{S}(\text{field},D)$, the fraction of stars out of the total number present that would actually be observed, accounting for survey observation strategies and the presence of dust. This important term is non-trivial to calculate and explained in more detail in Section $\ref{sec:method:selfunc}$.

For practical application in this study, we derive the best-fit parameters $\theta$ for each stellar population bin using a Markov-Chain Monte Carlo (MCMC) sampling of the likelihood function (equation \ref{eq:likelihood}), described more in Section \ref{sec:method:application}.

\subsection{APOGEE Selection Function} \label{sec:method:selfunc}

A selection function is broadly defined as the fraction of stars (or more generalized "targets") observed by an astronomical survey out of the total intrinsic number of targets existing, often presented as a function of spatial position. There are two main complications to account for in the calculation of the selection function for APOGEE: the targeting strategies of the survey, and the presence of interstellar dust in the Galaxy which obscure potential observations. The targeting strategies of the survey produce what will hereafter be referred to as the \textit{raw selection function}, the fraction of stars observed in APOGEE as a function of Galactic location $(l,b)$ (or field in APOGEE), apparent magnitude $H$, and $(J-K)_{0}$ color. The quantity needed to calculate the effective volume in Equation \ref{eq:effvolume} is called the \textit{effective selection function}, which is the fraction of stars observed as a function of Galactic location $(l,b)$ (field), and distance along the line-of-sight. The raw selection function is essentially a 2D sky projection, whereas the effective selection function is a 3D Galactic map. The calculation of the effective selection function requires the raw selection function, a three-dimensional dust map, and some assumption of the expected distribution of stars from stellar evolution theory (i.e., from a set of isochrones).

To evaluate the raw selection function of APOGEE, the targeting procedures of APOGEE must be well understood. Full details on the targeting strategies and motivations can be found in \citet{Zasowski2013,Zasowski2017} for previous data releases of APOGEE, \citet{Beaton_2021} for the final data release from the northern hemisphere (APOGEE-N), and \citet{Santana_2021} for the southern hemisphere (APOGEE-S). A brief overview of APOGEE's targeting procedure, as relevant to this study, is described here.

Any application of the raw selection function inherently assumes stars were randomly targeted, which is only true for the "main survey sample" in APOGEE. Other observations which include ancillary program targets observed for a specific science purpose (e.g., satellite or dwarf galaxy targets, star cluster member candidates, {\it Kepler} Objects of Interest) were removed from our sample previously with the cuts described in Section \ref{sec:data:selection}. Main-survey APOGEE targets are randomly selected from the 2MASS catalog \citep{2MASS_Skrutskie_2006}, separated into different lines-of-sight across the sky (referred to as fields) and in bins of apparent magnitude $H$, and dereddened $(J-K)_{0}$ color. 

The magnitude bins are referred to as $cohorts$ in APOGEE. Each field can have up to 3 cohorts, referred to as the short, medium, and long cohort, usually corresponding to the brightest (short) to faintest (long) groups of stars respectively, which need different amounts of telescope time for adequate signal-to-noise. The exact $H$-band magnitude limits that define a cohort can vary on a field-to-field basis. 

After binning by magnitude, the sample is further split into bins of dereddened $(J-K)_{0}$ color to select targets for APOGEE. Each field can have up to 2 bins in color space, which vary between fields depending on which Galactic component (disk, halo, or bulge) is sampled by the direction of the field. In APOGEE-1 (observations from 2014 and prior), a single color limit of $(J-K)_{0} \geq 0.5 $ was used for fields that targeted the Galactic disk (defined by latitude $|b| \leq 25^{\circ}$). In APOGEE-2, two different limits of $ 0.5 \leq (J-K)_{0} \leq 0.8 $ and $(J-K)_{0} \geq 0.8$ were imposed for the disk to increase the number of distant stars sampled. In both iterations of the survey, fields in the direction of the Galactic halo (latitude $|b| \geq 25^{\circ}$) were selected with a bluer color limit of $(J-K)_{0} \geq 0.3$. Halo-defined fields frequently include foreground disk stars which are included in this study. Bulge fields towards the center of the galaxy have a single color limit of $(J-K)_{0} \geq 0.5 $. For every APOGEE field, in each of these bins in color-magnitude space, targets were randomly selected from a list of all possible sources in the 2MASS catalog \citep{2MASS_Skrutskie_2006}. 

We evaluate the raw selection function for each field in the same color-magnitude bins as used for targeting. The formulation of the raw selection function $S(\text{field},k)$ is straightforward, as the ratio of number of observed stars $N_{\text{APOGEE}}$ to the number of all possible targets $N_{\text{2MASS}}$:

\begin{equation}
    \label{eq:rawsel}
    S(\text{field},k) = \frac{N_{\text{APOGEE}}}{N_{\text{2MASS}}}
\end{equation}

\noindent where $k$ is a color-magnitude bin defined by the targeting strategies on a field-by-field basis, $N_{\text{APOGEE}}$ is the number of stars for which APOGEE spectra were obtained, and $N_{\text{2MASS}}$ is the total number of possible targets from the 2MASS photometry catalog, assumed to be complete within the APOGEE magnitude limits. We calculate this fraction for every field, color, and magnitude combination using the publicly available {\texttt{apogee}}\footnote{\url{https://github.com/jobovy/apogee}} Python module, originally developed in \citet{Bovy2016a} and adapted for DR17 in this work.\footnote{\url{https://github.com/astrojimig/apogee/tree/dr17-selection}}. {The code updates for DR17 include:}

\begin{itemize}
    \item{{Incorporating all new plates, fields, and location designs observed in DR17. Many of these new fields cover the important inner regions of the galaxy $R<3$ kpc}}
    \item{{Updating column names and file paths that were changed in DR17 (for example: the ALL\_VISITS column was replaced by VISIT\_PK, used for cross-matching the allStar and allVisit files.)}}
    \item{{Increasing the limit on the number of plates per location as defined in the survey (previously, a single location in the sky could have up to 20 plates assigned, but that limit has been increased to 50).}}
\item{{Miscellaneous efficiency improvements and bug fixes.}}
    \item{{Adding the distmass VAC as a new option for stellar ages, with new functions to download and read from the distmass VAC.}}
\end{itemize}

The APOGEE raw selection function is used as one ingredient in the effective selection function, which transforms the selection fraction in color-magnitude bins into a selection fraction purely based on position in the Galaxy (field location and distance). For each field, the effective selection function is evaluated at many distances along the line-of-sight, and assumed to be constant across the solid angle area covered by the field. The effective selection function is evaluated by summing across all targeting color-magnitude bins as:

\begin{equation}
    \label{eq:effsel}
    \mathfrak{S}(\text{field},D) = \sum_{k}  S(\text{field},k) \int{\frac{\Omega_{k}(M_H)}{\Omega_{f}} dM_{H}}
\end{equation}

\noindent where $S(\text{field},k)$ is the raw selection function described previously. $M_{H}$ is the expected distribution of absolute magnitudes for a stellar population at distance $D$, estimated from stellar evolution theory using a set of isochrones. In this work, we use a set of the PARSEC isochrones \citep{Bressan_2012, Chen_2014}, using appropriate age and metallicity selections for each stellar population bin we wish to fit and adopting a Kroupa initial mass function \citep[][]{Kroupa_2002}. The metallicity and age range of the isochrones is the same as our stellar population bins, with a metallicity range of $-0.7 \leq [M/H] \leq 0.5$ and an age range of  $9.0 \leq \log (\textrm{age}) \leq 10.2$ in step size of $\Delta [M/H]  = \Delta \log (\textrm{age}) = 0.1$. Finally, dust is accounted for in the last term, $\frac{\Omega_{k}}{\Omega_{f}}$, which is the area percentage on the plate that a star would be observable given the distribution of dust in that part of the Galaxy. $\Omega_{f}$ is the total solid angle covered by the field, and $\Omega_{k}$ is the observable area of the field not obscured by dust for a given distance, defined in more detail as:

\begin{equation}
    \label{eq:obsarea}
    \begin{split}
    \Omega_{k}(M_H) = \Omega( H_{min} < m_{H} < H_{max}) \\
    m_{H} = M_{H} + \mu(D) + A(l,b,D)
    \end{split}
\end{equation}

\noindent where $H_{min,max}$ denotes the apparent magnitude limits for each APOGEE cohort and color bin $k$ from targeting. The distribution of apparent magnitude $m_{H}$ is calculated from absolute magnitudes $M_{H}$ of the stellar population from the isochrones, combined with the distance modulus $\mu(D)$ and the extinction value $A_{H}(l,b,D)$ retrieved from a dust map.

Putting this into plain words, the $\frac{\Omega_{k}}{\Omega_{f}}$ term of the effective selection function  tests where a hypothetical star (or a distribution of isochrone points) would be observable within APOGEE by moving it along line-of-sight through the dust map. If the star is too close to the Sun, its apparent magnitude would be brighter than the APOGEE targeting limit ($m_H < H_{\textrm{min}}$) and it would not have been selected for observation. If the star is too far away or obscured by dust, it would also not be observable ($m_H < H_{\textrm{max}}$), as the magnitude or color would once again fall outside APOGEE's targeting range. Because the dust maps are typically resolved within the area of an APOGEE field (solid angle $\Omega_f$), this "test" results in a fraction of observability (i.e., the star could have been observed if it was on this 50\% of the plate) instead of the binary used in this simplified example (i.e., the hypothetical star either was or was not observed).

This "observability test" is repeated for the full distribution of isochrone points and summed over the integral in equation \ref{eq:effsel}. This integral is deceptively complex, as each isochrone point has to be weighted by its contribution to the stellar population. This includes the choice of an assumed initial mass function (IMF), for which we elect to use the two-part power law Kroupa IMF \citep{Kroupa_2002}. The integration over the full population goes as:

\begin{equation}
\int \frac{\Omega_{k}(M_H)}{\Omega{f}} dM_{H} =  \sum_{i} \frac{\Omega_{k}(M_H)}{\Omega{f}} *
 \xi(m_{i}) \label{eq:int_imf}
\end{equation}

where the sum over $i$ corresponds to each isochrone point. The initial mass function comes in $\xi(m_{i})$, which is the number of expected observations of each star per unit stellar population mass, evaluated for the stellar mass of each isochrone point. Practically, this quantity is provided by the difference between two consecutive values in the {\texttt{INT$\_$IMF}} column in the PARSEC isochrones that we use \citep{Bressan_2012, Chen_2014}. This sum weights each isochrone point by its contribution to the IMF: lower mass stars are more numerous, so they will contribute more to the effective selection function than their high-mass siblings.

The dust map for determining extinction values $A_{H}(l,b,D)$ is another important choice to make when computing the effective selection function. No single 3D extinction map covers the whole sky, therefore a combination of several maps is necessary to sample the entire Galaxy. We use a combination of three dust maps \citep{Drimmel_2003, Marshall_2006,Green_2019} distributed as the "Combined 2019" map in the {\texttt{mwdust}}\footnote{\url{https://github.com/jobovy/mwdust}} package, first described in \citet{Bovy2016a}. For readers interested in how the choice of dust map or the uncertainties in the extinction values may influence the selection function, we point to \citet{Bovy2016a} (their Section 4.2 and Figure 6). Typical uncertainties for the effective selection function due to the choice of dust map are on the order of $\sim 10^{-3}$.

The final effective selection fraction $\mathfrak{S}(\text{field},D)$ gives the ratio of observed stars to the intrinsic number of stars in the Milky Way, as a function of APOGEE field and distance along each line-of-sight. 

\begin{figure}
    \centering
    \includegraphics[width=0.5\textwidth]{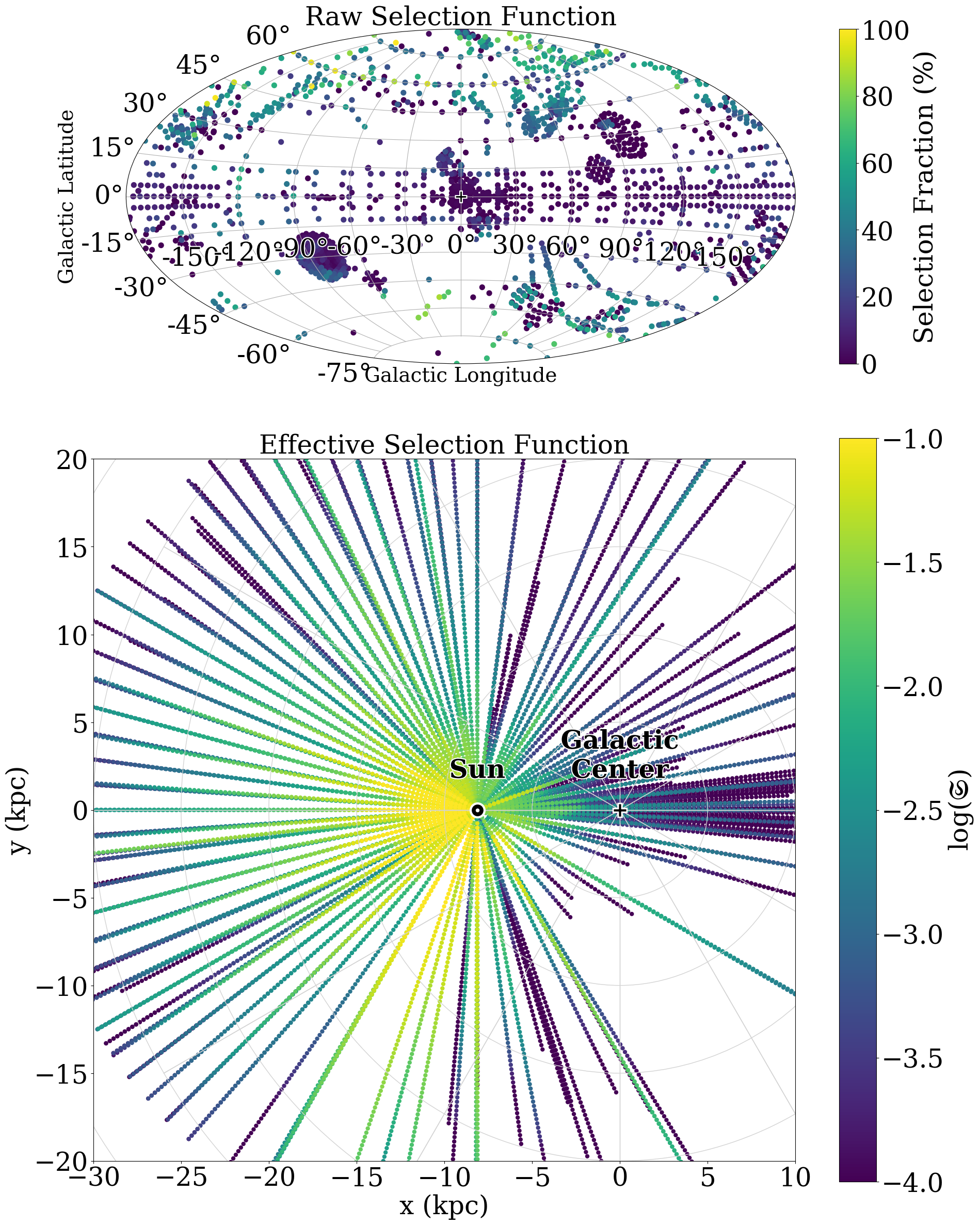}
    \caption{The APOGEE selection functions. \textbf{Top:} The raw Selection Function, reflecting the targeting strategies of APOGEE, displayed as the selection fraction (color) based on the APOGEE field position on the sky. \textbf{Bottom:} The Effective Selection Function, reflecting APOGEE's sampling of the intrinsic distribution of stars in the Milky Way, shown for within 10 deg of disk midplane.}
    \label{fig:selection_func}
\end{figure}

Both selection functions are shown in Figure \ref{fig:selection_func}. The raw selection function is shown in the top panel. Observation fields close to the disk plane generally have a lower raw selection fraction simply because there are more stars in the disk. The effective selection function is shown in the bottom panel. The effective selection fraction is mainly correlated with distance, where stars closer to the Sun generally have a higher selection fraction than those farthest away, with significant variation due to the inhomogeneous distribution of dust. For example, lines-of-sight towards the Galactic center have lower effective selection fractions due to the higher concentration of dust.

\subsection{Stellar Density Profiles} \label{sec:method:densitymodels}

\begin{figure*}
    \centering
    \includegraphics[width=\textwidth]{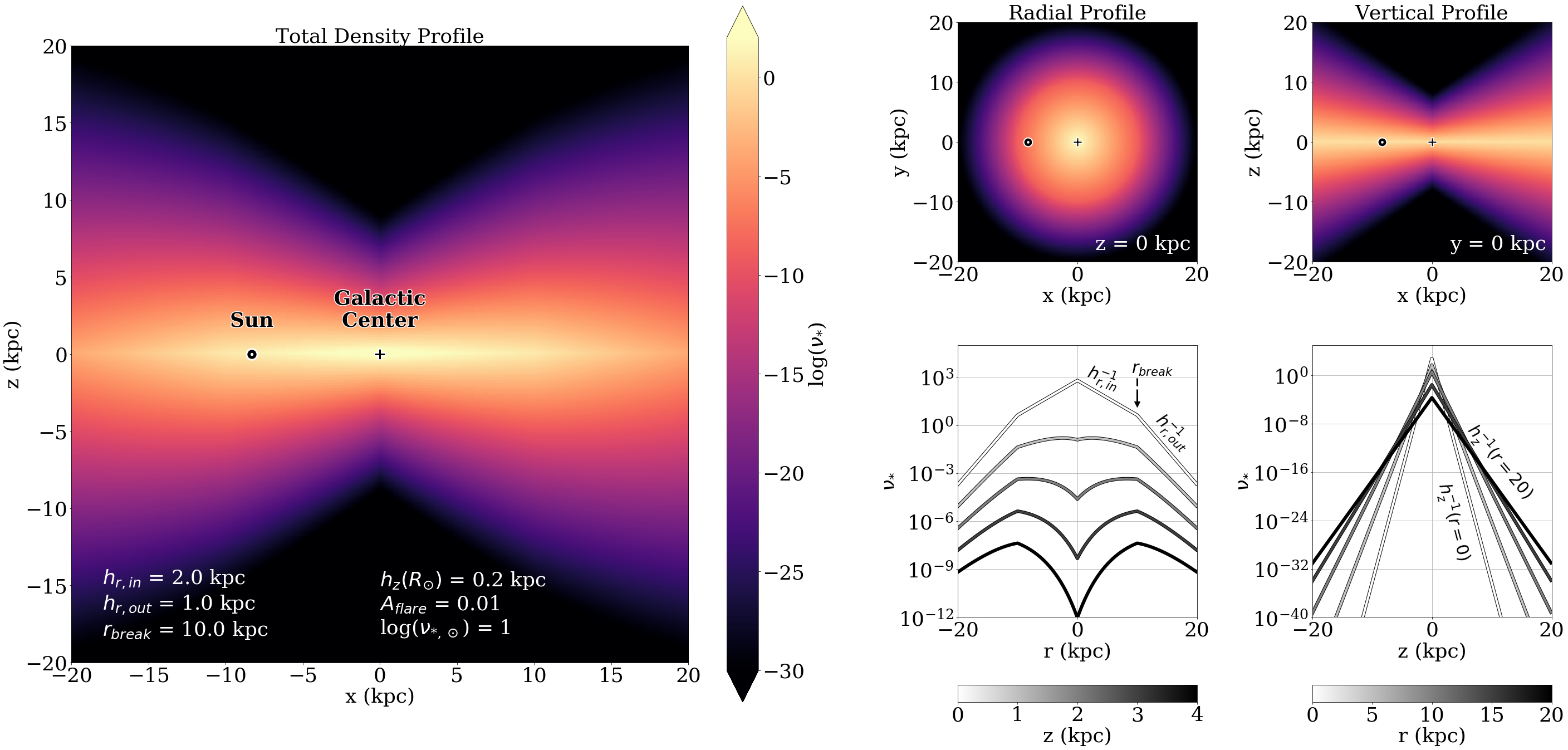}
    \caption{A visual example of the stellar number density model described in Section \ref{sec:method:densitymodels}, which we use to fit each of the stellar populations in the sample. \textbf{Large Left Panel:} the total number density profile (Equation \ref{eq:massprofile_all}) in the edge-on view $x-z$ plane. The total number density profile is the product of the radial profile (Equation \ref{eq:massprofile_r}) and vertical profile (Equation \ref{eq:massprofile_z_linear}). \textbf{Middle-Top:} The radial density profile is modeled as a broken exponential (Equation \ref{eq:massprofile_r}), shown here face-on in the $x-y$ plane.  \textbf{Middle-Bottom:} The total radial density profile (same as left panel) shown in different slices of height above the plane $z$. \textbf{Top-Right:} The vertical profile, a single exponential with radial flaring (Equation \ref{eq:massprofile_z_linear}), in the $x-z$ plane. \textbf{Bottom-Right:} The total vertical density profile (same as left panel) shown at different radii. The flaring term causes the scale height to increase with radius.}
    \label{fig:model_example}
\end{figure*}

\begin{table*}
    \centering
    \begin{tabular}{lllll}
         Parameter & Units & Description & Priors & Eq. \\
         \hline \hline
$\nu_{\odot}$& $N$ kpc$^{-3}$ & scaled amplitude, the stellar number density at $R_{\odot}$ & None (calculated from star counts) & \ref{eq:massprofile_all}, \ref{eq:nu_0}\\
         $h_{R,\textrm{in}}$ & kpc & scale length of inner disk & $-5 \leq h_{R,\textrm{in}}^{-1} \leq 5$ &  \ref{eq:massprofile_r}\\
         $h_{R,\textrm{out}}$ & kpc & scale length of outer disk & $0 \leq h_{R,\textrm{out}}^{-1} \leq 5$ & \ref{eq:massprofile_r}\\
         $R_{\text{break}}$ & kpc & radius of the break in the exponential &  $0 \leq R_{\text{break}} \leq 30.0 $ & \ref{eq:massprofile_r}\\
         $h_{Z\odot}$ & kpc & scale height of the disk at the solar radius & $0 \leq h_{Z\odot} \leq 2.5$ & \ref{eq:massprofile_z_linear} \\
$A_{\text{flare}}$ & none & disk flaring term; slope of variation in $h_{Z}$ with radius & $0 \leq A_{\text{flare}} \leq 0.1$; $h_Z(R=0) \geq 0$& 
         \ref{eq:massprofile_z_linear}\\

    \end{tabular}
    \caption{Summary of the six parameters characterizing the density profile.}
    \label{tab:parameter_summary}
\end{table*}

In the previous sections, the density fitting methodology was formulated as a maximum likelihood problem using a generic density profile $\nu_{*}(X,Y,Z)$ with arbitrary parameters $\theta$. In this section, we describe the exact form of the density profiles adopted to fit each population of the Milky Way, and describe the relevant structural parameters.

We formulate the total stellar number density of the disk $\nu_{*}$ as the product of the radial profile $\Sigma(R)$, the vertical density profile $\xi(R,Z)$, and a normalization factor $\nu_{\odot}$:

\begin{equation}
    \label{eq:massprofile_all}
    \nu_{*}(R, \phi, Z) = \nu_{\odot} \Sigma(R) \xi(R,Z)
\end{equation}

The normalization factor $\nu_\odot$ is defined as the stellar number density at $R_{\odot}$ and is the overall amplitude of the density profile. The likelihood equation (equation \ref{eq:likelihood}) is independent to this parameter, as it drops out of the equation because it is included in both the model and the effective volume terms in equation \ref{eq:likelihood}. Instead, this value can be derived using the observed star counts in APOGEE for each stellar population bin corrected for the selection function, following the approach from \citet[][their Appendix A]{Bovy_2012} and \citet[][their Section 3.3]{Mackereth2017}:

\begin{equation}
    \label{eq:nu_0}
    \nu_{\odot} = \frac{N_{\textrm{APOGEE}}}{\int dO d\lambda(O|\theta)}
\end{equation}

where $N_{\textrm{APOGEE}}$ is the number of stars in our sample for a given stellar population bin and $\int dO d\lambda(O|\theta)$ is the effective volume as described in Equation \ref{eq:effvolume}. The effective volume is the integral of the selection function over the survey volume for a given density model, and is dependent on the other parameters $\theta$.

\noindent This profile assumes azimuthal symmetry in the disk depending only on Galactocentric radius $R$ and height above the plane $Z$, and contains no terms for the spiral arms, bar, or bulge.

When applicable, all parameters are scaled to the value at the Solar neighborhood ($R_{\odot}$), the natural center of the observations and where we have the most data. As before, we define the solar position as $R_{\odot} = 8.3$ kpc and $Z_{\odot} = 27$ pc \citep{BlandHawthorn2016}.

For the radial profile $\Sigma(R)$, an exponential form is commonly adopted for disk galaxies \citep[e.g.,][]{Freeman1970,Gilmore1983,Robin2003,Pohlen2006}. Previous studies on the Milky Way have found that for many stellar populations (particularly the low-$\alpha$ stars in relatively narrow age bins), a broken exponential profile is a better fit to the data \citep[e.g.,][]{Bovy2012,Bovy2016b,Mackereth2017,Lian2022}. Here, we adopt a broken exponential profile for the radial distribution:

\begin{equation}
    \label{eq:massprofile_r}
    \ln{\Sigma(R)} \propto 
    \begin{cases}
    -h_{R,\textrm{in}}^{-1}(R-R_{\odot}) - C & \text{where } R \leq R_{\text{break}}\\
    -h_{R,\textrm{out}}^{-1}(R-R_{\odot}) & \text{where } R > R_{\text{break}}\\
    \end{cases}
\end{equation}

\noindent where $h_{R,\textrm{in}}$ is the scale length of the inner disk, $h_{R,out}$ is the scale length of the outer disk, and  $R_{\textrm{break}}$ is the radius at which the break occurs. $C$ is a normalization factor to ensure that the profile is continuous at the break radius, such that $C = (h_{R,\textrm{out}}^{-1}-h_{R,\textrm{in}}^{-1})*(R_{\text{break}}-R_{\odot})$. This profile does not exclude the possibility of a single exponential profile being the best fit, which would be the case if $h_{R,\textrm{in}} = h_{R,\textrm{out}}$.

We note that the inner disk scale length, $h_{R,\textrm{in}}$, is the only parameter allowed to be negative in its priors. A negative $h_{R,\textrm{in}}$ in Equation \ref{eq:massprofile_r} corresponds to a positively-increasing density profile which peaks at $R_{\text{break}}$ before declining in the outer galaxy with $h_{R,\textrm{out}}$. A profile with a negative value for $h_{R,\textrm{in}}$ can be thought of as a "ring" or "donut"-shaped profile rather than a traditional exponential disk.

The vertical profile is modeled as a single exponential in $Z$, but allowed to vary with radius $R$, accounting for a "flared" population that has a larger scale height at large radii than near the center of the disk, as observed previously in the Milky Way for different populations \citep[e.g.,][]{Bovy2012,Bovy2016b,Mackereth2017,Lian2022, Robin_2022}.

While some studies \citep{Bovy2012,Bovy2016b,Mackereth2017} have used an exponentially flaring profile (such that the scale height varies exponentially with radius characterized by its own scale length), more recent studies have shown that a profile which flares \textit{linearly} with radius better approximates the vertical distribution of stars in the Milky Way \citep{Yu2021,Lian2022}. The corresponding linear flaring term, $A_{\text{flare}}$, is the slope of the scale height varying with radius and represents the strength of the flaring. 

\begin{equation}
    \label{eq:massprofile_z_linear}
    \begin{split}
    & \ln{\xi(R,Z)} \propto -h_{Z}(R)^{-1} |Z| \\ & h_{Z}(R) = h_{Z\odot} + A_{\text{flare}}(R - R_{\odot})
    \end{split}
\end{equation}

\noindent where $h_{Z\odot}$ is the scale height of the disk at $R_{\odot}$ and $A_{\textrm{flare}}$ is the slope of the flaring varying with radius.

To summarize, a total of six free parameters characterize our density model, summarized in Table \ref{tab:parameter_summary}; three parameters for the radial profile ($h_{r,\textrm{in}}$, $h_{r,\textrm{out}}$, and $R_{\textrm{break}}$), two parameters for the vertical profile ($h_{Z\odot}$ and $A_{\textrm{flare}}$), and one scaling parameter ($\nu_{\odot}$) as the number density at the Sun. Figure \ref{fig:model_example} provides a visualization for an example model with arbitrary (but reasonable) parameters.

\subsection{Practical Application \& MCMC} \label{sec:method:application}

\begin{figure}
    \centering
\includegraphics[width=0.45\textwidth]{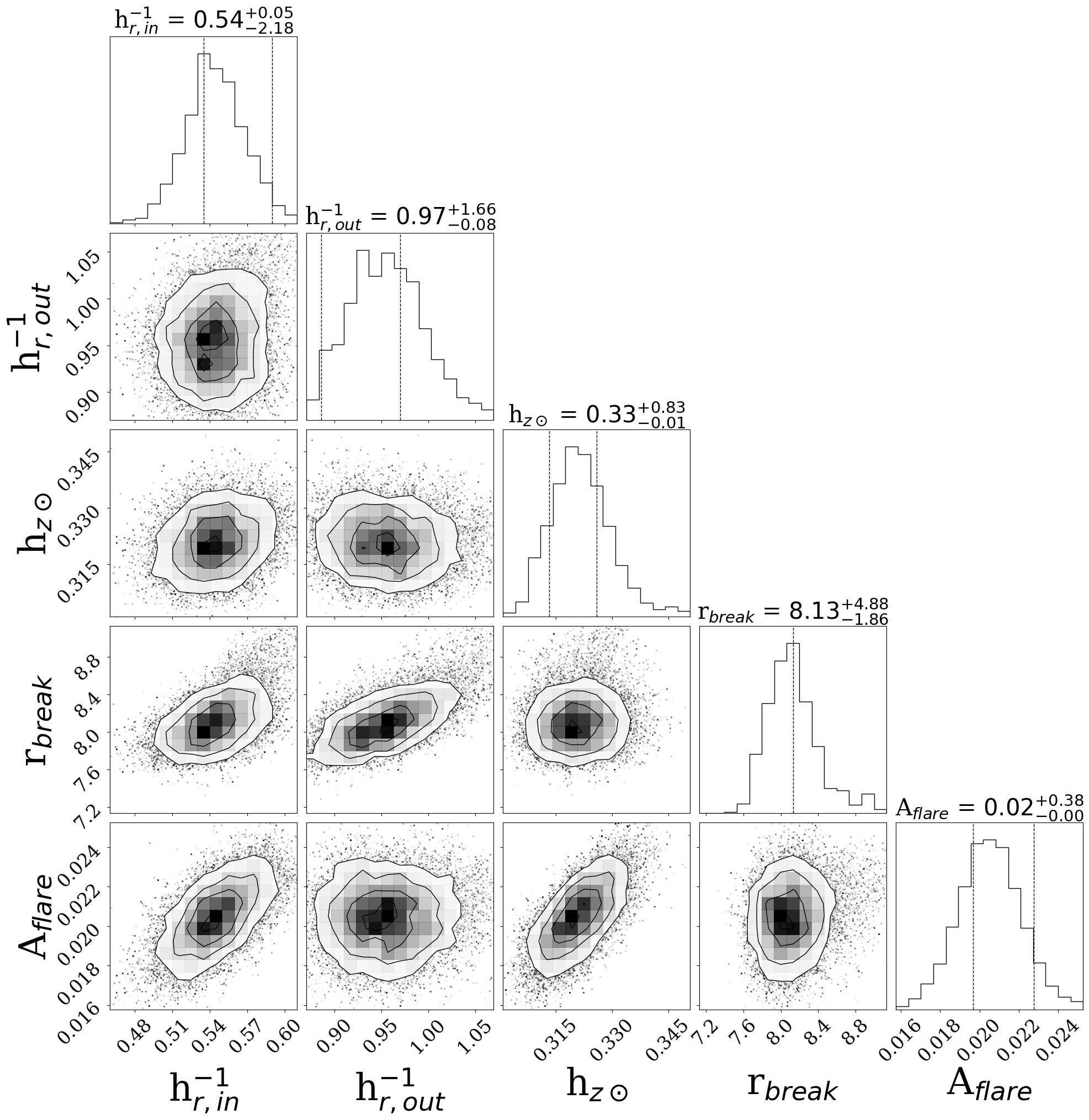}
    \caption{Example corner plot showing the MCMC results for the low-$\alpha$ stellar population bin centered at $\rm [M/H] = 0.15$ and $\log(\textrm{age}) = 9.6$.}
    \label{fig:cornerplot}
\end{figure}

As outlined in Section \ref{sec:method:densityfitting}, deriving the structural parameters of each stellar population becomes a maximum likelihood problem, with Equation \ref{eq:likelihood} predicting the likelihood $\mathcal{L}(\theta)$ that a model with parameters $\theta$ would produce the observations.

We sample the likelihood function using a Markov-Chain Monte Carlo (MCMC) algorithm to derive the best-fit structural parameters for each stellar population. We implement this optimization using the Python {\texttt{emcee}} module, setting up the six-parameter problem using \textcolor{black}{100} walkers and \textcolor{black}{1,000} steps (after an initial burn-in period of \textcolor{black}{500} steps), totaling \textcolor{black}{100,000} iterations considered for each stellar population fit. If the results have not converged after 1,000 steps, the chain is allowed to continue for another 1,000 steps until it does.

We restrict the region of parameter space the MCMC chain is allowed to sample using the priors noted in Table \ref{tab:parameter_summary}. The priors are flat, with the likelihood function automatically returning $\ln{\mathcal{L}(\theta)} = -\infty$ if the parameter falls outside the designated range. These priors were chosen motivated by results from previous studies, but are generous enough to allow the model to sample a wide range of physically possible values. The starting guess for each parameter is selected randomly within the range of the priors.

Computationally, both the radial density profile (equation \ref{eq:massprofile_r}) and the vertical density profile (equation \ref{eq:massprofile_z_linear}) are treated as equalities in the model, which is equivalent to setting the scaling of each such that $\Sigma(R_{\odot}) = 1$ and $\xi(R_{\odot}, 0) = 1$ at the Sun. They are scaled to a true number density together in the combined profile (equations \ref{eq:massprofile_all}) using the derived number density outlined in equation \ref{eq:nu_0}. 

The effective selection function ($\mathfrak{S}$; equation \ref{eq:effsel}) is pre-calculated along a grid of heliocentric distances prior to the likelihood fitting for computational efficiency. For each stellar age and metallicity bin, the effective selection function is computed for each APOGEE field for a distance range $0 \leq d \leq 25$ kpc and saved. This allows for a quicker computation of the effective survey volume (Equation \ref{eq:effvolume}) for any given set of model parameters, greatly reducing the computation time of the likelihood calculation.

The best-fitting set of parameters and associated uncertainties for each population of stars is presented in Section \ref{sec:results}, defined as the median value 
and standard deviation ($\sigma$) of the last $10\%$ of each MCMC chain. An example corner plot showing the best-fit parameters and spread in the MCMC chain for one of the stellar population bins is shown in Figure \ref{fig:cornerplot}.

All code required for reproducing the APOGEE selection function, the density profile MCMC fits, and the figures used in this study is provided on GitHub  at \url{https://github.com/astrojimig/mw_density_imig2025} {and also on Zenodo under DOI:}{\dataset[10.5281/zenodo.16423243]{\doi{10.5281/zenodo.16423243}}.}

\section{Results} \label{sec:results}

For each stellar population separated by age, metallicity, and $\alpha$-element abundances (Figure \ref{fig:bincounts}), we independently fit the density profile described in Section \ref{sec:method:densitymodels} to derive the six best-fit parameters (Table \ref{tab:parameter_summary}) for each population. A full table listing our results for each stellar population bin is available in Appendix Table \ref{tab:best_fit_params}. This section presents these results, organized by the radial parameters in \ref{sec:results:radial}, the vertical parameters in \ref{sec:results:vertical}, and the overall mass contribution in \ref{sec:results:mass}. The subsequent implications for the formation of the Milky Way are reserved for discussion in Section \ref{sec:discussion:implications}.

\subsection{Radial Density Profile} \label{sec:results:radial}

\begin{figure}
    \centering
    \includegraphics[width=0.45\textwidth]{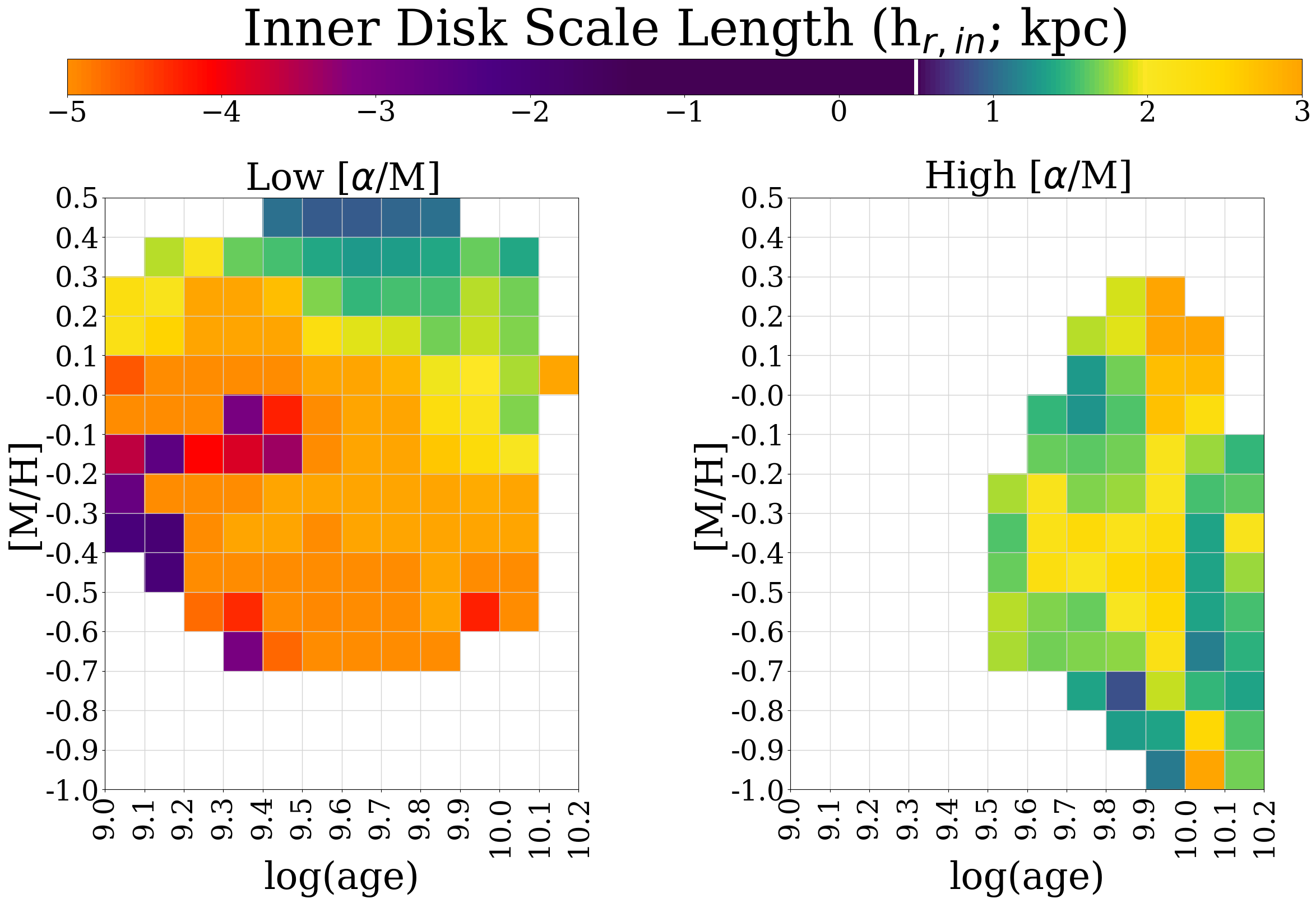}
    \includegraphics[width=0.45\textwidth]{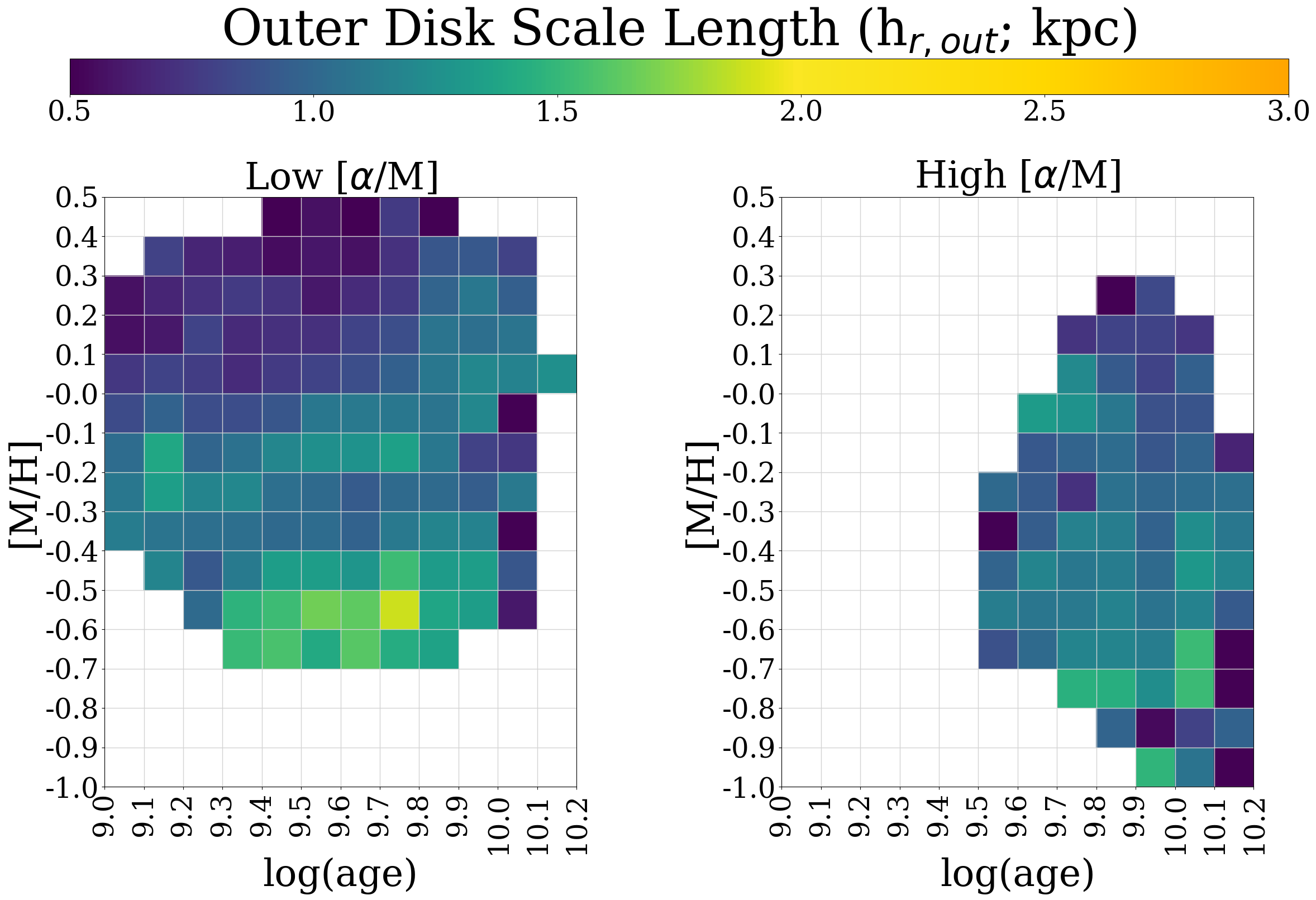}
    \includegraphics[width=0.45\textwidth]{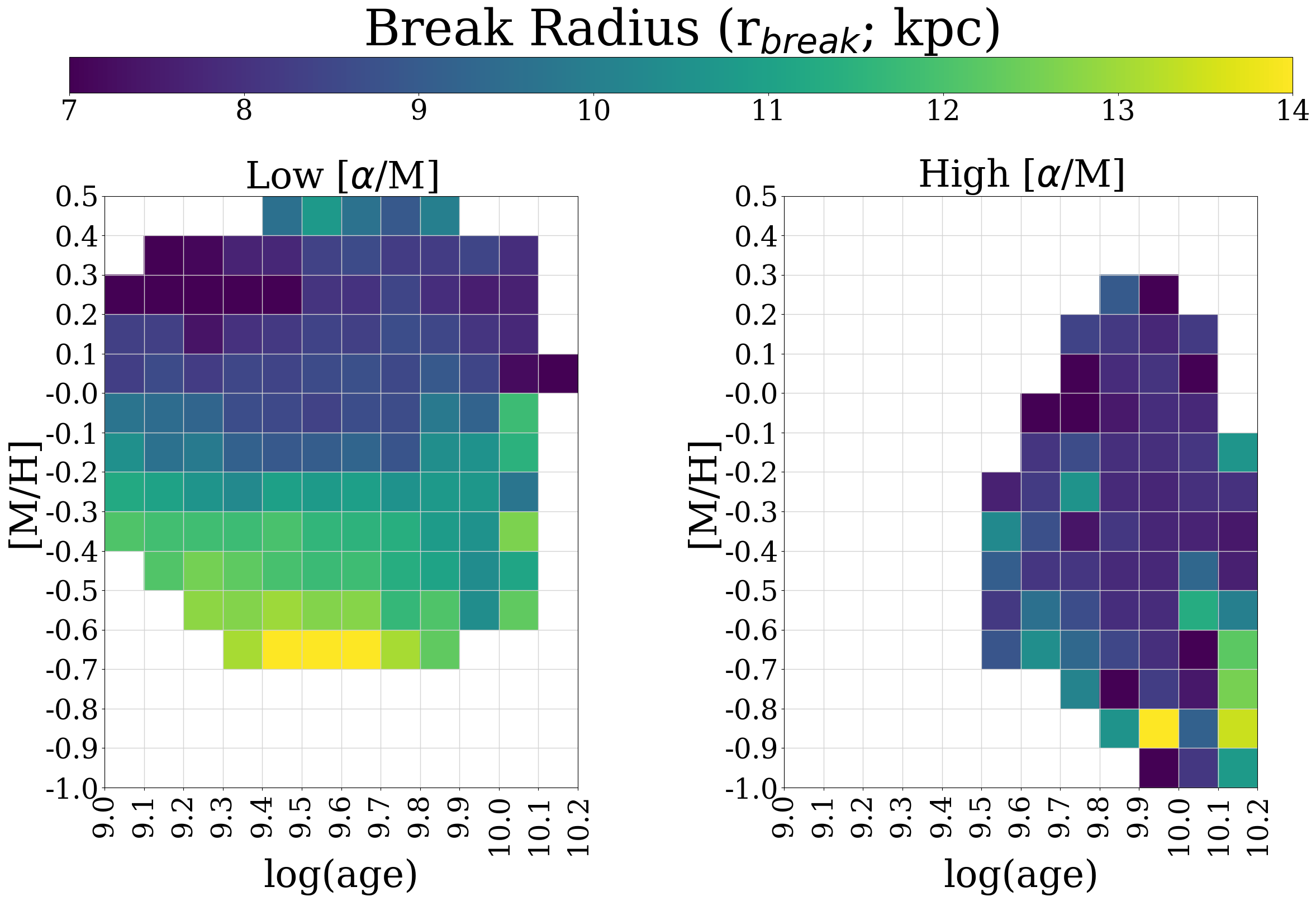}
    \caption{Radial profile parameters best-fit results, including inner scale length ($h_{r,in}$; top), outer scale length ($h_{r,out}$; middle), and break radius ($r_{break}$; bottom). Each row is set up with the same grid layout as Figure \ref{fig:bincounts} to show the trends across stellar populations: low-$\alpha$ (left) and high-$\alpha$ (right), split into corresponding bins of metallicity (rows) and stellar age (columns) in the grid. The color indicates the parameter, with the noted values on the color scale. A white color indicates that there were not enough stars in that population bin to perform a fit. {The $h_{r,in}$ (top) and $h_{r,out}$ (middle) panels share the same color scaling for values $h > 0.5$ (noted by the white vertical line in the top panel's color bar), enabling a direct comparison between them.}}
    \label{fig:params_radial}
\end{figure}

\begin{figure}
    \centering
    \includegraphics[width=0.45\textwidth]{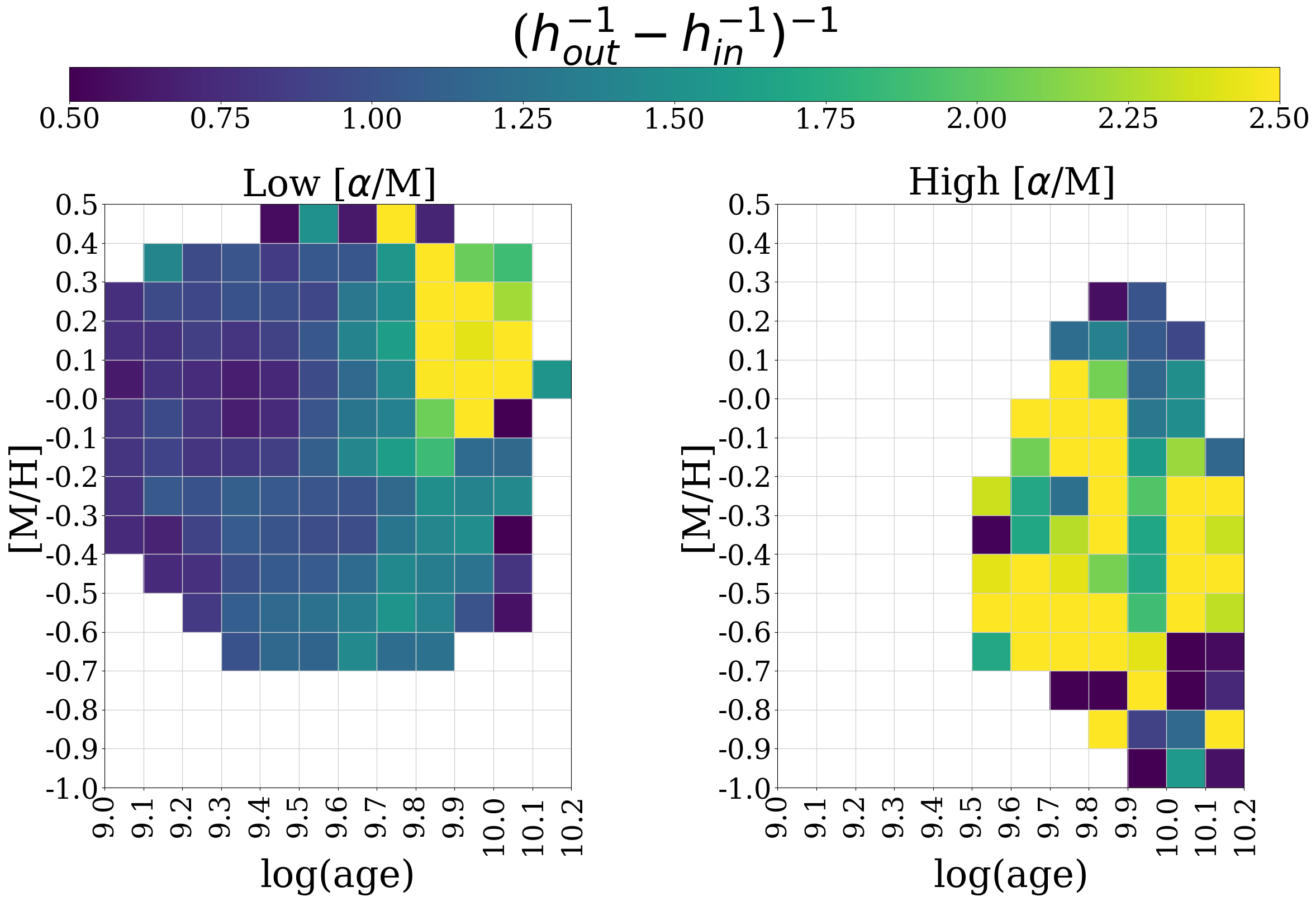}
    \caption{Quantifying the "broadening  metric" with the same grid setup as Figure \ref{fig:params_radial}. The broadening metric quantifies the difference between the inner scale length $h_{r,in}$ and outer scale length $h_{r,out}$. A small broadening metric (blue colors) corresponds to a sharply-peaked "donut"-shaped profile, where a large broadening metric (yellow colors) approaches a single exponential profile.}
    \label{fig:params_peak}
\end{figure}

\begin{figure*}
    \centering
    \includegraphics[width=\textwidth]{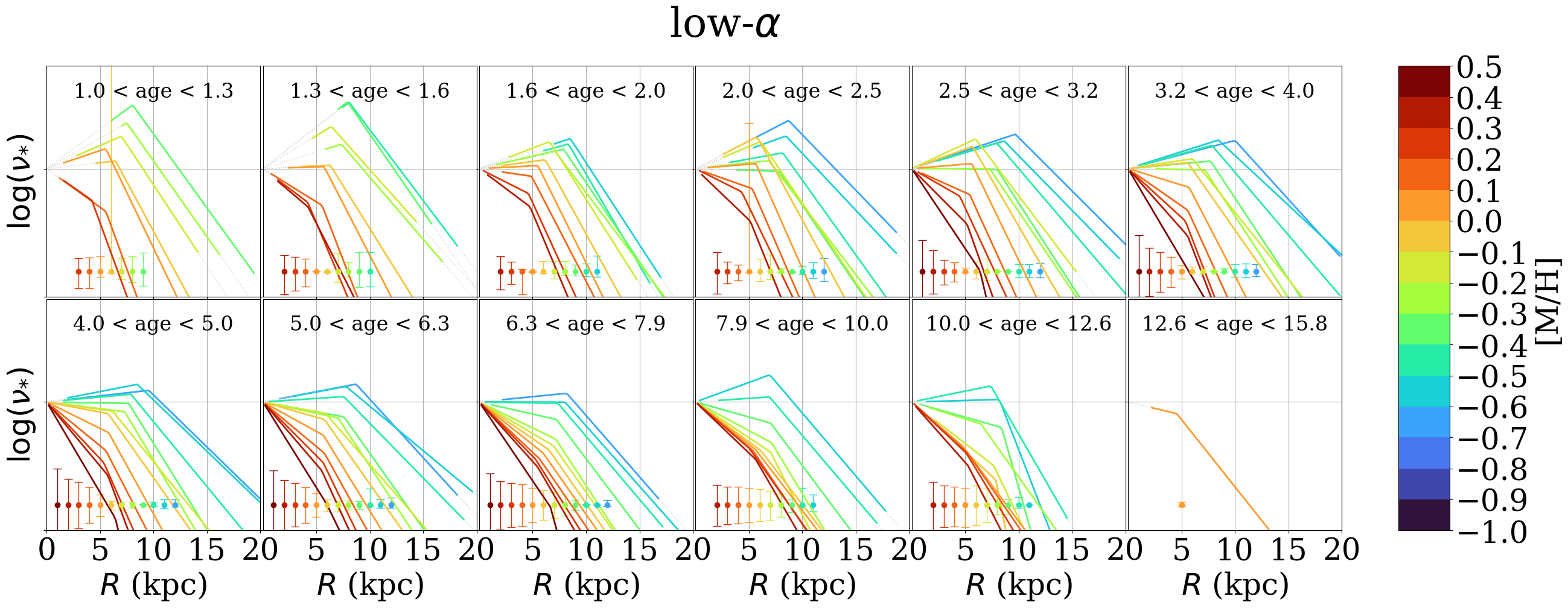}
    \includegraphics[width=\textwidth]{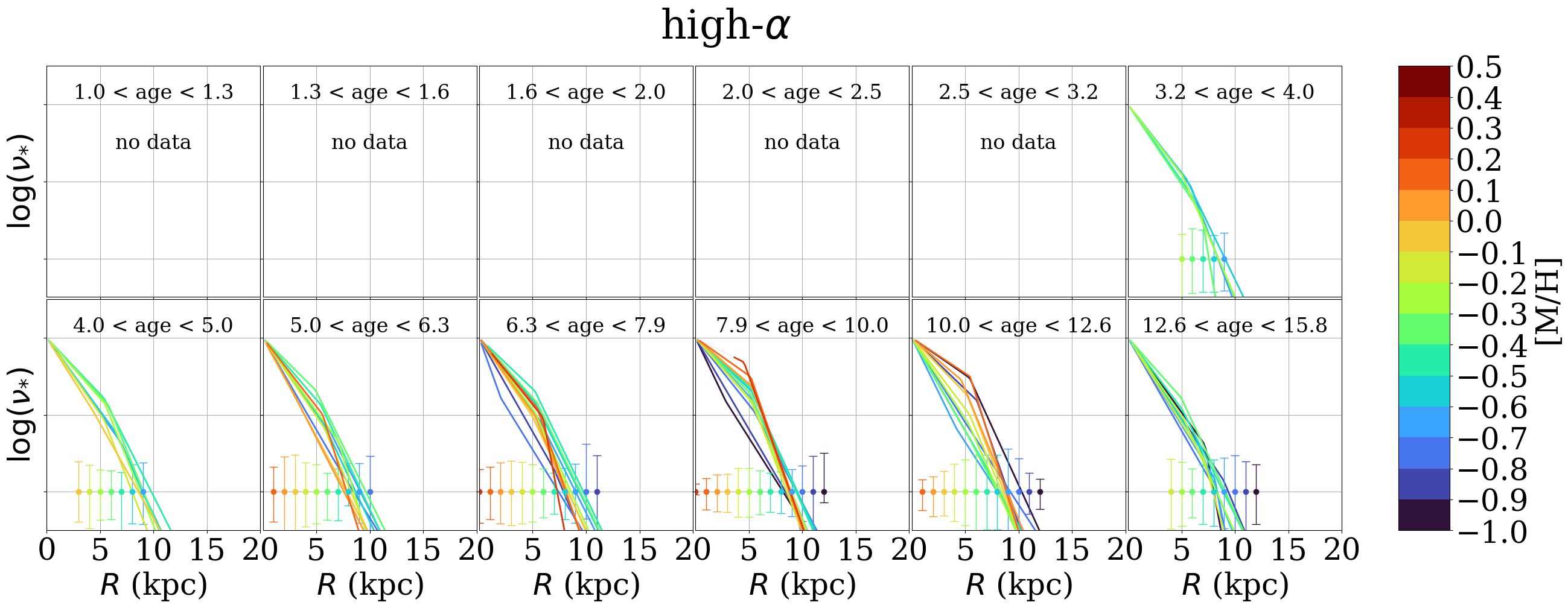}
    \caption{Radial density profiles of the stellar populations in the Milky Way disk, split by low-$\alpha$ (top) and high-$\alpha$ (bottom), stellar ages (panels, increasing left to right), and metallicity (line color). All profiles have been normalized to $\nu_{*} = 1$ at $R=0$ for easier comparison between shapes. The uncertainties on each profile are shown in the error bars along the bottom of the plot.}
    \label{fig:radial_profiles}
\end{figure*}

The derived best-fit parameters for the broken exponential radial profile (Equation \ref{eq:massprofile_r}), including the inner-disk scale length ($h_{r,in}$), outer-disk scale length ($h_{r,out}$), and the radius of the break ($r_{break}$) are shown in Figure \ref{fig:params_radial} as a function of the different stellar populations.

\textbf{Inner Scale Length:} The inner-disk scale length increases with age and metallicity in the low-$\alpha$ disk (Figure \ref{fig:params_radial}, top row). At young ages and low metallicities, the derived scale length is negative, corresponding to a sharply-peaked "donut" shaped profile in the broken exponential in Equation \ref{eq:massprofile_r} and Figure \ref{fig:model_example}, with a peak density at the break radius of $r_{break}$ instead of a monotonically-declining disk profile. As age and metallicity increase, the density profile of the inner disk eventually flattens ($h_{r,in}^{-1} = 0$ being perfectly horizontal), and reverses, until the density consistently decreases with radius everywhere.

\textbf{Outer Scale Length:} The outer-disk scale length increases with age and decreases with metallicity in the low-$\alpha$ disk (Figure \ref{fig:params_radial}, middle row), with trends perpendicular to the gradient seen with the inner disk-scale length (the inner scale length increases from bottom left to top right, the outer scale length gradient increases from top left to bottom right). {The $h_{r,in}$ (top) and $h_{r,out}$ (middle) panels share the same color scaling for $h > 0.5$ kpc, enabling a direct comparison between the inner and outer scale length values in these two panels. In the high-$\alpha$ disk, the inner disk generally has a larger scale length than the outer disk, although the differences are typically less than $\leq < 1$ kpc.}

\textbf{Break Radius:} The break radius $r_{break}$ seems independent of age but correlates with metallicity in the low-$\alpha$ populations. The metal-poor populations have a larger break radius ($r_{break} \sim 12$ kpc ) than those of the metal-rich populations ($r_{break} \sim 6$ kpc).

For the high-$\alpha$ group, $h_{r,in}$ and $h_{r,out}$ tend to be of similar magnitude (note the different scales on the color bars). This means that shape of the density profile of these populations approximates a single-exponential profile rather than the broken exponential profile of the low-$\alpha$ stars, and $r_{break}$ has little meaning. The high-$\alpha$ populations have a shorter scale length than the low-$\alpha$ stars of comparable age and metallicity, in both $h_{r,in}$ and $h_{r,out}$.

\textbf{Profile Broadening:} Quantifying the difference between $h_{r,in}$ and $h_{r,out}$ is a useful test to measure how sharply peaked or broadened the radial profile is. Figure \ref{fig:params_peak} highlights the "broadening metric", defined as $(h_{out}^{-1} - h_{in}^{-1})^{-1}$, which quantifies how broad each profile is. A high value of the broadening metric indicates that there is little difference between $h_{r,in}$ and $h_{r,out}$, and that particular population could be well modeled by a single-exponential profile. A small value of $(h_{r,out}^{-1} - h_{r,in}^{-1})^{-1}$ represents a strongly peaked profile, or a donut-shaped locus of stars around the Milky Way. The broadening metric has little correlation with metallicity or $\alpha$-element abundances, but is strongly correlated with age for all populations: old populations tend to be very broad, and young populations tend to be sharply peaked.

\textbf{Radial Profiles:} The full radial profiles for each population are plotted in Figure \ref{fig:radial_profiles} as  {stellar number density $\nu_{*}$ vs radius}. The top set of panels show the profiles of the low-$\alpha$ populations, and the bottom set of panels shows the profiles of the high-$\alpha$ populations, with each panel therein corresponding to a different age and line color denoting metallicity. These profiles reflect the trends seen in the parameter results described above, including the sharply-peaked profiles of the young, low-$\alpha$ stars and the trend of break radius decreasing with metallicity.

The time-dependent evolution of the total radial density profile will be shown later in Section \ref{sec:integrated:movie}.

{Our findings are generally consistent with previous results. \mbox{\citet{Bovy2016b}}, \mbox{\citet{Mackereth2017}}, \mbox{\citet{Yu2021}}, and \mbox{\citet{Lian2022}} all find that the high-$\alpha$ thick disk stars are generally well fit by a single exponential, while the low-$\alpha$ stars tend to have a broken exponential profile. The peak radius of the break is a function of metallicity, with metal-rich populations peaking at $\sim 7$ kpc and metal-poor populations peaking farther out at $\sim 12$ kpc, reflecting the inside-out growth of the disk and subsequent metallicity gradient. The high-$\alpha$ thick disk stars are characterized by smaller scale lengths than the low-$\alpha$ thin disk. \mbox{\citep[e.g.,]{Bensby_2011, Cheng_2012}}}

{We note that \mbox{\citet{Bovy2016b}}, \mbox{\citet{Yu2021}} and \mbox{\citet{Lian2022}} do not use stellar ages in their analysis, and only groups stars by metallicity and $\alpha$-element abundances; our analysis is most similar to \mbox{\citet{Mackereth2017}} in that aspect. While age is known to be correlated with $\alpha$-element abundances and high-$\alpha$ stars are generally older, the age-$\alpha$ relation varies significantly over galactic position (e.g., \mbox{\citet{Aguirre_2018}}, \mbox{\citet{Feuillet2018}}, \mbox{\citet{Vazquez_2022}}) and for most populations mono-$\alpha$ does not correspond to mono-age due to the radial dependence on chemical evolution in the disk \mbox{\citep[][]{Minchev_2017}}.}

{Our trends with age are similar to that of \mbox{\citet{Mackereth2017}}. The scale height increases with age for both populations. The break radius, $R_{break}$ only depends on metallicity and is largely independent of stellar age. The difference between the inner disk scale length and outer disk scale length, a measure of how sharp the break is, correlates well with stellar age and little to no correlation with metallicity, with the most sharply peaked being the youngest populations.}

\subsection{Vertical Density Profile} \label{sec:results:vertical}

\begin{figure}
    \centering
    \includegraphics[width=0.45\textwidth]{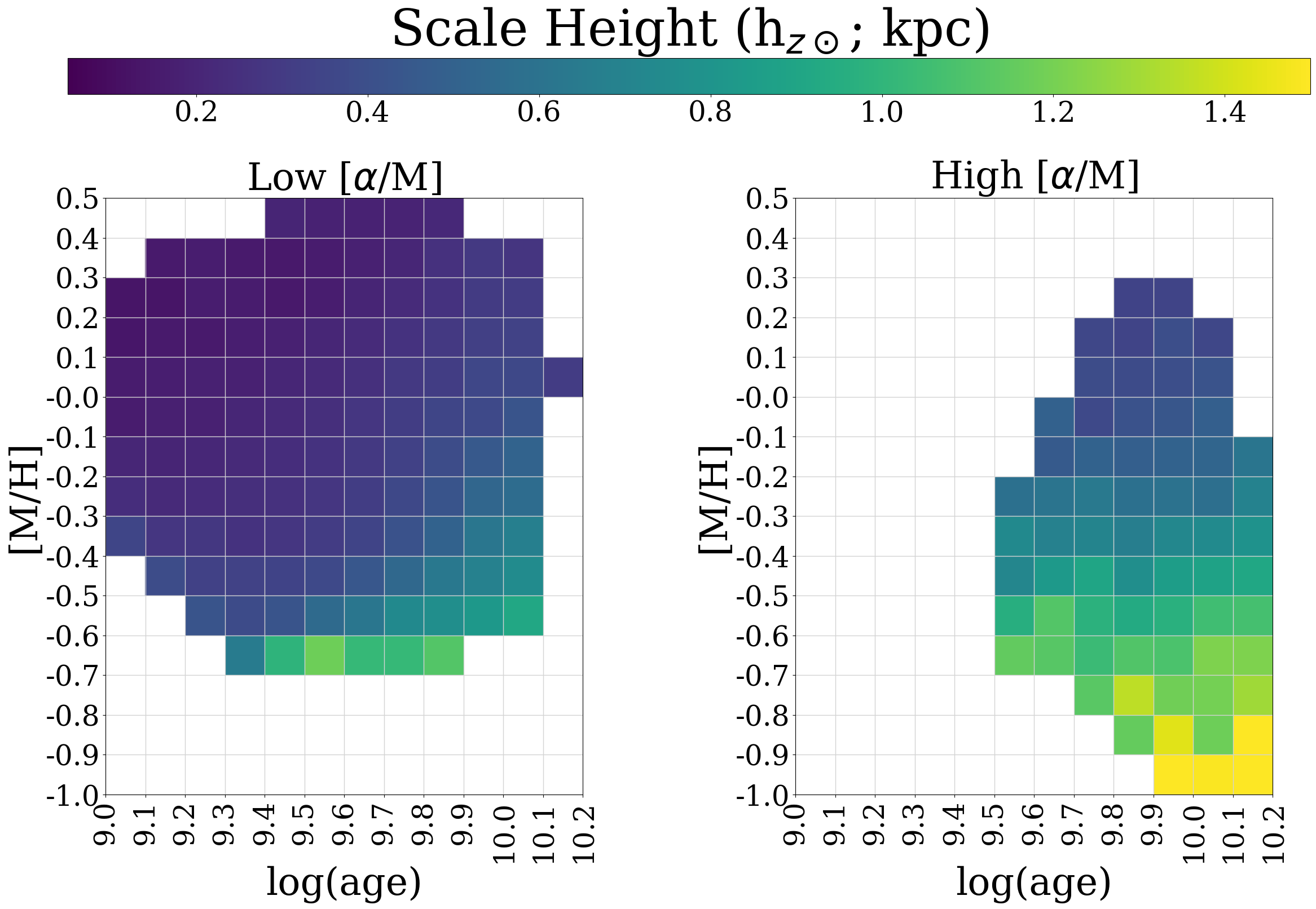}
    \includegraphics[width=0.45\textwidth]{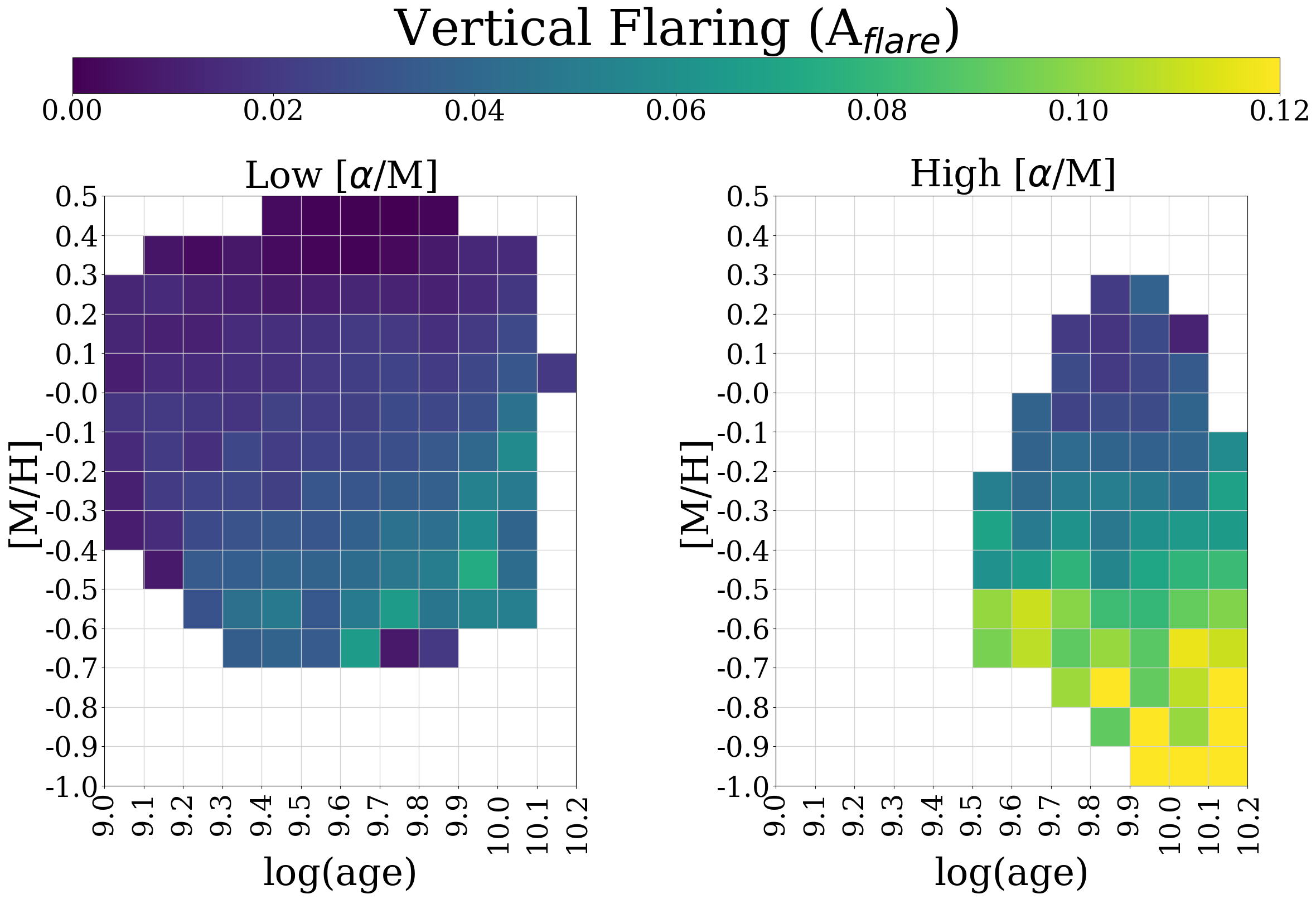}
    \caption{Vertical parameters best-fit results for our density profile, the scale height at the Solar radius $h_{z,\odot}$ and the flaring parameter $A_{flare}$, or the slope of $h_{z}$ as a function of R. Each row is set up with the same grid layout as Figure \ref{fig:bincounts} to show the trends across stellar populations: low-$\alpha$ (left) and high-$\alpha$ (right), split into corresponding bins of metallicity (rows) and stellar age (columns) in the grid. The color indicates the parameter value, with the noted values on the color scale. A white color indicates that there were not enough stars in that population bin to perform a fit.}
    \label{fig:params_vertical}
\end{figure}

The derived best-fit parameters relating to the vertical mass density profile (Equation \ref{eq:massprofile_z_linear}) are shown in Figure \ref{fig:params_vertical}, which include the scale height at the solar neighborhood $(h_{z\odot}$; top panel) and the flaring parameter $(A_{flare}$; bottom panel). 

\textbf{Scale Height:}  The scale height $(h_{z\odot})$ is significantly higher for the high-$\alpha$ population ($h_{z\odot} \sim 0.75$ kpc) compared to the low-$\alpha$ group ($h_{z\odot} \sim 0.25$ kpc). This is consistent with the known chemical bimodality in the disks, where the geometric thick disk {\mbox{\citep[e.g.,][]{Gilmore1983}}} is made primarily of older high-$\alpha$ {populations} and the thin disk is made of younger low-$\alpha$ populations \citep[e.g.,][]{Furhmann1998,Bensby2005,Juric_2008}. 

In the low-$\alpha$ populations, the scale height $(h_{z\odot})$ increases with age and decreases with metallicity. Older and metal poor stars have a higher scale height, although the low-$\alpha$ population is still thinner than the high-$\alpha$ populations everywhere. {In the high-$\alpha$ population, the scale height decreases with metallicity is generally consistent across stellar age.}

\textbf{Disk Flaring:} The flaring term, $(A_{flare})$, determines the slope of the scale height as a function of radius $h_{z}(R)$. Like the scale height, the flaring parameter increases with age and {decreases with} metallicity {in the low-$\alpha$ populations}. The oldest, metal-poor populations are the most strongly flared whereas the young, metal-rich populations have a slope close to horizontal ($A_{flare} = 0$).

\begin{figure*}
    \centering
    \includegraphics[width=\textwidth]{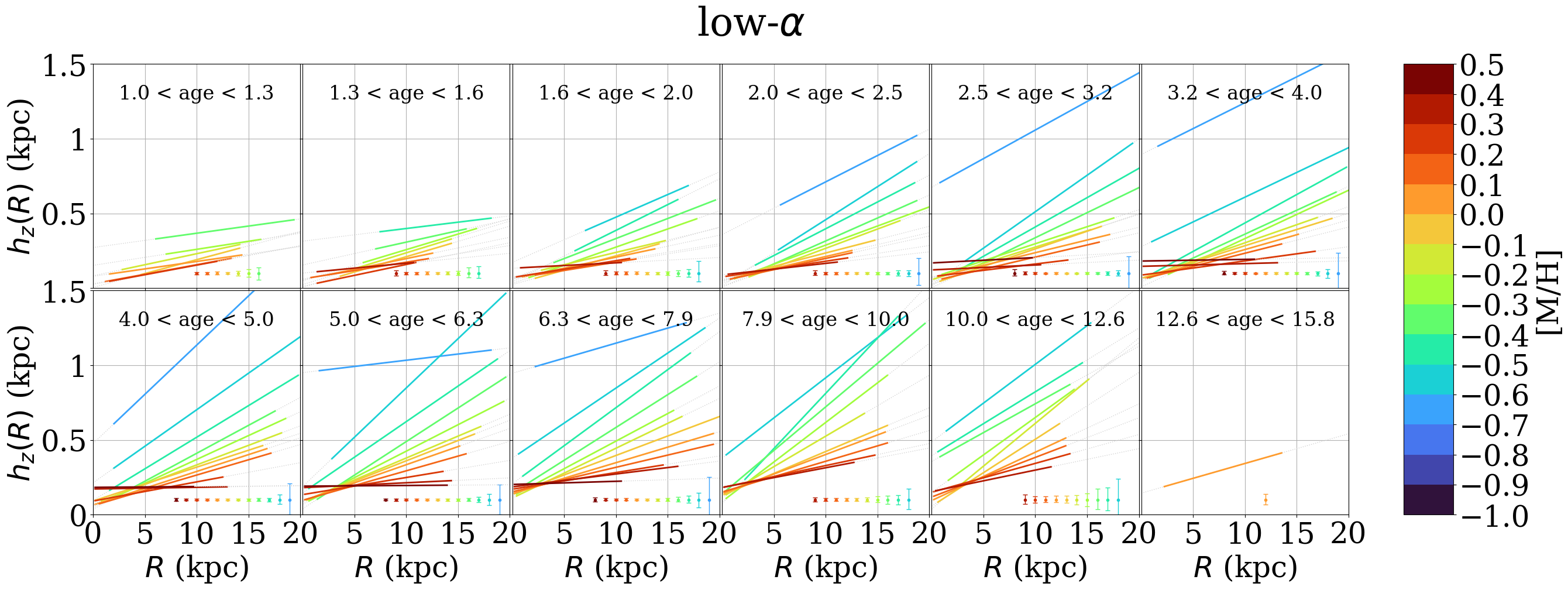}
    \includegraphics[width=\textwidth]{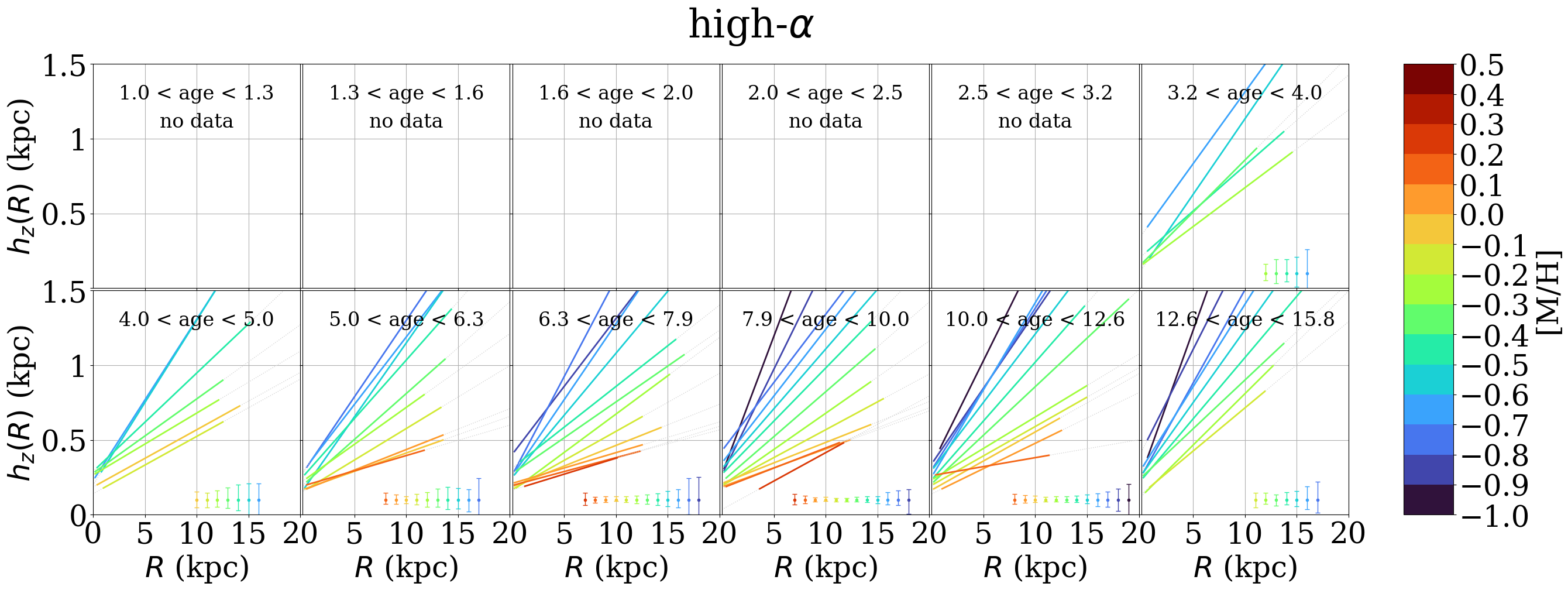}
    \caption{Vertical density profiles showing scale height $h_{z}$ as a function of radius of the stellar populations in the Milky Way disk, split by low-$\alpha$ (top) and high-$\alpha$ (bottom), stellar ages (panels, increasing left to right), and metallicity (line color). The uncertainties on each profile are shown in the error bars along the bottom of the plot.}
    \label{fig:vertical_profiles}
\end{figure*}

\textbf{Vertical Profiles:} A demonstration of the vertical density profile is shown in Figure \ref{fig:vertical_profiles} as the scale height as a function of galactic radius $h_{z}(R)$, a combination of the two vertical parameters $h_{z\odot}$ and $(A_{flare})$ analogous to an edge-on view of the disk. The older, metal-poor populations have the largest scale height at $R_\odot$ and the most vertical flaring. The high-$\alpha$ populations show stronger flaring than the low-$\alpha$ populations.

{Comparing to previous results, we find that the vertical flaring parameter, $A_{flare}$ is strongest in the old high-$\alpha$ stars. This is firstly consistent with the model predictions made by \mbox{\citet{Minchev_2015}}, who showed that flaring of mono-age populations is unavoidable in the cosmological context and increases for older ages. Observationally, this is consistent with the recent results of \mbox{\citet{Yu2021}} and \mbox{\citet{Lian2022}}, and with the older models of \mbox{\citet{Minchev_2015}}, but in conflict with older studies who report the strongest flaring in the low-$\alpha$ disk and little to no flaring in the high-$\alpha$ populations \mbox{\citep[e.g.,][]{Bovy2016b,Mackereth2017}}. This apparent tension is plausibly caused by the difference in spatial sampling between the different data sets. The sixteenth data release of APOGEE in 2019 \mbox{\citep{Jonsson_2020}} included the first data from the southern hemisphere APOGEE-S survey, which significantly increased the spatial sampling towards the Galactic center. The older APOGEE samples from \mbox{\citet{Bovy2016b}} ranges $4 \leq R \leq 15$ kpc and \mbox{\citet{Mackereth2017}} ranges $3 \leq R \leq 15$ kpc, both missing the crucial parts of the inner galaxy where the high-$\alpha$ disk dominates the mass. Our estimates of scale height $h_{Z}$ agree with both of these studies within that radial range, but the trends in the flaring parameter $A_{flare}$ differ. \mbox{\citet{Bovy2016b}} notably do not include stellar ages in their analysis, while \mbox{\citet{Mackereth2017}} includes ages but had a significantly smaller stellar sample overall. This work and \mbox{\citet{Lian2022}} include APOGEE-S data in the analysis. \mbox{\citet{Yu2021}} uses a red clump sample from the LAMOST survey, which although is also limited in radial distribution in the inner disk ($ 5 \leq R \leq 20$ kpc) has an overall larger sample size than \mbox{\citet{Bovy2016b}} and \mbox{\citet{Mackereth2017}}. The flaring trends seen in this work, \mbox{\citet{Yu2021}}, \mbox{\citet{Lian2022}}, where the high-$\alpha$ mono-age populations flare most strongly, are all consistent with the models presented in \mbox{\citet{Minchev_2015}}, attributing the disk flaring in mono-age populations to the natural inside-out growth of the Galaxy.}

\subsection{Surface Mass Density \& Total Mass}
\label{sec:results:mass}

\begin{figure}
    \centering
    \includegraphics[width=0.45\textwidth]{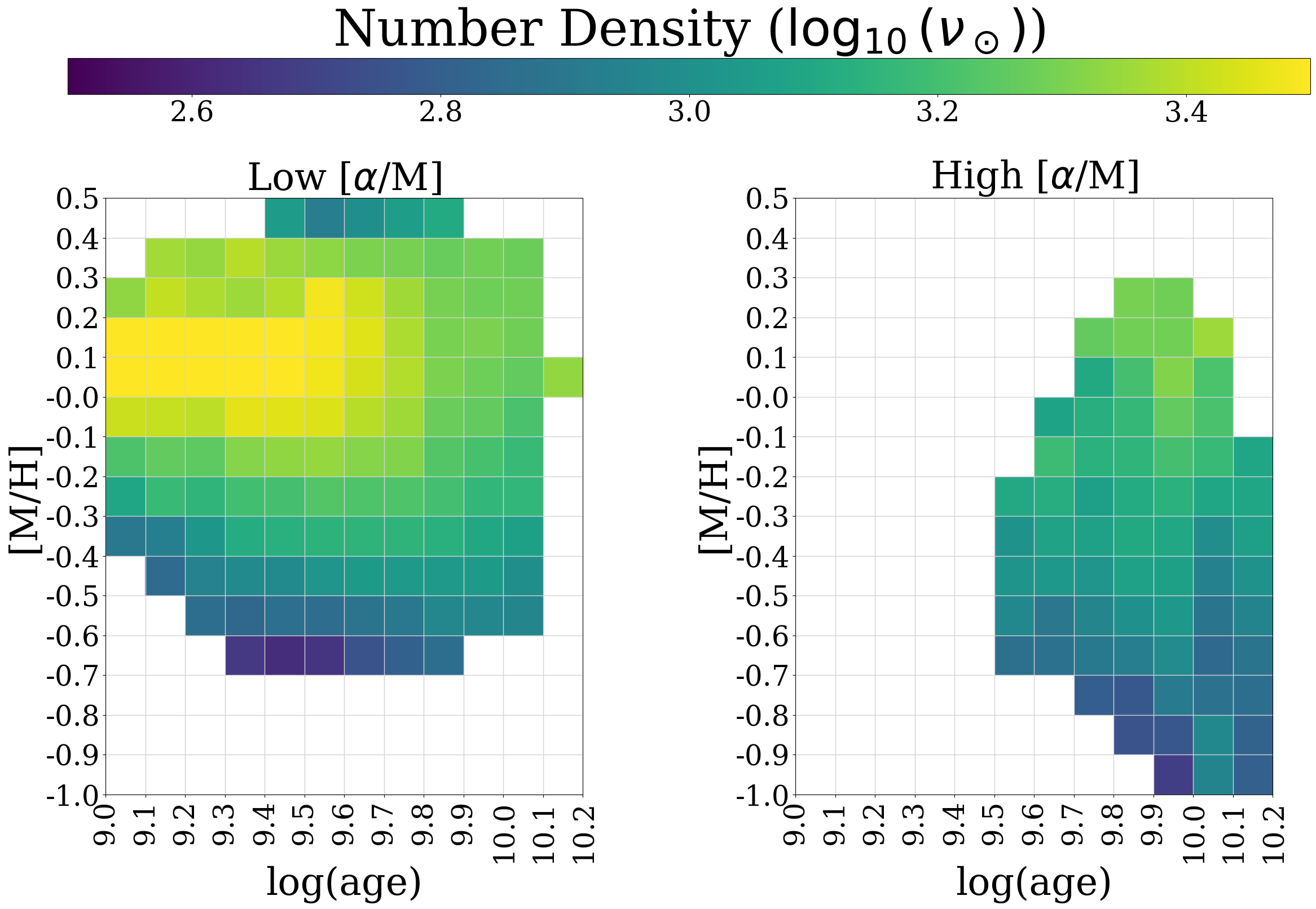}
    \includegraphics[width=0.45\textwidth]{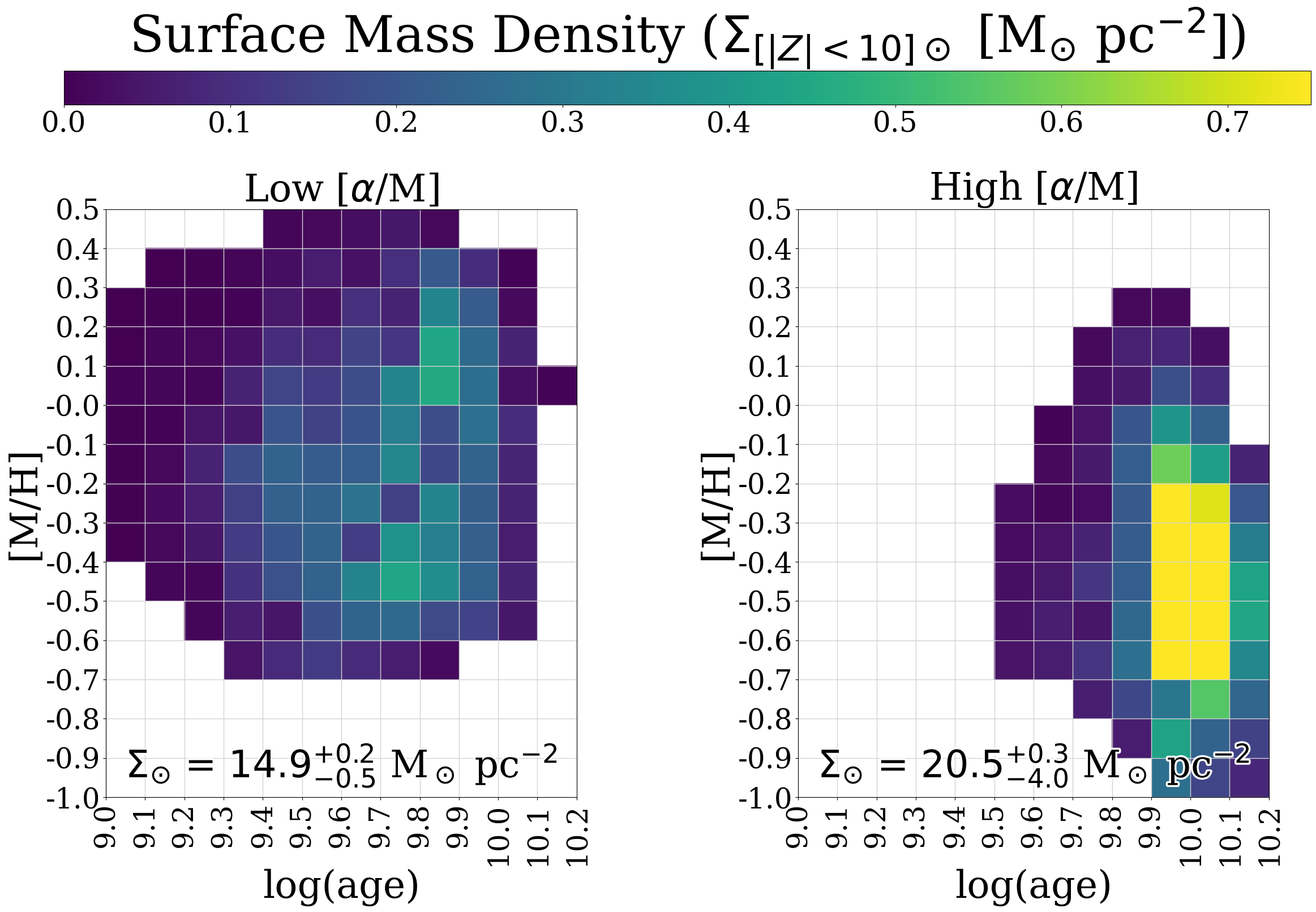}
    \includegraphics[width=0.45\textwidth]{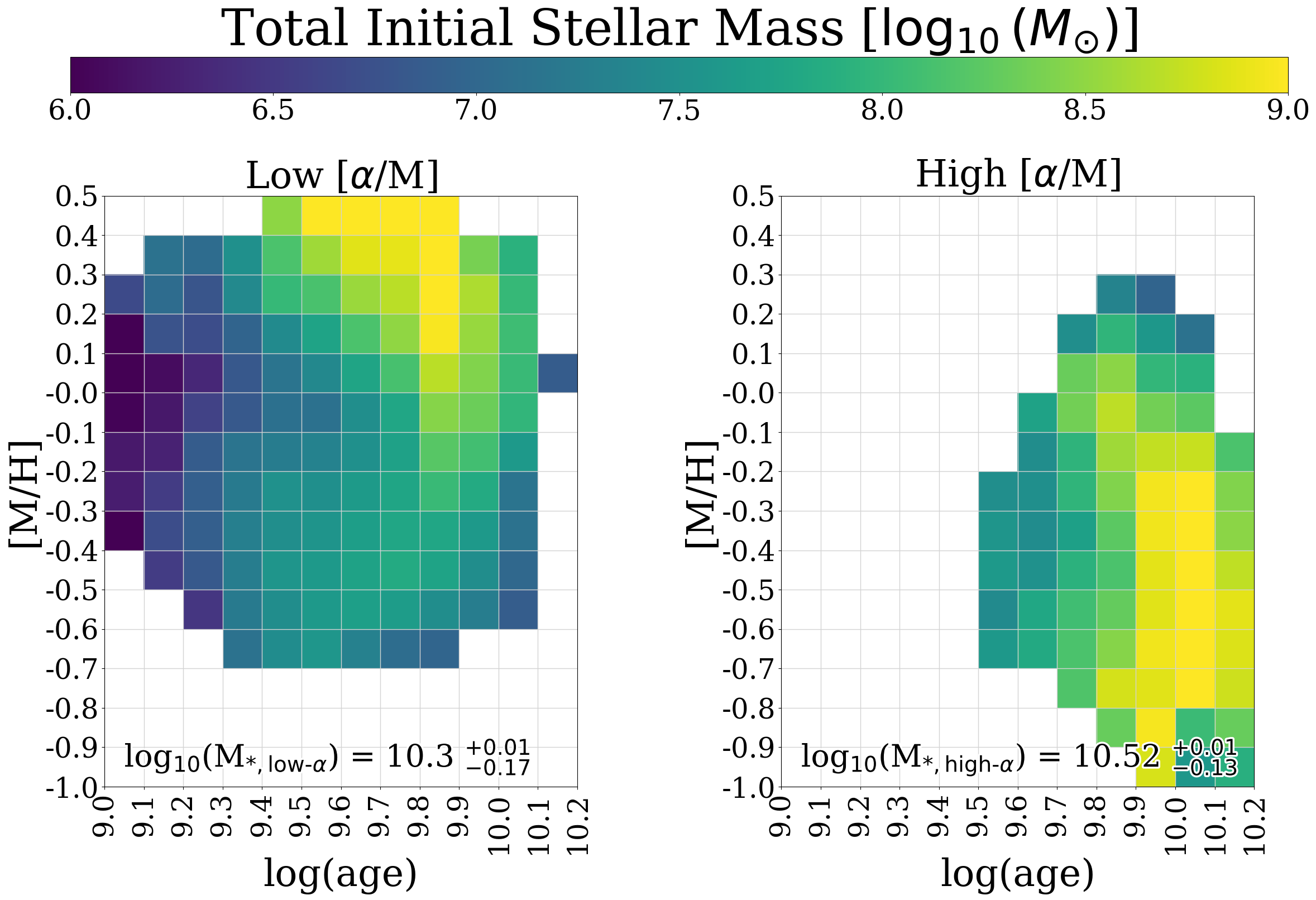}
    \caption{The mass parameter from our density profile, the number density at the Solar Neighborhood ($\nu_{\odot}$; top), and the subsequent calculation 
    of surface mass density ($\Sigma_{odot}$, middle) and total mass contribution ($M_{*}$; bottom) of each stellar population calculated by integrating the best-fit density profiles over space. Each row is set up with the same grid layout as Figure \ref{fig:bincounts} to show the trends across stellar populations: low-$\alpha$ (left) and high-$\alpha$ (right), split into corresponding bins of metallicity (rows) and stellar age (columns) in the grid. The color indicates the parameter value, with the noted values on the color scale. A white color indicates that there were not enough stars ($<100$) in that population bin to perform a fit.}
    \label{fig:mass_params}
\end{figure}

The sixth and final parameter in our density model, $\nu_{\odot}$, is the scaled amplitude of the profile or the stellar number density at $R_{\odot}$. This parameters is not fit as a free parameter in the MCMC chain, but rather calculated from the observed star counts as described in Equation \ref{eq:nu_0} (Section \ref{sec:method:densitymodels}).
This is primarily used as the overall normalization factor in each profile (Equation \ref{eq:massprofile_all}), but can be converted into the surface mass density $\Sigma(R_{\odot})$ and subsequently the total mass of each stellar population $M_{*}$ by integrating over volume. Figure \ref{fig:mass_params} presents these three quantities.

\textbf{Stellar Number Density:} The best-fit result of $\nu_{\odot}$ for each stellar population is shown in the top panels Figure \ref{fig:mass_params}. The stellar populations with the highest density are slightly higher than solar metallicity $[M/H] = 0.1$, low-$\alpha$, and young. This is consistent with other studies, that find stars near the Solar neighborhood tend to be slightly more metal-rich and younger than the Sun \citep[e.g.,][]{Haywood_2013,Mackereth2017,Miglio_2021,Lian2022,Imig_2023}, and the low-$\alpha$ disk represents the majority of the mass at this radius. We find that the low-$\alpha$ thin disk comprises around 65\% of the total stellar density at the solar location. 

\textbf{Surface Mass Density:} The number density of the RGB sample ($\nu_{*}$) can be converted into a surface mass density with a few additional steps. First, the number counts of the RGB stars are converted into mass of the total population using corrections from our isochrones \citep[the PARSEC isochrones; ][]{Bressan_2012, Chen_2014} and assuming a Kroupa IMF \citep{Kroupa_2002}. In a younger population, a smaller number of massive stars populate the giant branch; In older populations, a higher number count of lower mass stars stars populates the giant branch. If two stellar populations have the same number density of RGB stars, they could have a significantly different mass density due to a difference in population age, thus the need for this correction. Then, the mass density ($M_\odot$ pc$^{-3}$) is converted into a surface mass density ($M_\odot$ pc$^{-2}$) by integrating all of our best-fit density profiles over height $Z$ for a range of $|Z| \leq 5$ kpc. The resulting surface mass density estimations are shown in the middle panel of Figure \ref{fig:mass_params}. We measure a total surface mass density at the Solar radius of the low-$\alpha$ disk to be $14.9{+0.2}_{0.5} M_\odot$ pc$^{-2}$ and the high-$\alpha$ disk as $20.5^{+0.3}_{-4.0} M_\odot$ pc$^{-2}$, for a total local surface density of $35.4^{+0.3}_{-4.1} M_\odot$ pc$^{-2}$. While the low-$\alpha$ disk contributes around $65\%$ of the number density at the solar neighborhood, we find that it only contributes $42\%$ of the local surface mass density due primarily to the increased scale height of the high-$\alpha$ disk.

Our estimate of $35.4 M_\odot$ is generally consistent with other studies. \citet{Bovy_2013} report a value of $38 \pm 4 M_\odot$ pc$^{-2}$. \citet{Mackereth2017} found a significantly smaller value of $20.1^{+2.4}_{-2.9} M_\odot$ pc$^{-2}$ and \citet{Lian2022} report $21.31 \pm 0.32 M_\odot$ pc$^{-2}$. The fractional contribution across the entire Galaxy between the surface mass density of the high-$\alpha$ and low-$\alpha$ disk varies between studies. \citet{Mackereth2017} report a value of $f_\Sigma = 18\%$, \citet{Lian2022} report a value of $f_\Sigma = 36\%$ and this work derives a comparatively larger ratio of $53\%$. This is likely due to our more complete sampling of the inner disk ($R \leq 3$ kpc), where the high-$\alpha$ populations contribute the majority of the mass.
 
\textbf{Total {Initial} Mass:} The total mass of each stellar population is calculated by integrating all of our density profiles over radius ($0 \leq R \leq 30$ kpc) and height ($ -10 \geq Z \leq 10$ kpc) with the mass correction from isochrones used previously. The mass contribution of each stellar population is shown in the bottom panel of Figure \ref{fig:mass_params} and varies with age and metallicity. The trend with age is consistent with a declining star formation history, and the trend with metallicity can be explained by chemical enrichment; at younger ages, there is less pristine gas with which to form metal-poor stars, resulting in a larger fraction of each population being metal-rich at fixed age. 

We measure the total stellar mass of the disk as $M_{*} = 5.27^{+0.2}_{-1.5} \times 10^{10} M_{\odot}$. With the low-$\alpha$ thin disk composing of 38\% of that total at $M_{*} = 1.99^{+0.1}_{-0.6} \times 10^{10} M_{\odot}$ and the high-$\alpha$ populations contributing $M_{*} = 3.28^{+0.1}_{-0.8} \times 10^{10} M_{\odot}$. The thin disk contributes most of the mass near the solar neighborhood, but the high-$\alpha$ population contributes more at smaller radii due to its characteristically smaller scale lengths. 

Our total value of $5.27^{+0.2}_{-1.5} \times 10^{10}$ M$_\odot$ is within uncertainties of several previous estimates from literature; \citet{Bovy_2013} estimates $4.6 \pm 0.3 \times 10^{10} M_\odot$ using a sample of stars from SEGUE, \citet{Cautun2020} derive a mass of $5.04^{+0.43}_{-0.52} \times 10^{10} M_\odot$ using Gaia data, and \citet{Licquia_2016} report a disk stellar mass of $4.8^{+1.5}_{-1.1} \times 10^{10} M_\odot$ from Bayesian Mixture modeling of other literature measurements. {\mbox{\citet{Nesti_2013}} estimate a total disk stellar mass of $5.5\pm0.5 \times 10^{10} M_\odot$ by fitting a dark matter halo profile to observed stellar velocities.}

{\textbf{Total Surviving Mass:} The number density, surface mass density, and total mass values reported above reflect the total \textit{initial} mass of each stellar population. This notably includes all stellar evolutionary stages, and is not limited to the red giant branch region used in our sample. This also includes the mass contributed by stellar remnants and evolved stars, and any mass lost from stellar winds and supernovae. The initial mass of a population can therefore be significantly greater than the present-day surviving mass of a stellar population as the massive stars die over time.}

{To demonstrate this, \mbox{\citet{zasowski2025}} calculated the fraction of luminous mass and remnant mass compared to the initial population mass using the PARSEC isochrones \mbox{\citep{Bressan_2012, Chen_2014}} and assuming a Kroupa IMF \mbox{\citep[][]{Kroupa_2002}}. For a population that is 10 Gyr old, only 55\% of the initial mass has survived as luminous stars; the other 45\% of the initial mass has been lost to supernovae, stellar winds, or stellar remnants. We adopt the mass fractions calculated from isochrones from their Section 2.7 to produce values for the \textit{aged} stellar mass, which consists of the surviving luminous mass ($M_{lum}$) and stellar remnants ($M_{rem}$) of each population. The polynomial relation between initial mass $M_{init}$ and surviving aged stellar mass $M_{lum+rem}$ for age $\tau$ (in years) as fit from \mbox{\citet{zasowski2025}}:}

\begin{equation}
\begin{split}
    \frac{M_{lum+rem}}{M_{init}} = 2.024 - 0.183\log_{10}(\tau) + 0.005(\log_{10}(\tau))^{2}
\end{split}
\end{equation}

{With this adjustment, we find that the present-day surviving stellar mass (luminous + remnant) ($M_{lum+rem}$) of the Milky Way is $3.7 \times 10^{10}  M_\odot$. As shown in \mbox{\citet{zasowski2025}}, this does not change any other qualitative trends in the results.}

\section{Integrated Properties} \label{sec:integrated}

Our view of the Milky Way is more detailed near the Solar vicinity due to obvious observational limitations, whereas the exponential density profile suggests that the majority of the mass in the Milky Way lies closer the Galactic center. Therefore a complete picture of our Galaxy has long been out of reach and obscured by dust, but our selection-function corrected density fits can offer insight on the integrated properties of the Milky Way.

In this section, we apply our results from Section \ref{sec:results} and sum the best fit density profiles together across all of the stellar population bins to portray the Milky Way in its entirety. This allows us to study the Milky Way in broader galactic context, comparing its density profile, scale length, star formation history, integrated spectrum, and photometric colors to other galaxies more directly for the first time. See Zasowski et al. ({\it in prep}) for some complementary analyses, including comparisons to simulated galaxies.

\subsection{Total Mass Profile \& Disk Scale Length} \label{sec:integrated:scale_length}

\begin{figure}
    \centering
    \includegraphics[width=0.5\textwidth]{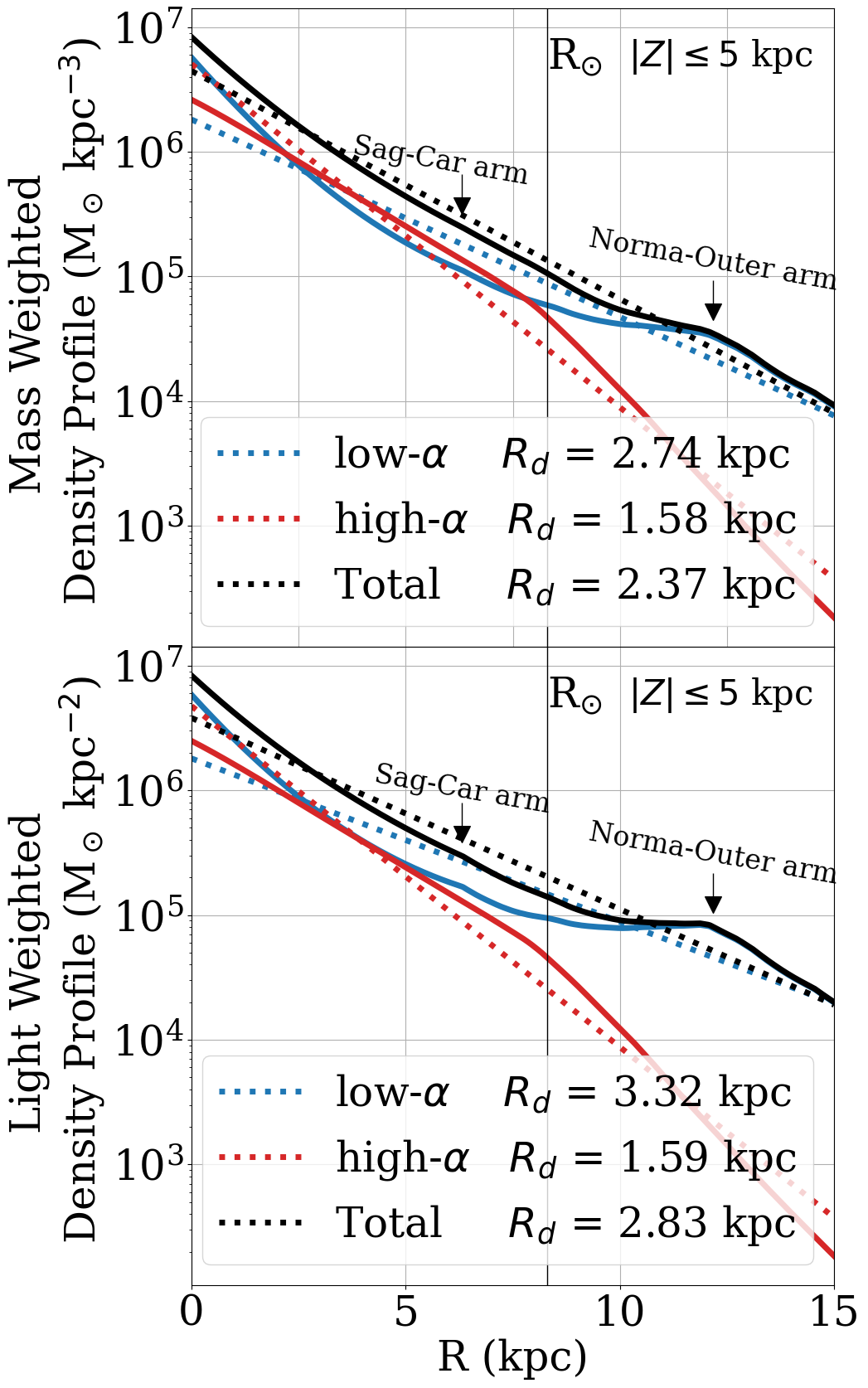}
    \caption{Total surface mass density profile of the Milky Way disk, in a mass-weighted (top) and light-weighted (bottom) profile. The high-$\alpha$ populations (red line),  low-$\alpha$ populations (blue line), and total profile (black line) are plotted for each. For each, the solid line denotes the empirical profile from the sum of all mono-age, mono-abundance populations, and the dotted line is a straight-line fit for a single "total" scale length value for each $\alpha$ group, the value of which is noted in the legend.}
    \label{fig:total_massprof}
\end{figure}

The characteristic scale length of the Milky Way disk is still uncertain, with estimates commonly ranging between $1.8$ and $6$ kpc \citep[e.g., the review in ][]{BlandHawthorn2016}. Despite the large range of radius estimates for the MW, several studies suggest that the Milky Way is too small for its mass \citep[e.g.,][]{Bovy_2013,Licquia_2016,Boardman_2020}, with a characteristic scale length shorter than the $2\sigma$ range of most of its galactic siblings, which generally fall in the $R_{d} = 3.2 - 5.7$ kpc range.

We obtain a total scale length measurement for the MW disk by summing the radial mass density profiles of every stellar population (Figure \ref{fig:radial_profiles}) together, integrating vertically over $|Z| \leq 5$ kpc and fitting for a single scale length of the total. The summed radial density profile is shown in the top panel of Figure \ref{fig:total_massprof}. Although the individual stellar populations are best modeled by broken exponential profiles with a continuous range of $r_{break}$, the total profile mostly resembles a single exponential with a scale length of R = 2.37 kpc. The locations of the Milky Way's most prominent spiral arms from \citet{Reid_2019} are also annotated on this figure, the inner Sag-Car arm (at $R \sim 6.04$ kpc) and the Norma-Outer arm (at $R \sim 12.24$ kpc), possibly explaining the excess in the profile at these locations. The density profile used (equations \ref{eq:massprofile_all}-\ref{eq:massprofile_z_linear}) is axisymmetric and independent of azimuthal angle $\theta$, but because the APOGEE footprint (Figure \ref{fig:data_sample}) covers only a limited azimuthal range in the Galaxy the spiral arms possibly emerge as "ring"-shapes structures extrapolated in our integrated fit.

This total profile can be reasonably approximated by a single exponential, which is in contract with the recent findings of \citet{Lian_2024} who found that the luminosity-weighted surface mass density profile flattens inwards of $R \lesssim 6 $ kpc in the Milky Way disk. 

The total high-$\alpha$ profile resembles a slightly broken exponential profile; we interpret this as the truncation of the thick disk near the Solar neighborhood ($R \sim 8$ kpc), consistent with the dearth of high-$\alpha$ stars observed beyond the Solar radius \citep[e.g.,][]{Hayden2015}. This is also close to the expected Outer Lindblad Resonance (OLR) of the Galactic Bar at $R \sim 9.4$ kpc \citep[e.g.,]{Halle_2015,Michtchenko_2016, Dias_2019, Khoperskov_2020b}, which may act as a boundary for radial migration past which no stars can migrate \citep[e.g.,][]{Halle_2015, Khoperskov_2020}.

The total mass-weighted scale length of the disk measured this way is {$R_{d} = 2.37 \pm 0.2$} kpc.
This is fairly consistent with other literature measurements for the scale length of the total disk, which vary from
$R_{d}=2.15 \pm 0.14$ kpc \citep{Bovy_2013},
 $R_{d} = 2.71 \pm 0.21$ kpc \citep{Licquia_2016},
 $R_{d} = 2.48 \pm 0.14$ kpc \citep{Fielder_2021},
 and $R_{d} = 1.977 \pm 0.01$ kpc \citep{Lian2022}.
 Our estimate may be larger than some previous estimates due to our increased sampling of the outer disk ($R \geq 15$ kpc) with the final data release of APOGEE. 

However, for comparison with external galaxies, one should consider the bolometric (in the SDSS bandpass) "light-weighted" integrated mass profile which we present in the bottom panel of Figure \ref{fig:total_massprof}, with each stellar population weighted by its relative light contribution calculated from the MaStar SSP models \citep{Maraston_2020} over wavelength range $3622-10,354$ \AA, such that the bright and young blue populations contribute more to the profile than the fainter, older, and redder stars. Notably, the total scale lengths measured from the light-weighted profile are larger than those from the mass-weighted profile, bringing the total scale length to $R_{d} = 2.83 \pm 0.2$ kpc. This increase suggests that part of the Milky Way's apparent compactness may be explained by differences in the techniques used to measure the scale length of the Milky Way (typically through mass) and other galaxies (weighted by light), although it is still slightly smaller than the typical analog range of $R_{d} = 3.2 - 5.7$ kpc. The spiral arms of the Milky Way are also more apparent in this light-weighted profile.

The steep decrease in the total light-weighted surface mass density profile at $R \geq 13$ kpc could be interpreted as a truncation of the disk, like the sudden decrease in surface brightness observed in other disk galaxies \citep{Freeman1970,vanderKruit_1979}. MW-like disk galaxies often have typical truncation radii of $\sim 14$ kpc \citep[e.g.,][]{Martin_Navarro_2012,Martinez_2018,Diaz_2022}, thought to be linked with a critical gas density threshold for star formation.

\subsection{Integrated Spectrum} \label{sec:integrated:spectrum}

\begin{figure*}
\centering
    \includegraphics[width=\textwidth]{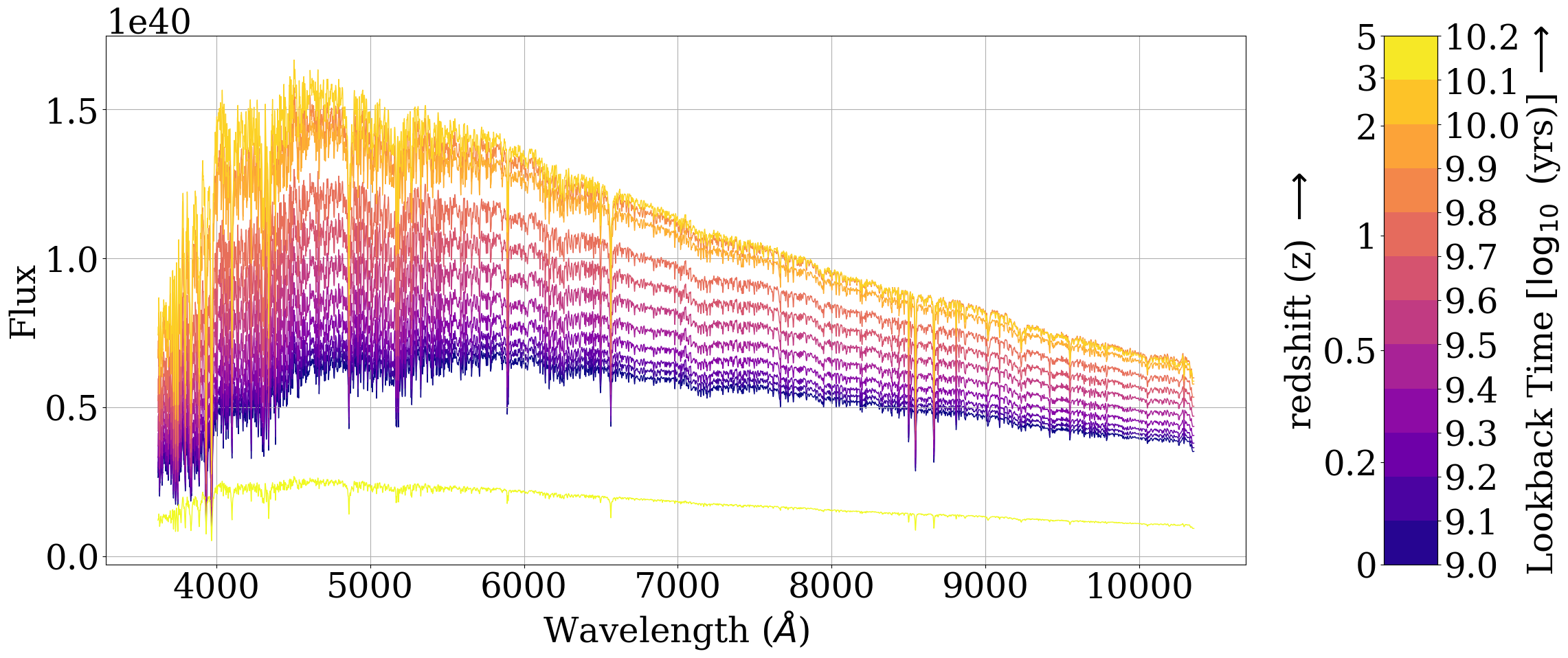}
    \caption{The integrated spectrum of the stellar populations of the Milky Way and its subsequent evolution over time, calculated from our mass estimates paired with the MaStar Simple Stellar Population spectra \citep{Maraston_2020}.}
    \label{fig:ispec_time}
\end{figure*}

\begin{figure*}
\centering
    \includegraphics[width=\textwidth]{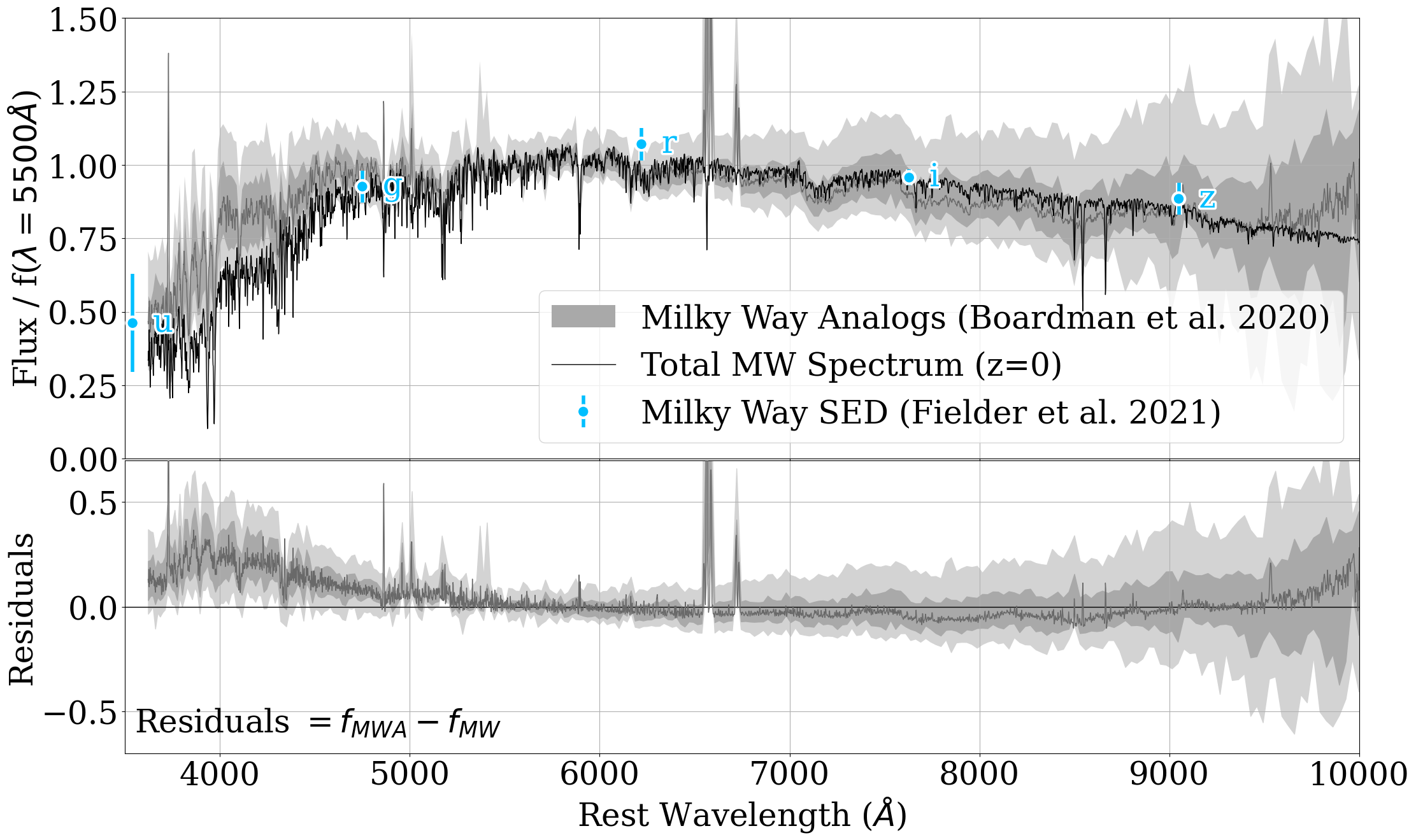}
    \caption{Top: Comparison between the present-day integrated spectrum of the Milky Way (black line), the median spectrum from the MW analog sample in MaNGA \citep[][dark gray line, with shaded regions denoting $\pm1\sigma$ and $\pm2\sigma$ spread among the spectra]{Boardman_2020} and estimates of the MW's SED from Gaussian process regression \citep[][blue points]{Fielder_2021}. Bottom: Residuals comparing the median MWA spectrum to the spectrum of the MW.}
    \label{fig:ispec_comparison}
\end{figure*}

\begin{figure}
\centering
    \includegraphics[width=0.45\textwidth]{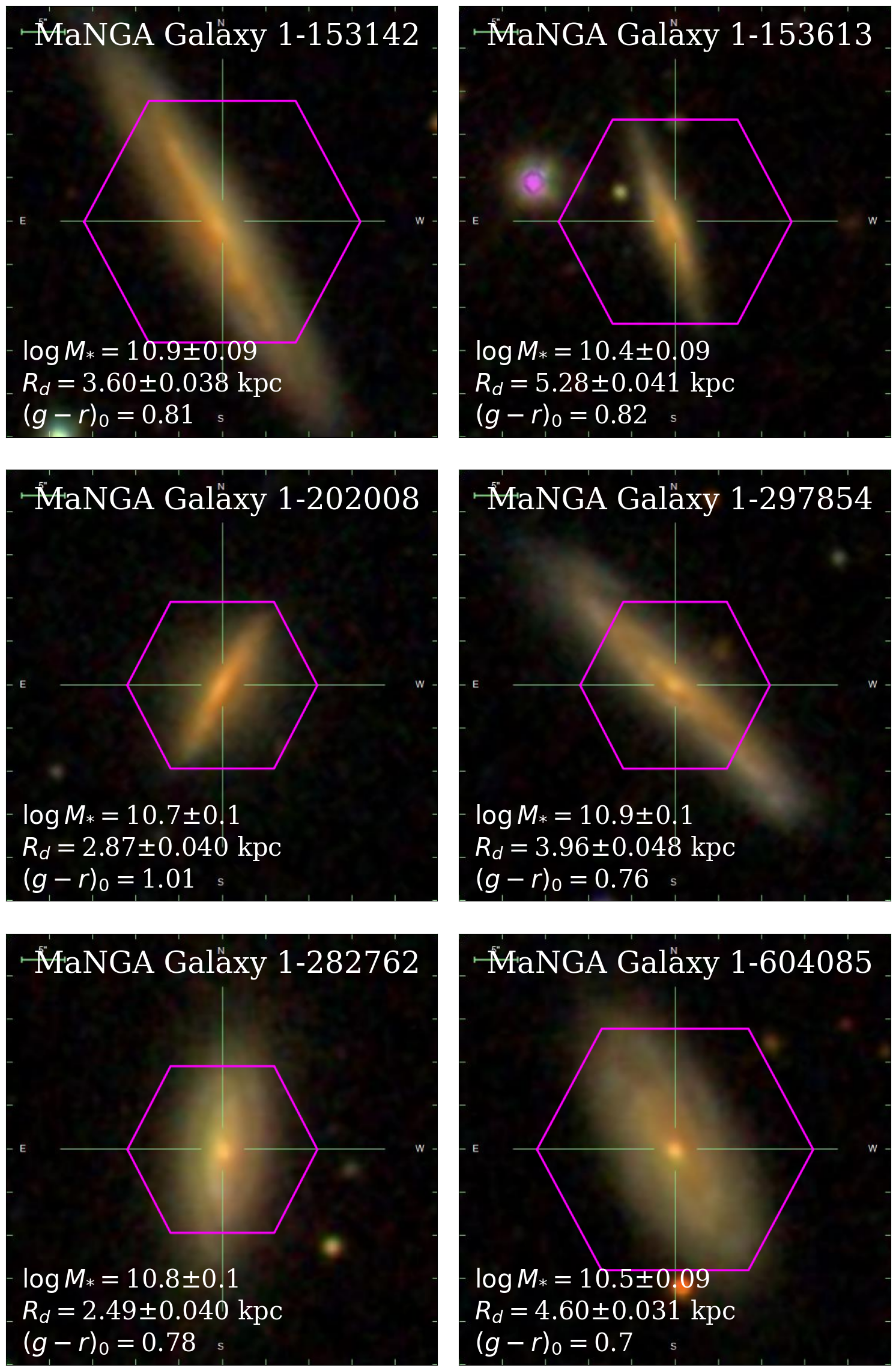}
    \caption{The six best Milky Way Analog galaxies in MaNGA determined by the similarity of their integrated spectra compared to the Milky Way spectrum in Figure \ref{fig:ispec_comparison}. Selected properties for each galaxy, including stellar mass ($\log M{*}$), disk scale length ($R_d$), and (g-r) color are listed from the NASA-Sloan Atlas \citep{Blanton_2011}.}
    \label{fig:best_analogs}
\end{figure}

The properties of other galaxies are often inferred through their integrated spectra, with large population statistics accessible through galaxy surveys including MaNGA \citep{MaNGAOverview}, CALIFA  \citep{CALIFA}, and SAMI \citep{Croom2021}. Here, we estimate the integrated spectrum of the Milky Way and qualitatively compare it to analog galaxies.

Figure \ref{fig:ispec_time} shows the integrated stellar spectrum of the Milky Way and its evolution over time. The present-day $(z=0)$ spectrum was calculated using the empirical Simple Stellar Population (SSP) models from \cite{Maraston_2020}, and summing the corresponding SSP spectrum of each stellar population in our sample weighted by the total mass contribution for each population (Figure \ref{fig:mass_params}). The MaStar SSP spectra cover a wavelength range of $3,622-10,354$ \AA, with resolution $R\sim 1800$, calculated using empirical spectra from the MaStar stellar library \citep{Yan_2019}.

Tracking the Milky Way's spectrum backwards through time involves the additional step of adjusting the snapshot age of our stellar populations to select the appropriate SSP model (i.e., the stellar population that is 6 Gyr old at present day would have been 1 Gyr old at a lookback time of 5 Gyr), and removing from the sum all populations younger than the snapshot time, which had not been born yet. 

The Milky Way's stellar continuum reached peak brightness during our second oldest age bin ($8.6 \leq$ age $\leq 10.7$ Gyr; $z \approx 2$), and has been steadily declining over time while shifting redder in color: the flux in blue wavelengths decline more quickly than the redder wavelengths. Note that this represents only the pure stellar continuum for the Milky Way; no dust attenuation or gas emission is present in this spectrum.

For a more realistic comparison, we adjust the Milky Way's spectrum for reddening using the extinction curve from \cite{Fitzpatrick_2019} and a simple dust screen model. Specifically, we adopt the curve corresponding to $R(V) = 3.1$, which they report as the typical value for the diffuse Milky Way environment. We adjust half of the light for reddening with $A_V = 1$, and keep the other half of the light in its original state in an oversimplified approximation of a face-on disk with a screen layer of dust in the mid plane.

Figure \ref{fig:ispec_comparison} compares the reddened spectrum at present day ($z=0$) to the estimated Milky Way SED constraints from \citet{Fielder_2021} (blue points) and a selection of 62 Milky Way Analog (MWA) galaxies in MaNGA from \citet{Boardman_2020}. Every spectrum has been normalized such that the flux at 5500\AA\, is equal to 1, for direct comparison on the same scale.

The Milky Way SED from \cite{Fielder_2021} was derived using Gaussian process regression on a large galaxy sample spanning GALEX, SDSS, 2MASS, and WISE coverage to predict the photometric properties of the Milky Way. We adopt the values for the optical Sloan $ugriz$ flux values \citep[][Table 1]{Fielder_2021}, plotted as the blue points with corresponding error bars on Figure \ref{fig:ispec_comparison}. This predicted SED matches our present-day Milky Way spectrum well, falling within the $1\sigma$ range for all 5 passbands.

The MaNGA MW Analog (MWA) sample from \cite{Boardman_2020} consists of 62 galaxies selected by stellar mass 
($4.6 \leq M_{*} \leq 7.2 \times 10^{10}$ M$_{\odot}$) and bulge-to-total mass ratio ($0.13 \leq B/T \leq 0.19$), i.e., their "bulge analog" subsample. For each MaNGA MWA, we constructed a single stacked spectrum by stacking spaxel spectra within 1.5 half-light radii ($R_e$). MaNGA galaxies are typically observed out to at least 1.5 $R_e$, though certain MWAs were observed out to less than this, as detailed in \citet{Boardman_2020}. We obtain spaxels' galactocentric radii from the MaNGA data analysis pipeline \citep[DAP;][]{Belfiore_2019, Westfall_2019}, which uses for $R_e$ the elliptical Petrosian half-light radius in the SDSS r-band. Prior to performing stacking, we masked out spaxels with DAP quality flags {\texttt{NOCOV}} (no coverage), {\texttt{LOWCOV}} (low coverage), {\texttt{DEADFIBER}}, {\texttt{FORESTAR}} (foreground star) or {\texttt{DONOTUSE}}.

The observed wavelengths for each MWA spectrum have been shifted to rest-frame using the redshift reported in the MaNGA {\texttt{dr}} catalog. The maximum redshift galaxy in this sample has $z=0.079$. The median flux at each wavelength across all 62 galaxies (the "typical" spectrum for a MWA) is plotted as the dark gray line in Figure \ref{fig:ispec_comparison}, with the $\pm1\sigma$ and $\pm2\sigma$ range in the sample as the medium gray and light gray ranges respectively. The bottom panel shows the residual differences between the MW spectrum and the MWA sample. Our Milky Way spectrum matches the MWA sample reasonably well, falling within the $\pm2\sigma$ at all but the bluest wavelengths ($\lambda \leq 5000$ \AA). There is more variation within the MWA sample at red wavelengths than blue. The MW spectrum deviates most from the MWA sample at blue wavelengths ($\lambda \leq 5000$ \AA), which is consistent with the fact that most of the MWAs are less massive than the Milky Way, and possibly amplified by the simplified prescription for dust extinction included in our spectrum.

Out of the 62 galaxies in the MWA sample, the six closest matches to the Milky Way's integrated spectrum (determined via a $\chi^2$ calculation) are shown in Figure \ref{fig:best_analogs}. Morphologically, they all resemble spiral galaxies with red colors, supporting the idea that the Milky Way belongs to the "red spiral" category of galaxies \citep[e.g.,][]{Licquia_2015, Fielder_2021}.
Red spirals are characterized by their red optical colors generally comparable to early-type galaxies, and are often massive galaxies that feature a prominent bar \citep{Masters_2010}.

\subsection{Spectral Line Indices}
\label{sec:integrated:indices}

Spectral line indices can be a useful tool for measuring the strength of absorption lines in a galaxy's spectrum, and are known to be correlated with stellar population properties such as age and metallicity \citep[e.g.,][]{Burstein_1984,Worthey_1994,Worthey_1997, Trager_1998}. In Table \ref{tab:lick_indices}, we report measurements of several absorption line indices for our present-day Milky Way spectrum in Figure \ref{fig:ispec_time} measured using the publicly available MaNGA Data Analysis Pipeline \citep[DAP\footnote{\url{https://sdss-mangadap.readthedocs.io/}};][]{Westfall_2019}. As index measurements are independent of stellar continuum, the un-reddened spectrum was used for these measurements. The indices were measured using the \texttt{AbsorptionLineIndices} class and the waveband definitions provided in the MaNGA DAP \citep[][see their Table 4]{Westfall_2019}, which concatenates index definitions from a variety of literature \citep[][]{Trager_1998,Worthey_1997,Serven2005,Cenarro_2001,Conroy_2012,Spiniello_2012,Spiniello_2014,LaBarbera_2013}. The index measurement $\mathcal{I}$ is defined as:

\begin{equation}
    \label{eq:spectral_index}
    \mathcal{I} =
    \begin{cases}
    S(1-f/C) & \text{for angstrom units}\\
    -2.5\log[\langle f/C \rangle] & \text{for magnitude units}\\
    \end{cases}
\end{equation}

where $f$ is the spectrum flux density, $C$ is the linear continuum approximation defined using two side bands (“blue” and “red”), and $S$ is a discrete sum over wavelength \citep[][their equations 8, 11, 22]{Westfall_2019}.

Selected index measurements for the present-day MW spectrum are presented in Table \ref{tab:lick_indices}, and the full table is available for download with the spectra as described at the end of Section \ref{sec:integrated:spectrum}. 

\begin{figure}
    \centering
    \includegraphics[width=0.45\textwidth]{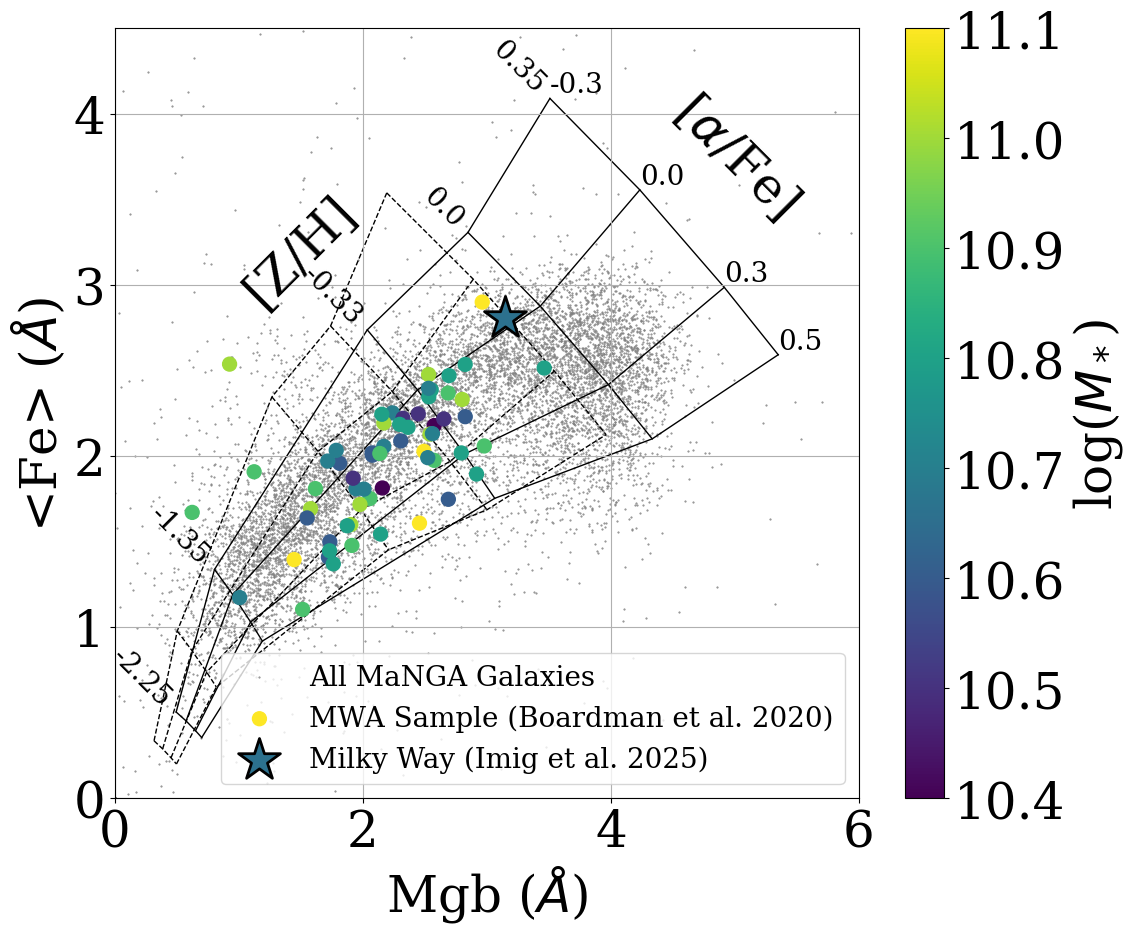}
    \caption{Spectral Line Indices Mgb and $<$Fe$> = $(Fe5270 + Fe5335)/2 for the present-day Milky Way spectrum (star point), the MWA sample \citep[large circles;][]{Boardman_2020}, and the entire MaNGA sample \citep[gray dots;][]{Westfall_2019}. Reference values from the SSP models of \citet{TMJ_2011} are overplotted as the grid lines for a population of age 10 Gyr (solid lines) and 1 Gyr (dashed lines).}
    \label{fig:indices}
\end{figure}

One such application of the spectral index measurements is plotted in Figure \ref{fig:indices}, showing the relationship between an $\alpha$-element index (MgB) and the average iron index ($<$Fe$> = $(Fe5270 + Fe5335)/2). This relation can be used to investigate the overall $\alpha$-enhancement of a galaxy, which is generally linked to star formation efficiency, and how it compares to other galaxies as a function of stellar mass \citep[e.g.,][]{Thomas_2005, Fraser_McKelvie_2018}. In Figure \ref{fig:indices}, our MW spectrum is plotted as the star point, the MW analog sample is plotted as the large circles, and the broader MaNGA sample as small circles. Index measurements for the MaNGA galaxies are retrieved from the \texttt{SPECINDEX$\_$1RE} column of the MaNGA Data Anaylsis Pipeline catalog, which provides the median index measurement across all spaxels within one effective radius for each galaxy. Reference values from the SSP models of \citet{TMJ_2011} are overplotted as the grid lines for a population of age 10 Gyr. 

Based on these measurements, the Milky Way appears both more metal-rich and more $\alpha$-enhanced than most of its peers. However, as with many of our results, the metallicity limit of [M/H] $\geq -0.7$ for low-$\alpha$ stars in our sample may bias these results.

\begin{table}
    \centering
    \begin{tabular}{l l l l}
    Index & $\lambda$ Window (\AA) & Units & Measurement \\
    \hline \hline
    CN1 & 4143.3-4178.3 & mag & 0.02 \\
    CN2 & 4143.3-4178.3 & mag & 0.06 \\
    Ca4227 & 4223.4-4235.9 & \AA & 1.43 \\
    G4300 & 4282.6-4317.6 & \AA & 5.16 \\
    Fe4383 & 4370.4-4421.6 & \AA & 4.7 \\
    Ca4455 & 4453.4-4475.9 & \AA & 1.53 \\
    Fe4531 & 4515.5-4560.5 & \AA & 3.56 \\
    C24668 & 4635.3-4721.6 & \AA & 4.13 \\
    Hb & 4849.2-4878.0 & \AA & 2.07 \\
    Fe5015 & 4979.1-5055.4 & \AA & 5.59 \\
    Mg1 & 5070.5-5135.6 & mag & 0.07 \\
    Mg2 & 5155.6-5198.1 & mag & 0.19 \\
    Mgb & 5161.6-5194.1 & \AA & 3.14 \\
    Fe5270 & 5247.1-5287.1 & \AA & 2.88 \\
    Fe5335 & 5313.6-5353.6 & \AA & 2.73 \\
    Fe5406 & 5389.0-5416.5 & \AA & 1.81 \\
    Fe5709 & 5698.2-5722.0 & \AA & 1.03 \\
    Fe5782 & 5778.2-5798.2 & \AA & 0.84 \\
    NaD & 5878.5-5911.0 & \AA & 2.8 \\
    TiO1 & 5938.3-5995.8 & mag & 0.03 \\
    TiO2 & 6191.3-6273.9 & mag & 0.06 \\
    HDeltaA & 4084.7-4123.4 & \AA & -1.49 \\
    HGammaA & 4321.0-4364.7 & \AA & -4.29 \\
    HDeltaF & 4092.2-4113.4 & \AA & 0.68 \\
    HGammaF & 4332.5-4353.5 & \AA & -0.66 \\
    CaHK & 3900.6-4004.6 & \AA & 18.29 \\
    CaII1 & 8486.3-8515.3 & \AA & 1.34 \\
    CaII2 & 8524.3-8564.4 & \AA & 3.48 \\
    CaII3 & 8644.4-8684.4 & \AA & 2.87 \\
    Pa17 & 8463.3-8476.3 & \AA & 0.28 \\
    Pa14 & 8579.4-8621.4 & \AA & 0.36 \\
    Pa12 & 8732.4-8774.4 & \AA & 0.42 \\
    MgICvD & 5165.0-5220.0 & \AA & 5.09 \\
    NaICvD & 8177.0-8205.0 & \AA & 0.47 \\
    MgIIR & 8801.9-8816.9 & \AA & 0.67 \\
    FeHCvD & 9905.0-9935.0 & \AA & 0.4 \\
    NaI & 8170.7-8236.4 & \AA & 1.0 \\
    bTiO & 4759.8-4801.3 & mag & 0.02 \\
    aTiO & 5446.5-5601.6 & mag & 0.01 \\
    CaH1 & 6359.3-6403.5 & mag & 0.0 \\
    CaH2 & 6776.9-6901.9 & mag & 0.02 \\
    NaISDSS & 8182.2-8202.3 & \AA & 0.61 \\
    TiO2SDSS & 6191.3-6273.9 & mag & 0.06 \\  
    \end{tabular}
    \caption{Selected spectral absorption line index measurements for the present-day Milky Way integrated spectrum in Figure \ref{fig:ispec_time}. The full collection of index measurements is available for download with the spectra as described at the end of Section \ref{sec:integrated:spectrum}.}
    \label{tab:lick_indices}
\end{table}

\subsection{Galaxy Color-Mass Diagram} 
\label{sec:integrated:colors}

The photometric colors of the Milky Way are impossible to observe directly from our inside perspective, and have historically only been estimated using analog galaxies or machine learning models \citep{Mutch_2011, Licquia_2015, Fielder_2021}. The general consensus from these studies is that the Milky Way falls in the relatively underpopulated "green valley" region of the optical Galaxy Color-Mass Diagram, thought to be short-lived transition state as a galaxy evolves from the Blue Cloud onto the Red Sequence \citep{Strateva_2001,Baldry_2004}.

Using the integrated spectra of the Milky Way in Figure \ref{fig:ispec_time}, reddened with the extinction law from \cite{Fitzpatrick_2019}, we can measure the evolution of the observed $(g-r)$ color of the Milky Way over time. The $(g-r)$ color is computed from the spectrum using:

\begin{equation}
    g = -2.5 \log_{10}\bigg(\frac{\int \lambda f_{\lambda} R_g d\lambda}{\int \lambda f_{AB} R_g d\lambda}\bigg)
    \label{eq:color_g}
\end{equation}

\begin{equation}
    r = -2.5 \log_{10}\bigg(\frac{\int \lambda f_{\lambda} R_r d\lambda}{\int \lambda f_{AB} R_r d\lambda}\bigg)
    \label{eq:color_r}
\end{equation}

\noindent where $f_{\lambda}$ is the flux of our MW integrated spectrum for each wavelength $\lambda$, and the filter response curves $R_g$ and $R_r$ are retrieved from \citet{Doi_2010} for the $g$ and $r$ SDSS photometric filters, respectively. The reference value $f_{AB}$ is defined in the AB photometric system as a constant source $f_{\nu} = 3631$ Jy such that $f_{AB} \propto f_{\nu}/\lambda$ \citep[e.g.,][]{Oke_1983,Blanton_2007}.

\begin{figure}
    \centering
    \includegraphics[width=0.5\textwidth]{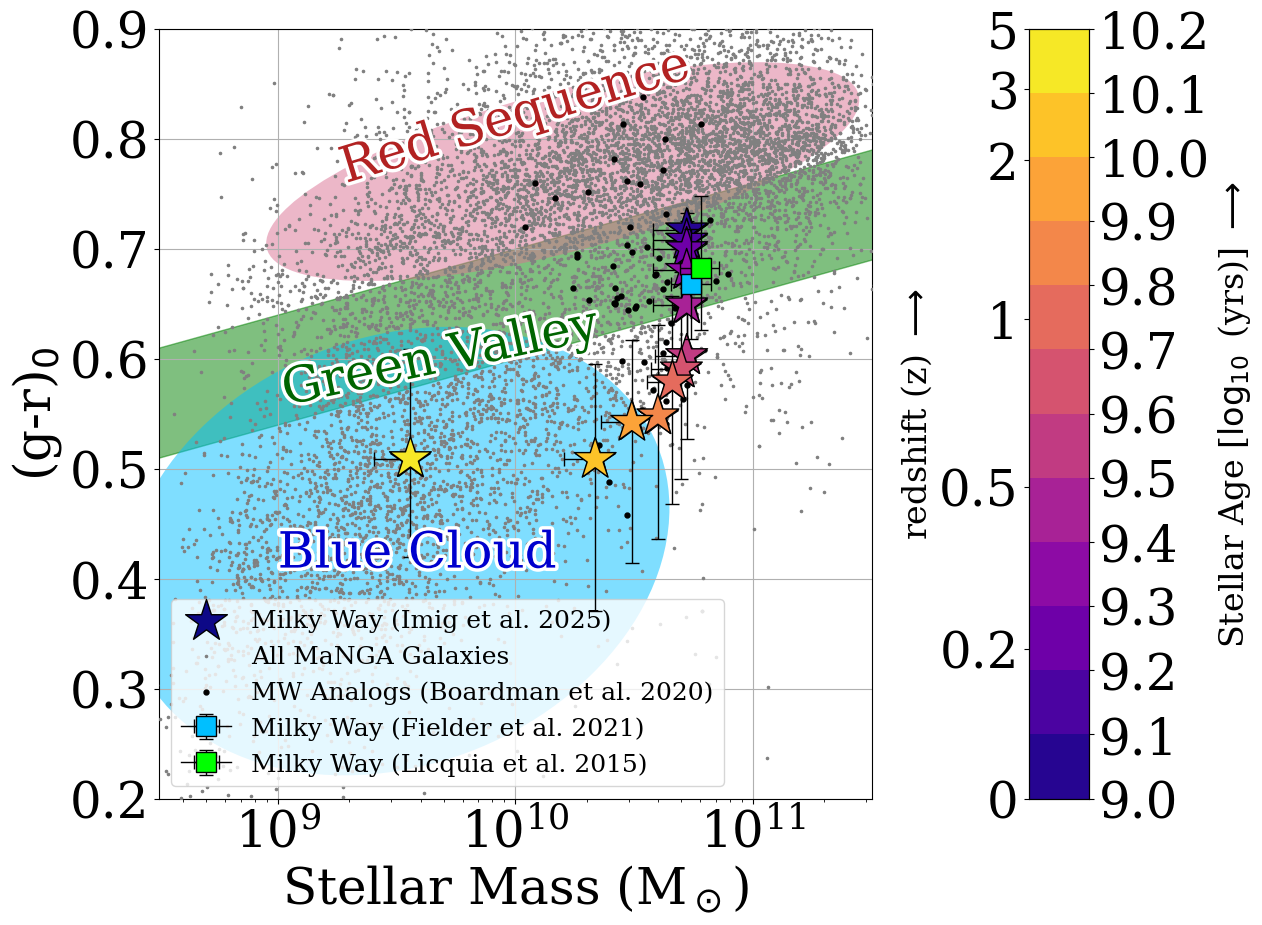}
    \caption{Galaxy color-mass diagram. The Milky Way's $(g-r)$ color and stellar mass $M_{*}$ are plotted as the star-shaped points, with their evolution over time (point color). Previous studies that have estimated the Milky Way's optical colors through observing Milky Way Analog galaxies, including \cite{Fielder_2021} ({blue} square) and \cite{Licquia_2015} ({green} square) are plotted for reference. The boundaries of the highlighted "green valley" zone are adopted from \cite{Mendel_2013}. The full MaNGA galaxy sample \citep[gray points;][]{MaNGAOverview} and the Milky Way Analog sample \citep[black points;][]{Boardman_2020} are plotted with their reported mass and color estimates from the NASA-Sloan Atlas \citep{Blanton_2011}.}
    \label{fig:cmd}
\end{figure}

In Figure \ref{fig:cmd}, we use this calculation to plot the evolution of the Milky Way on the galaxy color-mass diagram. The highlighted "green valley" region is adopted from the definition set in \cite{Mendel_2013}, corresponding to boundary lines of $(g-r)_{0} = 0.6 + 0.06(\log(M_{*}) - 10)$ and $(g-r)_{0} = 0.6 + 0.06(\log(M_{*}) - 10)+0.1$. The gray background points are the full sample of MaNGA galaxies; the colors and masses for the MaNGA galaxies are adopted from the NASA-Sloan Atlas catalog \citep[NSA;][]{Blanton_2011}, taken from the columns {\texttt{NSA$\_$SERSIC$\_$ABSMAG}} and {\texttt{NSA$\_$SERSIC$\_$MASS}} from the MaNGA {\texttt{drpall}} catalog. 

Previous estimates of the Milky Way's present-day color \citep[][]{Licquia_2015, Fielder_2021} are plotted for comparison as square points. {Using a sample of Milky Way analogs from SDSS defined by mass and star formation rates, \mbox{\cite{Licquia_2015}} estimate a color of $(g-r) = 0.682 \pm 0.061$. \mbox{\citet{Fielder_2021}} employed machine learning techniques on a large galaxy sample spanning GALEX, SDSS, 2MASS, and WISE catalogs to predict a value of $(g-r) = 0.668 \pm 0.05$ for the Milky Way. The color measurements in this study are unique in that they are driven by the stellar populations of the Milky Way itself, rather than using analog galaxies; despite the different approaches,} we measure a present-day color of $(g-r) = 0.72^{+0.02}_{-0.01}$, consistent with the prior estimates.

The MaNGA sample of Milky Way Analogs from \citet{Boardman_2020} are highlighted as the larger black points in the background. The MWA sample contains a wide variety of colors from $ 0.46 \leq (g-r) \leq 1.0$, and 33 out of 62 galaxies ($55\%$) fall within the green valley.

Unique to our method, the time evolution of the Milky Way's observed colors can be traced backwards in time using the ages of our different stellar populations and corresponding snapshot spectrum in Section \ref{sec:integrated:spectrum}. The galaxy color-mass bimodality has been observed up to a redshift of $z = 1.1$ \citep{Bell_2004,Weiner_2005}, although the sequence has been observed to evolve with redshift. Approaching present-day, the density peak of each population transitions to redder colors, consistent with the aging of stellar populations, and the relative number fraction of red sequence to blue cloud galaxies increases, suggesting that galaxies evolve from one state to the other over time through different possible quenching processes, passing through the green valley in the interim \citep{Salim_2007,Schawinski_2014,Smethurst_2015}.

The time evolution of the MW's $(g-r)$ color is shown in Figure \ref{fig:cmd} as the star points, transitioning from present day (purple) to our oldest stellar age bin (yellow). As before, each snapshot follows our stellar age bins, meaning they are logarithmically spaced in bins of $\Delta \log$(Age) $= 0.1$. The mass evolution with time is taken from our parameter results presented in Section \ref{sec:results:mass}. Our oldest snapshot (yellow point) lies in the blue cloud, and as time increases (decreasing redshift), the Galaxy gains stellar mass and shifts redder in color. This is consistent with the the evolution of this diagram observed at different redshifts \citep{Bell_2004,Weiner_2005}. In our results, the MW has only been in the green valley for the last $\approx 3$ Gyr.
A comparable figure is shown in Zasowski et al. ({\it in prep}), using isochrone-based integrated magnitudes, and leads to very similar conclusions.

We note that the simple reddening approximation described previously does not evolve with time when used here. The dust content in a galaxy does vary significantly over time \citep[e.g.,][]{Galliano_2021}, so our color estimates for older snapshots should be regarded as an upper limit. 

Table \ref{tab:colormass} presents the Milky Way's stellar mass and color as a function of time used in Figure \ref{fig:cmd}.

\begin{table}
    \centering
    \begin{tabular}{cccc}
    log(age) & redshift (z) & $(g-r)$ & $M_{*} (\times 10^{10} M_{\odot}$) \\\hline \hline
    9.0& 0.0& 0.72$^{+0.02}_{-0.01}$ & $5.27^{+0.1}_{-1.5}$\\
    9.2& 0.12& 0.71$^{+0.02}_{-0.0}$ & $5.27^{+0.1}_{-1.5}$\\
    9.2& 0.15& 0.7$^{+0.02}_{-0.0}$ & $5.26^{+0.1}_{-1.5}$\\
    9.4& 0.18& 0.68$^{+0.03}_{-0.02}$ & $5.26^{+0.1}_{-1.5}$\\
    9.4& 0.24& 0.65$^{+0.04}_{-0.04}$ & $5.24^{+0.1}_{-1.5}$\\
    9.5& 0.33& 0.6$^{+0.08}_{-0.08}$ & $5.16^{+0.1}_{-1.4}$\\
    9.6& 0.43& 0.59$^{+0.09}_{-0.1}$ & $4.9^{+0.1}_{-1.3}$\\
    9.7& 0.6& 0.58$^{+0.08}_{-0.11}$ & $4.5^{+0.1}_{-1.0}$\\
    9.8& 0.87& 0.55$^{+0.08}_{-0.11}$ & $3.91^{+0.09}_{-0.9}$\\
    9.9& 1.43& 0.54$^{+0.07}_{-0.13}$ & $3.03^{+0.07}_{-0.8}$\\
    10.0& 2.73& 0.51$^{+0.09}_{-0.14}$ & $2.13^{+0.04}_{-0.6}$\\
    10.1& 5.0& 0.51$^{+0.08}_{-0.09}$ & $0.35^{+0.0}_{-0.5}$\\
    \end{tabular}
    \caption{Evolution of the Milky Way's $(g-r)$ color and total stellar mass over time from Figure \ref{fig:cmd}.}
    \label{tab:colormass}
\end{table}

\subsection{Total Star Formation History} \label{sec:integrated:global_sfh}

Earlier in Section \ref{sec:results:mass}, we present the {total initial stellar} mass from each stellar population in our sample as a function of age, metallicity, and $\alpha$-element abundances summed over the whole Galactic disk. This can easily be displayed as a star formation history by summing over the bins in metallicity, resulting in the mass contribution as a function of only stellar age.

\begin{figure}
    \centering
    \includegraphics[width=0.5\textwidth]{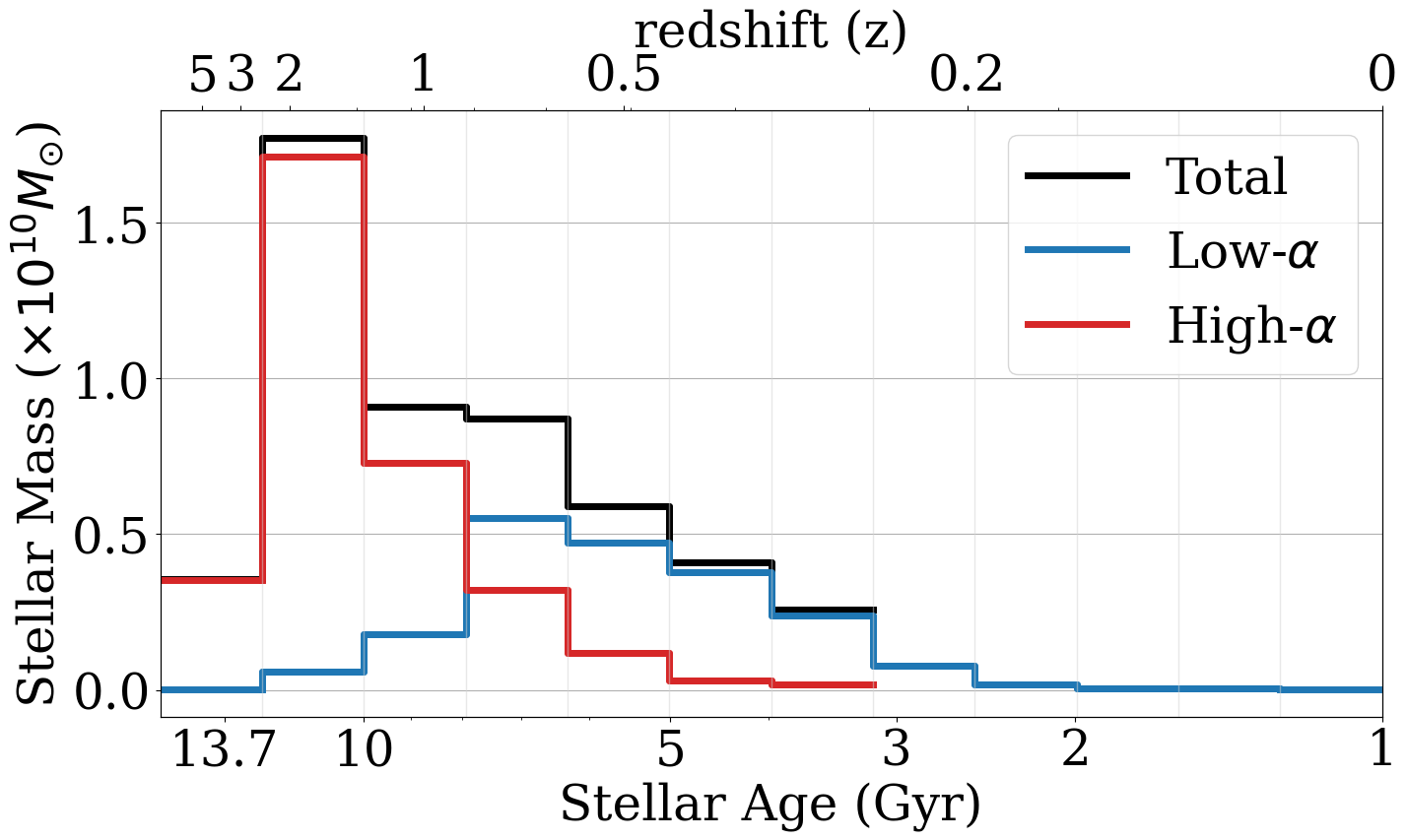}
    \includegraphics[width=0.5\textwidth]{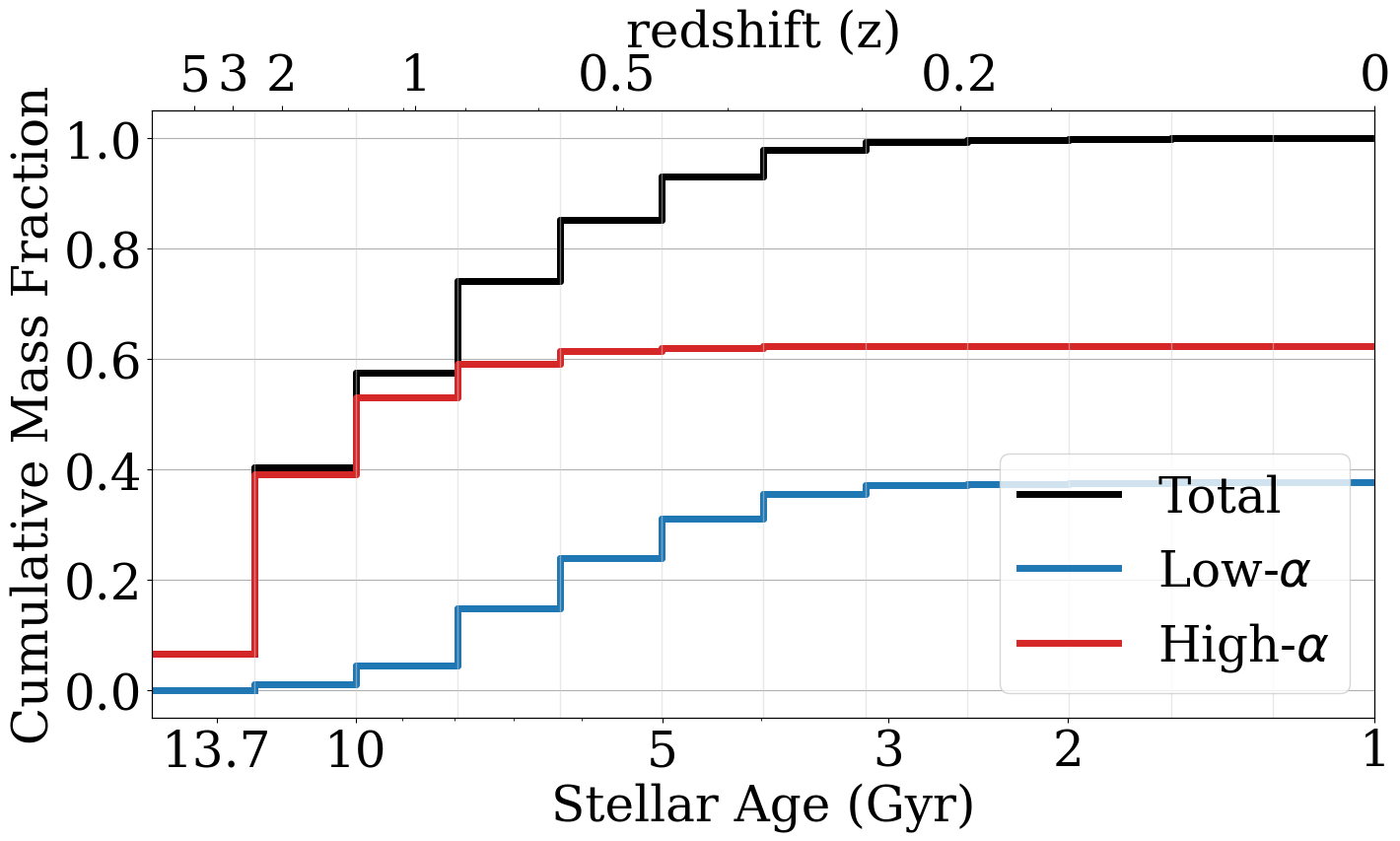}
    \caption{{Top:} The star formation history of the Milky Way, for the high-$\alpha$ population (red line), low-$\alpha$ population (blue line), and total population (black line) calculated from our total mass results and summed over all stellar metallicities. {Bottom: The cumulative star formation history, showing the cumulative fraction of mass as a function of stellar population age.} }
    \label{fig:sfh}
\end{figure}

The star formation history of the Milky Way is shown in {the top panel of} Figure \ref{fig:sfh}, split into the high-$\alpha$ population (red line), the low-$\alpha$ population (blue line) and the total population (black line) as mass contribution over time (see Zasowski et al. ({\it in prep}) for a similar figure split by Galactic radius).  {The bottom panel of Figure \mbox{\ref{fig:sfh}} shows the cumulative mass fraction as a function of time.} By the end of our oldest age bin ({$\log(\rm{age}) \geq$ 10.1}), the Milky Way had only assembled a small fraction of its mass of $3.54 \pm 0.47 \times 10^{9} M_{\odot}$. {Within the first 3 Gyr of the Universe, it had assembled around 40\% of its current-day mass, the majority (but not all) of this being high-$\alpha$ stars.} The total star formation rate declines over time. 

The high-$\alpha$ stars were formed early on during an efficient burst and quickly declining star formation rate, whereas the low-$\alpha$ population formed more steadily over time peaking $\sim 6$ Gyr ago. This has been observed before in the Milky Way \citep[e.g.,][]{Haywood_2013,Fantin_2019}, and is similar to the characteristic star formation history of simulated EAGLE galaxies that show chemical bimodality at present day, generally ignited by an infall of gas from a merger event kick starting the formation of the low-$\alpha$ population \citep{Mackereth_2018}. These galaxies represent only $5\%$ of the total population at Milky Way comparable masses, implying that the chemical bimodality may be a rare phenomenon.

We observe significant age overlap between the low-$\alpha$ and high-$\alpha$ star formation history, which has been documented before \citep[e.g.,][]{Haywood_2013,Hayden_2017,Aguirre_2018,Gent2022, Imig_2023}. This scenario implies that the the low-$\alpha$ disk started forming stars while the high-$\alpha$ disk was still developing in different places in the Galaxy, although the age uncertainties contribute to this overlap.

There is no meaningful way to discriminate between stars formed \textit{in situ} and accreted populations in this figure, as it is all based on the present-day data and we made no orbital selection cuts to our sample. The fraction of \textit{ex situ} stars in the disk is expected to be minimal; while accreted populations represent up to $30\%$ of the mass in the Galactic halo \citep[e.g.,][]{Bonaca_2017,Mackereth_2020}, the fraction drops to less than $5\%$ in the disk \citep[e.g.,][]{Hawkins_2015,Boecker_2022,Conroy_2022}.

Our star formation history is {consistent with} previous results. {We find that the MW had assembled around 40\% of its current-day mass prior to 10 Gyr ago. The chemical evolution models of \mbox{\citet{Snaith_2015,Snaith_2022}} report a fraction of $\sim44\%$ and \mbox{\citet{Khoperskov_2024}} report 40\% for that same time frame.} Using globular clusters, \citep{Kruijssen_2019} estimated that the Milky Way assembled $25\%$ of its mass by $z=3$ and $50\%$ by $z=1.5$, which is consistent with our {interpolated} estimates of $36\%$ and $56\%$ respectively. \citet{Bernard_2018} used HST photometry and a CMD-fitting technique to measure the star formation rate of the central Milky Way, and report that $80\%$ of the stars formed prior to $8$ Gyr ago. This is larger than our estimate of $57\%$ of its total mass before 8 Gyr, but consistent given the fact that their analysis only focused on the central Galaxy while ours is integrated over the entire disk. They find only 10\% of stars are younger than 5 Gyr, compared to 7\% from our results.  

Comparison to extragalactic populations may help explain some of these discrepancies. The presence of a strong bar, like that which is seen in the Milky Way, is often linked to a lower star formation rate and early quenching of star formation in nearby spiral galaxies \citep{Fraser_McKelvie_2019,Fraser_McKelvie_2020}{.} Following this trend, the Milky Way would be expected to have a lower star formation rate than its peers of similar mass. 

\subsection{Evolution with time} \label{sec:integrated:movie}

\begin{figure*}
    \begin{interactive}{animation}{MW_movie.mp4}
\includegraphics[width=0.9\textwidth]{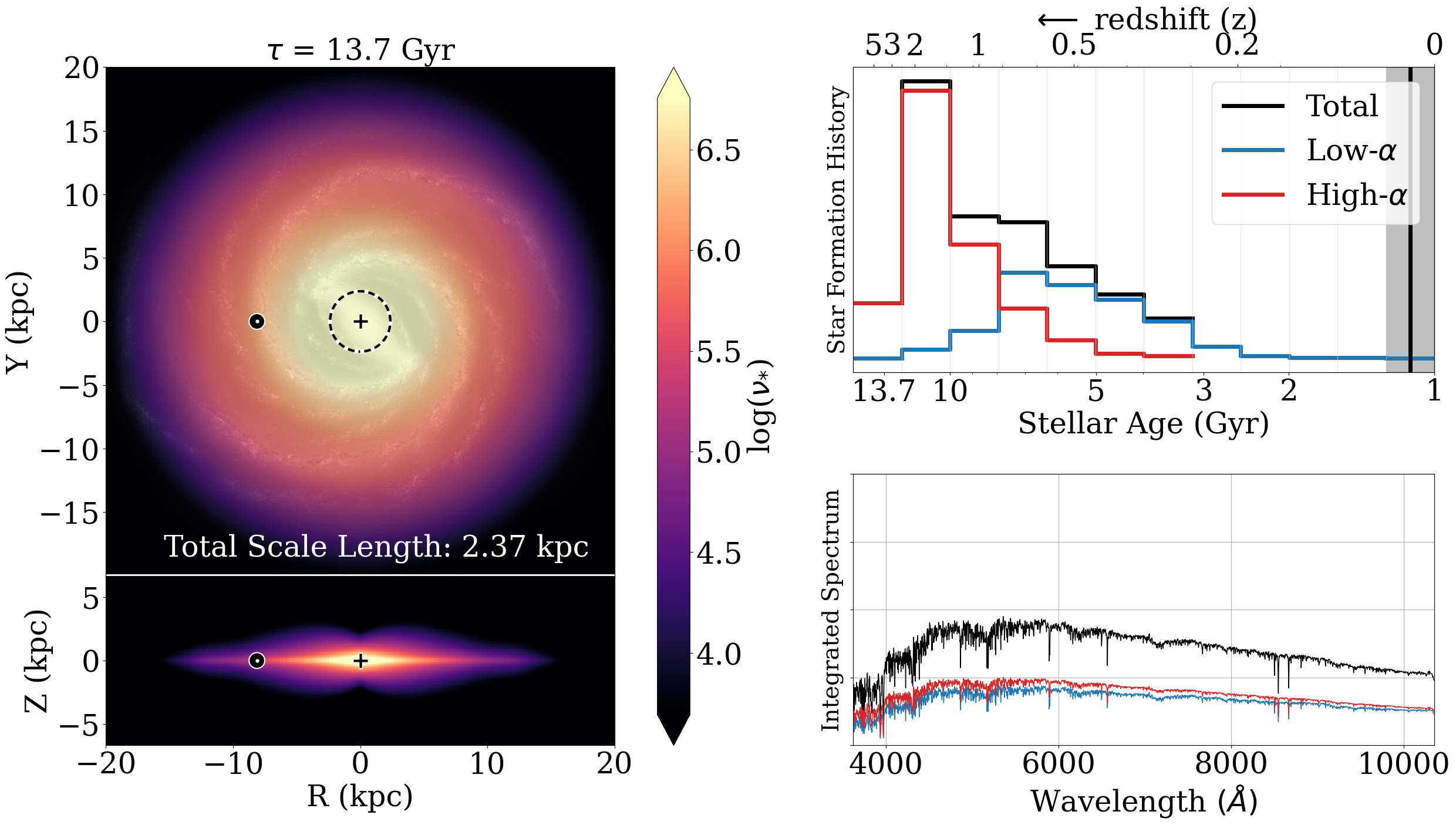}
    \end{interactive}
    \caption{An animation showing the time evolution of the density structure of the Milky Way disk {over time}. {Left:} The {stellar number} density {distribution} as a face-on view ($X-Y$ plane; top) and edge on view ($R-Z$ plane, bottom) of the disk. The location of the Sun is marked by the solar symbol ($\odot$) and the center of the Galaxy is marked by the plus sign ($+$). Strictly for visual interest, Robert Hurt's (SSC/Caltech) artistic rendition of the present-day Milky Way is overplotted to visualize the approximate present-day location of the Milky Way's bar and spiral arms. 
    {In the face-on view, the black dashed circle annotates the total disk scale length at the given time.}
    {Top right:} The star formation history {(same as Figure \mbox{\ref{fig:sfh}}), which serves} as a timeline for the animation. {Bottom right:} The integrated spectrum of the high-$\alpha$ population (red line), low-$\alpha$ population (blue line), and total disk (black line) is shown on the bottom-right. {The animation progresses from early times ($\tau=1.11$ Gyr after the big bang) to present-day ($\tau=13.7 Gyr$) with one frame representing each of our age bins. The Milky Way starts as a small, thick disk with a scale length of 1.38 kpc in the first frame and gradually grows an extended thin disk with a total scale length of 2.37 kpc in the present day. The integrated spectrum reaches peak brightness around $\tau=6.6$ Gyr and grows dimmer and redder over time as the stellar populations age.} {The full animated version of this figure is available in the HTML version of this article and also at:} {\url{https://github.com/astrojimig/mw_density_imig2025/blob/main/results/figures/mw_movie.gif}}}.
    \label{fig:mass_movie}
\end{figure*}

The total evolution of the Milky Way's density profile, combined with the star formation history, is shown in an animation in Figure \ref{fig:mass_movie}. 

The inside-out growth of the Galactic disk is obvious, as the scale length starts small and grows over time. In the edge-on view, the juxtaposition of the thin and thick disk shows how the thick disk dominated the Milky Way's mass at early times, but the thin disk grows as time progresses. At present-day, the thick disk only extends to $R \sim 9$ kpc, characterized by a smaller scale length than the thin disk. {The total scale length of the disk, measured in the same way as in Section \mbox{\ref{sec:integrated:scale_length}}, grows from $R_{d} = 1.58$ kpc in the oldest age bin to $R_{d} = 2.37$ kpc at present day.}

The integrated spectrum reaches peak brightness in the second snapshot, $t = 4.3$ Gyr after the Big Bang. For most of the Galaxy's history, the high-$\alpha$ population dominates the Milky Way's light, until the most recent $\sim 3$ Gyrs when the low-$\alpha$ stars begin to contribute more of the light as the high-$\alpha$ stars evolve and die.

Although this visualization aims to represent the Milky Way's evolution over time, we emphasize that this is still only a reflection of the present-day disk. This visualization does not account for the movement and redistribution of stars, through processes like radial migration or dynamical heating, that may shape the disk over time.

\section{Discussion} \label{sec:discussion}

\subsection{Implications for the formation of the MW} \label{sec:discussion:implications}

By binning stars by metallicity ([M/H]), $\alpha$-element abundances ([$\alpha$/M]), and stellar age, we have split the Milky Way disk into its component stellar populations and explored how its structural parameters and integrated properties vary over time and by chemical abundance.

The implications for how the Milky Way disk formed and evolved are revealed within these trends. For any of the parameter results, trends with stellar age, metallicity, and $\alpha$-element abundances can have different implications for the formation and evolution of the Milky Way. A trend in stellar age suggests an underlying process that evolves with time, such as a dynamical heating, stellar migration, gas inflows or outflows, or ongoing chemical enrichment. A correlation with chemical abundances, on the other hand, could imply a more intrinsic property of the disk and that the population was "born that way" - although chemistry is certainly not independent of time as different parts of the disk have different enrichment histories \citep[e.g., changes in the age-metallicity relation across the disk observed in ][]{Aguirre_2018,Feuillet2018,Vazquez_2022,Imig_2023}. Exploring differences among the $\alpha$-element groups is key for determining the origins of the thick disk, and whether the two groups have different physical origins \citep[e.g.,][]{Chiappini1997,Spitoni_2019,Xiang2022,Robin_2022} or if they are part of one continuous structure \citep[e.g.,][]{Bensby_2007,Bovy_2012,Kawata_2016,Hayden_2017,Anders_2018}.

The two measures of scale length, $h_{r,in}$ and $h_{r,out}$ vary with both time and metallicity (Figure \ref{fig:params_radial}). The break radius, $r_{break}$ appears independent of time and varies only with metallicity, with metal-poor populations characterized by a larger break radius (Figure \ref{fig:params_radial}; bottom). These trends depict that young, metal-poor populations are more radially extended than old, metal-rich population, which can only be a result of radially-dependent chemical evolution in the disk and supports an inside-out formation process \citep[e.g.,][]{Larson_1976,Matteucci_1989,Bird_2013}. The inner galaxy assembled first and enriched quickly due to the higher concentration of gas, a trend which expanded outwards to present day where large radii are forming stars from metal-poor still relatively pristine gas. This results in a negative age and chemical gradient in the disk, which is reflected in our results and in many other observations.

The radial profile broadening metric, $h_{r,in}$ and $h_{r,out}$ varies only with time (Figure \ref{fig:params_peak}). This implies that while any population is born in a sharply peaked ring-like profile, the shape of the peak broadens with time. This has previously been attributed to stellar radial migration \citep{Bovy2016b,Mackereth2017}, explaining the trend through stars moving away from their birth radius through dynamical interactions with transient density perturbations in the disk like the spiral arms or the galactic bar \citep[e.g.,][]{Schonrich_2009}. The older stars have had more time to migrate, and therefore migrate farther on average, broadening the peak of the profile. The high-$\alpha$ populations, despite starting more centrally concentrated, experienced comparable amounts of migration as the low-$\alpha$ populations of the same age.

The vertical parameters, scale height ($h_{z\odot}$) and disk flaring ($A_{flare}$) both increase with age and decrease with metallicity (Figure \ref{fig:params_vertical}). This implies that some combination of time-dependent process, such as dynamical heating, and \textit{in situ} structure determines the thickness of a disk. Consistent with the predictions of \citet{Minchev_2015}, we find the largest scale height and strongest flaring in the old, metal-poor populations. Interaction with a satellite galaxy or a merger event can lead to the heating of a disk \citep[e.g.,][]{Quinn_1993}, and such merger events contribute significantly more to disk flaring than radial migration does \citep{Minchev2012}. Radial migration is actually believed to suppress disk flaring, as migrator stars end up cooling the outer disk when mergers are present \citep{Minchev_2014,Minchev_2018}.

The trend with metallicity, on the other hand, reflects the vertical chemical gradient in the disk \citep[e.g.,][]{Carrell_2012,Hayden2014} and points towards a vertically "upside-down" disk assembly scenario where stars formed earlier are more subject to dynamical scattering, and as time goes on the disk stabilizes and becomes more rotationally supported, resulting in younger populations being thinner geometrically \citep[e.g.,][]{Bird_2013,Bird_2021}. Our results suggest a combination of effects; both dynamical heating and that older stars were "born thicker" being responsible for the scale height varying with age and metallicity.

We find the strongest flaring in the old, $\alpha$-rich populations, contrary to some previous studies \citep[e.g.,][]{Bovy2016b,Mackereth2017} but in good agreement with more recent observations \citep[e.g.,][]{Yu2021,Lian2022} and the models of \citet{Minchev_2015}. Radial migration has the effect of suppressing disk flaring in older populations \citep{Minchev_2014,Minchev_2018}, so these results are most consistent with the disk flaring produced by merger events \citep[e.e.,g]{Garcia2020} or the natural inside-out growth of galactic disks \citet[e.g.,]{Minchev_2015}. Early star formation in efficient gas "clumps" have also been shown to result in flared low-$\alpha$ disks in simulations \citep{Silva_2020}, which also reproduce the observed chemical bimodality in the disk \citep[e.g.,][]{Clarke_2019,Amarante_2020}. {Reconstructions of the Milky Way's star formation history inferred from an initial mass function measured at different locations in the disk predict that the MW would have had these star-forming clumps at early times \mbox{\citep{Zonoozi_2019}}.}

{The total star formation history peaks in our second oldest age bin, between $10 \leq \log(\textrm{age}) \leq 10.1$ (10 to 12.6 Gyr). This peak is driven almost exclusively by the high-$\alpha$ population. Comparatively, we find that the low-$\alpha$ star formation history peaks between $9.8 \leq \log(\textrm{age}) \leq 9.9$ (6.3 to 7.9 Gyr). The timing of these peaks could coincide with other important events in the Milky Way's history. For example, the Milky Way's accretion of the Gaia-Enceladus dwarf galaxy is estimated to have happened around 10 Gyr ago \mbox{\citep{Helmi_2018,Vincenzo_2019}}. Some simulations of the orbits in the local group suggest that the MW and M31 also experienced a close flyby between 7-11 Gyr ago \mbox{\citep{Bilek_2018,Banik_2022}}. Merger events and interactions like these could supply fresh gas reservoirs to the Milky Way and reignite star formation.}

The overall mass assembly and star formation history of the Milky Way are similar to that predicted by a merger event in simulated galaxies \citep[e.g.,][]{Mackereth_2018} where fresh gas is supplied to a galaxy by an infalling satellite. The age overlap between the low-$\alpha$ and high-$\alpha$ disk, however, implies that the disks evolved at least partially in parallel, which disfavors some "two-infall" sequential formation models \citep[e.g.,][]{Chiappini1997,Spitoni_2019}, but is expected in radially-dependent chemical evolution with stellar migration 
\citep[e.g.,][]{Schonrich_2009,Minchev_2013,Nidever2014}, and one of the main predictions in the star formation clump scenario (\citet{Clarke_2019}, {\mbox{\citet{eSilva_2021}}}). An age overlap could also arise under a two-infall scenario in which the accreted gas preferentially falls in at large radii \citep[e.g.,][]{chiappini_2008,Andrews_2017,Sharma_2021}. Current stellar age estimates are not yet precise enough to disentangle this history further. 
See {\mbox{\citet{zasowski2025}}} for a complementary comparison of mass assembly history inferred from simulations.

\section{Conclusions} \label{sec:conclusions}

Using the final data release of APOGEE \citep{Majewski2017} and stellar age estimates from the {\texttt{distmass}} catalog \citep{StoneMartinez_2023}, we present new measurements of the structural parameters of the Milky Way disk and explore how they vary with metallicity ([M/H]), $\alpha$-element abundances ([$\alpha$/M]) and stellar age. Our main conclusions are summarized as:

\begin{itemize}
    \item \textbf{Radial and Vertical Density Profiles}: Most high-$\alpha$ populations reasonably approximate a single exponential radial profile, while the low-$\alpha$ populations are best fit by a "donut"-shaped broken exponential radial profile, with the peak radius moving outward with decreasing metallicity. These findings are consistent with previous studies (e.g., {\mbox{\cite{Bensby_2011}}, \mbox{\cite{Cheng_2012}}}, \cite{Mackereth2017,Yu2021,Lian2022}).
    \item \textbf{Profile Broadening}: The difference between the inner-disk scale length and outer-disk scale length quantifies the broadening of a density profile. We find that the broadening is independent of metallicity but strongly correlates with stellar age, indicating that a time-dependent process like radial migration is responsible for the evolution of the density structure for each stellar population with time \citep[e.g.,][]{Mackereth2017,Lian2022}.
    \item \textbf{Vertical Flaring}: We find the strongest vertical flaring in the old, high-$\alpha$ populations, which is in apparent conflict with some studies \citep[e.g.,][]{Bovy2016b,Mackereth2017} but consistent with more recent results \cite[e.g.,][]{Yu2021,Lian2022} and the models of \citet{Minchev_2015}, likely due to the better spatial sampling of both the inner and outer galaxy in the newer data sets. Radial migration typically works to suppress flaring in the outer disk for the oldest populations \citep{Minchev_2014,Minchev_2018}, so our results are more consistent with flaring produced by a merger event in the disk \citep[e.g.,][]{Minchev_2014,Garcia2020}.
    \item \textbf{Total Mass}: We measure the total {initial} stellar mass of the disk as $M_{*} = 5.27^{+0.2}_{-1.5} \times 10^{10} M_{\odot}$. The low-$\alpha$ populations contribute $M_{*} = 1.99^{+0.1}_{-0.6} \times 10^{10} M_{\odot}$ and the high-$\alpha$ populations contribute $M_{*} = 3.28^{+0.1}_{-0.8} \times 10^{10} M_{\odot}$. {Of that total, $3.7 \times 10^{10}  M_\odot$ is contributed by luminous stars and stellar remnants while the rest is mass that was lost to stellar winds or supernovae over time.}
    \item \textbf{Total Scale Length}:  We measure the total mass-weighted scale length to be $2.74 \pm 0.2$ kpc for the low-$\alpha$ disk, $1.58 \pm 0.2$ kpc for the high-$\alpha$ disk and $2.37 \pm 0.2$ kpc for the sum population, although we emphasize that a single scale length fit oversimplifies the complex profiles of different stellar populations. In a light-weighted profile, the total scale length increases to $R_{d} = 2.83 \pm 0.2$ kpc as a result of the youngest (brightest) populations having longer scale lengths in general. This helps close the observational discrepancy suggesting that the Milky Way is too compact for its mass compared to other galaxies \citep[e.g.,][]{Bovy_2013, Licquia_2016, Boardman_2020}, although this should still be regarded as a lower limit because the youngest stars are excluded from our sample.
    \item \textbf{Integrated Colors}: We measure a present day integrated color of the Milky Way of $(g-r)_0$ = $0.72 \pm 0.02$. To our knowledge, this is the first estimate of the Milky Way disk's integrated colors directly from its stellar populations. This value is consistent with studies using Milky Way analogs or machine learning models as proxies \citep[e.g.,][] {Mutch_2011,Licquia_2015,Fielder_2021} and places the Milky Way in the "green valley" zone of the Galaxy color-mass diagram. Our results support that the present-day Milky Way is best classified as a red spiral galaxy \citep[e.g.,][]{Masters_2010}. Our estimate of the MW's colors become bluer with increasing lookback time, to a value of $(g-r)_0$ = $0.52 \pm 0.12$ at our oldest age bin more consistent with blue cloud galaxies. This is expected with the observed evolution of this relationship at higher redshifts \citep{Bell_2004,Weiner_2005}, and suggests that the Milky Way has only been in the green valley for the last $\approx$ 3 Gyr. 
    \item \textbf{Total Star Formation History}: The star formation history of the Milky Way disk increases over early times, peaks {prior to} $\geq 10$ Gyr ($z \geq 1.5$) ago and gradually declines over time. The high-$\alpha$ population formed quickly early on, peaking in our second-oldest age bin and declining quickly over time. The star formation history of the low-$\alpha$ population peaks around $7$ Gyr and decreases more slowly over time. This is similar to the behavior seen in simulations \citep{Mackereth_2018,Gebek_2022}, where chemical bimodality arises from a gas-rich merger event. We also find significant age overlap between the the low-$\alpha$ and high-$\alpha$ disks, consistent with other studies \citep[e.g.,][]{Haywood_2013, Hayden_2017, Aguirre_2018, Gent2022, Imig_2023} suggesting the co-evolution of the two disks, although the age uncertainties may contribute to this overlap.
\end{itemize}

In summary, our findings suggest that the present-day Milky Way is a red spiral galaxy characterized by a relatively small scale length. The low-$\alpha$ and high-$\alpha$ components of the disks, while overlapping in age, show different trends among their density profile parameters, reflecting the complex evolution history of our Galaxy.

\section*{Acknowledgements}

{We would like to thank the anonymous referee for their time and expertise reviewing our manuscript and for providing constructive feedback that improved this manuscript.}

J.I. gratefully acknowledges support from NSF grant AST-1909897 and STScI DDRF project D0001.82530.

J.A.H. also gratefully acknowledges support from NSF grant AST-1909897.

{N.F.B. acknowledges Science and Technologies Facilities Council (STFC) grant ST/V000861/1.}

{T.C.B acknowledges partial support from grant PHY 14-30152; Physics Frontier Center/JINA Center for the Evolution of the Elements (JINA-CEE), and from OISE-1927130: The International Research Network for Nuclear Astrophysics (IReNA), awarded by the US National Science Foundation.}

{DB is partly supported by RSCF grant 22-12-00080.}

{I.M. acknowledges support by the Deutsche Forschungsgemeinschaft under the grant MI 2009/2-1.}

{SM has been supported by the LP2021-9 Lend\"ulet grant of the Hungarian Academy of Sciences, and by the NKFIH excellence grant TKP2021-NKTA-64.}

Funding for the Sloan Digital Sky Survey IV has been provided by the Alfred P. Sloan Foundation, the U.S. Department of Energy Office of Science, and the Participating Institutions. SDSS acknowledges support and resources from the Center for High-Performance Computing at the University of Utah. The SDSS web site is \url{www.sdss.org}.

SDSS is managed by the Astrophysical Research Consortium for the Participating Institutions of the SDSS Collaboration including the Brazilian Participation Group, the Carnegie Institution for Science, Carnegie Mellon University, Center for Astrophysics | Harvard \& Smithsonian (CfA), the Chilean Participation Group, the French Participation Group, Instituto de Astrof{\'i}sica de Canarias, The Johns Hopkins University, Kavli Institute for the Physics and Mathematics of the Universe (IPMU) / University of Tokyo, the Korean Participation Group, Lawrence Berkeley National Laboratory, Leibniz Institut f{\"u}r Astrophysik Potsdam (AIP), Max-Planck-Institut f{\"u}r Astronomie (MPIA Heidelberg), Max-Planck-Institut f{\"u}r Astrophysik (MPA Garching), Max-Planck-Institut f{\"u}r Extraterrestrische Physik (MPE), National Astronomical Observatories of China, New Mexico State University, New York University, University of Notre Dame, Observat{\'o}rio Nacional / MCTI, The Ohio State University, Pennsylvania State University, Shanghai Astronomical Observatory, United Kingdom Participation Group, Universidad Nacional Aut{\'o}noma de M{\'e}xico, University of Arizona, University of Colorado Boulder, University of Oxford, University of Portsmouth, University of Utah, University of Virginia, University of Washington, University of Wisconsin, Vanderbilt University, and Yale University.

\section*{Data Availability}
The SDSS DR17 data are publicly available, and more information including data access instructions can be found at \url{https://www.sdss4.org/dr17/}. The {\texttt{distmass}} value-added catalog is available for download through the SDSS Science Archive Server (\url{https://data.sdss.org/sas/dr17/apogee/vac/apogee-distmass/}). All code required for reproducing the APOGEE selection function, the {file containing the best-fit parameter results of our density profiles}, and {all} figures in this study is provided on GitHub \url{https://github.com/astrojimig/mw_density_imig2025} {which is also on Zenodo under DOI:}{\dataset[10.5281/zenodo.16423243]{\doi{10.5281/zenodo.16423243}}.}.

\bibliography{new.ms}{}
\bibliographystyle{aasjournal}

\appendix
\setcounter{table}{0}
\renewcommand{\thetable}{A\arabic{table}}

\section{Machine-Readable Results}

The best-fit structural parameters for each stellar population are available as a downloadable FITS file on GitHub\footnote{\url{https://github.com/astrojimig/mw_density_imig2025}} and in the online Journal as a Machine Readable Table (MRT). The MRT version shown in Table  \ref{tab:best_fit_params} provides the best-fit parameters for every stellar population bin, while the FITS version contains extensively more information including the detailed bin definitions, the derived total stellar mass, the density profiles calculated at the disk midplane and the SSP spectrum for each population.

\begin{longtable*}[h!]{ccc|rrrrrr}
    \caption{Best-fit structural parameters for each mono-age-abundance stellar population. Descriptions of each parameter are outlined in Table \ref{tab:parameter_summary}. Table \ref{tab:best_fit_params} is published in its entirety in the machine-readable format. A portion is shown here for guidance regarding its form and content. }
    \label{tab:best_fit_params}
    \\ 
    \multicolumn{3}{c|}{Stellar population bin} & \multicolumn{6}{c}{Best-fit parameters} \\
    \hline \hline
    $\alpha$-group & [M/H] & $\log_{10}(\textrm{age})$ & $h_{R,\textrm{in}}$ & $h_{R,\textrm{out}}$ & $R_{\text{break}}$ & $h_{Z\odot}$ & $A_{\text{flare}}$ & $\log(\nu_{\odot})$\\
    \hline \hline
   low-$\alpha$ & $-0.65$ & $9.35$ &$-3.03^{+14.32}_{-13.87}$&$1.51^{+11.92}_{-11.21}$&$0.65^{+0.08}_{-0.06}$&$13.08^{+0.27}_{-0.16}$&$0.04^{+0.02}_{-0.01}$&$2.66^{+0.1}_{-0.14}$\\
   low-$\alpha$ & $-0.65$ & $9.45$ &$-4.72^{+22.33}_{-22.4}$&$1.57^{+9.83}_{-9.52}$&$0.99^{+0.09}_{-0.1}$&$14.55^{+0.36}_{-0.44}$&$0.04^{+0.02}_{-0.02}$&$2.63^{+0.07}_{-0.14}$\\
    \ldots & \ldots & \ldots & \ldots & \ldots & \ldots & \ldots & \ldots & \ldots \\
    low-$\alpha$ & $0.45$ & $9.75$ &$0.98^{+46.4}_{-53.3}$&$0.76^{+2.52}_{-3.54}$&$0.19^{+0.01}_{-0.01}$&$8.97^{+1.7}_{-0.85}$&$0.0^{+0.0}_{-0.0}$&$3.06^{+0.16}_{-0.84}$\\
    low-$\alpha$ & $0.45$ & $9.85$ &$1.06^{+28.75}_{-38.2}$&$0.42^{+0.6}_{-1.1}$&$0.22^{+0.01}_{-0.01}$&$10.01^{+2.17}_{-1.02}$&$0.0^{+0.0}_{-0.0}$&$3.11^{+0.1}_{-5.37}$\\
    \ldots & \ldots & \ldots & \ldots & \ldots & \ldots & \ldots & \ldots & \ldots \\
    high-$\alpha$ & $-0.95$ & $9.95$ &$1.12^{+4.04}_{-4.51}$&$1.48^{+0.51}_{-14.03}$&$1.8^{+0.13}_{-0.13}$&$4.27^{+13.63}_{-2.11}$&$0.18^{+0.02}_{-0.02}$&$2.69^{+0.05}_{-0.31}$\\
    high-$\alpha$ & $-0.95$ & $10.05$ &$3.39^{+18.6}_{-12.1}$&$1.07^{+10.0}_{-9.49}$&$1.49^{+0.08}_{-0.12}$&$8.13^{+0.26}_{-0.29}$&$0.14^{+0.02}_{-0.03}$&$2.95^{+0.25}_{-0.6}$\\
    \ldots & \ldots & \ldots & \ldots & \ldots & \ldots & \ldots & \ldots & \ldots \\
    high-$\alpha$ & $0.25$ & $9.85$ &$1.88^{+10.68}_{-5.7}$&$0.44^{+0.62}_{-0.94}$&$0.34^{+0.04}_{-0.03}$&$8.99^{+0.47}_{-1.7}$&$0.02^{+0.01}_{-0.01}$&$3.29^{+0.05}_{-8.55}$\\
    high-$\alpha$ & $0.25$ & $9.95$ &$4.66^{+2.17}_{-6.42}$&$0.84^{+7.14}_{-2.98}$&$0.35^{+0.03}_{-0.03}$&$6.82^{+0.29}_{-5.07}$&$0.04^{+0.01}_{-0.01}$&$3.29^{+0.26}_{-1.67}$\\       
\end{longtable*}

\end{document}